%% file: ML_Networking_INFOCOM20.tex
\documentclass[10pt,conference,letterpaper]{IEEEtran}

\usepackage{cite}
\usepackage{latexsym}
\usepackage{amsmath}
\usepackage{amsthm}
\usepackage{amssymb}
\usepackage{mathrsfs}
\usepackage[dvips,dvipdf]{graphicx}
\usepackage{algorithmicx}
\usepackage{multirow}
\usepackage{bbm}
\usepackage{arydshln}
\usepackage{url}
\usepackage{pict2e,picture} % picture
\usepackage{tikz}   % tikz
\usepackage{pifont} % ding
\usepackage{enumitem}
\usepackage{mathtools}
\usepackage[noend]{algpseudocode}
\usepackage{comment}
\usepackage{xcolor}
\usepackage[ruled,vlined,linesnumbered,noend]{algorithm2e}
\usepackage{caption}

\include{Complete.bib}

\input{Supporting_Preambles/symbols_commands}

\IEEEoverridecommandlockouts

\begin{document}

\title{A Sum-of-Ratios Multi-Dimensional-Knapsack Decomposition for DNN Resource Scheduling}

\author{Menglu Yu$^{1}$ \mbox{\hspace{0.4cm}} Chuan Wu$^{2}$ \mbox{\hspace{0.4cm}} Bo Ji$^{3}$ \mbox{\hspace{0.4cm}} Jia Liu$^{4}$
\\ $^{1}$Department of Computer Science, Iowa State University
\\ $^{2}$Department of Computer Science, The University of Hong Kong
\\ $^{3}$Department of Computer Science, Virginia Tech
\\ $^{4}$Department of Electrical and Computer Engineering, The Ohio State University
\thanks{
This work has been supported in part by NSF grants CAREER CNS-1943226, CCF-1758736, ONR grant N00014-17-1-2417, a Google Faculty Research Award, NSF CNS-1651947, and Hong Kong RGC grants HKU 17204619, 17208920.
}
}

%\author{Paper ID: 1570666673}

\maketitle

\input{Abstract/Abstract}
\input{Sec1_Intro/Sec1_Intro}

\input{Sec2_Related/Sec2_Related}
%\input{Sec7_Prelim/Sec7_Prelim}
\input{Sec3_Model/Sec3_Model}

\input{Sec4_Algorithm/Sec4_Algorithm}

\input{Sec5_Numerical/Sec5_Numerical}
\input{Sec6_Conclusion/Sec6_Conclusion}
\appendices
%\input{Appdx_A/Appdx_A}
%\input{Appdx_B/Appdx_B}
%\input{Appdx_C/Appdx_C}

%%%%%%%%%%%%%%%%%%%%%%%%%%%%%%%%%%%%%%%%%%
%\bibliography{IEEEabrv,./BIB/Complete,./BIB/HeavyBall,./BIB/Kevin_All,./BIB/CSMA}
%\clearpage
\bibliography{IEEEabrv,BIB/JobScheduling,BIB/DMLF,BIB/Kevin_All,BIB/Complete,BIB/ApproxAlg}

\end{document}

%% file: BIB/Complete.bib
@inproceedings{huo2017asynchronous,
	Author = {Huo, Zhouyuan and Huang, Heng},
	Booktitle = {AAAI},
	Date-Added = {2018-07-30 13:48:03 -0500},
	Date-Modified = {2018-07-30 13:48:03 -0500},
	Pages = {2043--2049},
	Title = {Asynchronous Mini-Batch Gradient Descent with Variance Reduction for Non-Convex Optimization.},
	Year = {2017}}

@inproceedings{lian2015asynchronous,
	Author = {Lian, Xiangru and Huang, Yijun and Li, Yuncheng and Liu, Ji},
	Booktitle = {Advances in Neural Information Processing Systems},
	Date-Added = {2018-07-30 13:23:31 -0500},
	Date-Modified = {2018-07-30 13:23:31 -0500},
	Pages = {2737--2745},
	Title = {Asynchronous parallel stochastic gradient for nonconvex optimization},
	Year = {2015}}

@book{Tignol_Book16,
	Author = {Jean-Pierre Tignol},
	Date-Added = {2017-07-18 18:34:56 +0000},
	Date-Modified = {2017-07-18 18:37:49 +0000},
	Edition = {2nd},
	Publisher = {World Scientific Publishing Company},
	Title = {Galois' Theory Of Algebraic Equations},
	Year = {2016}}

@article{Ekici11:Vehicular_Net,
	Author = {Georgios Karagiannis and Onur Altintas and Eylem Ekici and Geert Heijenk and Boangoat Jarupan and Kenneth Lin and Timothy Weil},
	Date-Added = {2017-07-18 07:56:38 +0000},
	Date-Modified = {2017-07-18 08:01:00 +0000},
	Journal = {IEEE Communications Surveys and Tutorials},
	Month = {April},
	Number = {4},
	Pages = {584 - 616},
	Title = {Vehicular Networking: A Survey and Tutorial on Requirements, Architectures, Challenges, Standards and Solutions},
	Volume = {13},
	Year = {2011}}

@article{Stolyar05:BackPressure,
	Author = {A. L. Stolyar},
	Date-Added = {2015-07-15 20:02:50 +0000},
	Date-Modified = {2015-07-15 20:04:44 +0000},
	Journal = {Queueing Systems},
	Number = {4},
	Pages = {401-457},
	Title = {Maximizing Queueing Network Utility Subject to Stability: Greedy Primal-Dual Algorithm},
	Volume = {50},
	Year = {2005}}

@article{Eryilmaz07:Backpressure,
	Author = {Atilla Eryilmaz and R. Srikant},
	Date-Added = {2015-06-26 14:34:31 +0000},
	Date-Modified = {2015-06-26 14:39:29 +0000},
	Journal = IEEE_J_NET,
	Month = dec,
	Number = {6},
	Pages = {1333-1344},
	Title = {Fair Resource Allocation in Wireless Networks Using Queue-Length-Based Scheduling and Congestion Control},
	Volume = {15},
	Year = {2007}}

@book{Bazaraa_Jarvis_Sherali_90:LP,
	Address = {New York},
	Author = {Mokhtar S. Bazaraa and John J. Jarvis and Hanif D. Sherali},
	Edition = {4},
	Publisher = {John Wiley \& Sons Inc.},
	Title = {Linear Programming and Network Flows},
	Year = {2010}}

@book{Bazaraa_Sherali_Shetty_93:NLP,
	Address = {New York, NY},
	Author = {Mokhtar S. Bazaraa and Hanif D. Sherali and C. M. Shetty},
	Edition = {3},
	Publisher = {John Wiley \& Sons Inc.},
	Title = {Nonlinear Programming: Theory and Algorithms},
	Year = {2006}}

@book{Benedetto06:UWB,
	Address = {New York, NY},
	Editor = {Maria-Gabriella Di Benedetto and Thomas Kaiser and Andreas F. Molisch and Ian Oppermann and Christian Politano and Domenico Procino},
	Publisher = {Hindawi Publishing Corporation},
	Title = {UWB Communication Systems: {A} Comprehensive Overview},
	Year = {2006}}

@book{Bertsekas_97:Distr,
	Address = {Belmon, MA},
	Author = {Dimitri P. Bertsekas and John N. Tsitsiklis},
	Publisher = {Athena Scientific},
	Title = {Parallel and Distributed Computation: Numerical Methods},
	Year = {1997}}

@book{Bertsekas_95:NLP,
	Address = {Belmon, MA},
	Author = {Dimitri P. Bertsekas},
	Publisher = {Athena Scientific},
	Title = {Nonlinear Programming},
	Year = {1995}}

@book{Bertsekas_Tsitsiklis_96:NDP,
	Author = {Dimitri P. Bertsekas and John N. Tsitsiklis},
	Edition = {1},
	Publisher = {Athena Scientific},
	Title = {Neuro-Dynamic Programming},
	Year = {1996}}

@book{Bertsekas_Nedic_Ozdaglar_03:Convex_Opt,
	Author = {Dimitri P. Bertsekas and Angelia Ndei\'{c} and Asuman E. Ozdaglar},
	Edition = {1},
	Publisher = {Athena Scientific},
	Title = {Convex Analysis and Optimization},
	Year = {2003}}

@book{Biglieri07:book,
	Author = {E. Biglieri and R. Calderbank and A. Constantinides and A. Goldsmith and A. Paulraj and H. V. Poor},
	Month = jan,
	Publisher = {Cambridge University Press},
	Title = {{MIMO} Wireless Communications},
	Year = {2007}}

@book{Borodin_ElYaniv98:OnlineCompet,
	Address = {Cambridge, New York, Melbourne},
	Author = {Allan Borodin and Ran El-Yaniv},
	Edition = {1},
	Publisher = {Cambridge},
	Title = {Online Computation and Competitive Analysis},
	Year = {1998}}

@book{Borwein_Lewis06:Cnvx_Opt,
	Address = {New York, NY},
	Author = {Jonathan Borwein and Adrian S. Lewis},
	Edition = {2},
	Publisher = {Springer},
	Title = {Convex Analysis and Nonlinear Optimization: Theory and Exampels},
	Year = {2006}}

@book{Boyd_Vandenberghe04:Cnvx_Opt,
	Address = {Cambridge, UK},
	Author = {Stephen Boyd and Lieven Vandenberghe},
	Publisher = {Cambridge University Press},
	Title = {Convex Optimization},
	Year = {2004}}

@book{Chung-Graham94:GraphLaplacian,
	Address = {Providence, RI},
	Author = {Fan R. K. Chung},
	Publisher = {American Mathematical Society},
	Title = {Spectral Graph Theory},
	Year = {1994}}

@book{Cottle92:MatrixSplitting,
	Address = {San Diego, CA},
	Author = {R. Cottle and J. Pang and R. Stone},
	Publisher = {Academic Press},
	Title = {The Linear Complementarity Problems},
	Year = {1992}}

@book{Cover91:Info_Thry,
	Address = {New York},
	Author = {Thomas M. Cover and Joy A. Thomas},
	Publisher = {John Wiley \& Sons, Inc.},
	Title = {Elements of Information Theory},
	Year = {1991}}

@book{Cox00:MDS,
	Address = {Boca Raton/London/New York/Washington, D.C.},
	Author = {Trevor F. Cox and Michael A. A. Cox},
	Edition = {2},
	Publisher = {Chapman and Hall/CRC},
	Title = {Multidimensional Scaling},
	Year = {2000}}

@book{Ehrgott05:MultiCriteria,
	Address = {Berlin/Heidelberg/New York},
	Author = {Matthias Ehrgott},
	Edition = {2},
	Publisher = {Springer},
	Title = {Multicriteria Optimization},
	Year = {2005}}

@book{Gao00:Duality,
	Address = {Dordrecht/Boston/London},
	Author = {David Yang Gao},
	Publisher = {Kluwer Academic Publishers.},
	Title = {Duality Principles in Nonconvex Systems: Theory, Methods and Applications},
	Year = {2000}}

@book{Haykin96:Adpt_Fltr,
	Address = {Englewood Cliffs, NJ},
	Author = {S. Haykin},
	Publisher = {Prentice-Hall},
	Title = {Adaptive Filter Theory},
	Year = {1996}}

@book{Hiriart-Urruty_Lemarechal01:Cnvx_Anl,
	Address = {Berlin},
	Author = {J.-B. Hiriart-Urruty and C. Lemar\'{e}chal},
	Publisher = {Springer-Verlag},
	Title = {Fundamentals of Convex Analysis},
	Year = {2001}}

@book{Hochbaum97:ApprxAlg,
	Address = {Boston},
	Author = {Dorit S. Hochbaum},
	Publisher = {PWS Publishing Company},
	Title = {Approximation Algorithms for {NP}-Hard Problems},
	Year = {1997}}

@book{Horn_Johnson90:Mtrx_Thry,
	Address = {New York, NY},
	Author = {Roger A. Horn and Charles R. Johnson},
	Publisher = {Cambridge University Press},
	Title = {Matrix Analysis},
	Year = {1990}}

@book{Horst_Tuy03:Global_Opt,
	Address = {Berlin/Heidelberg/New York},
	Author = {Reiner Horst and Hoang Tuy},
	Edition = {3},
	Publisher = {Springer-Verlag},
	Title = {Global Optimization: Deterministic Approaches},
	Year = {2003}}

@book{Gradshteyn_Ryzhik00:Int_Table,
	Address = {San Diego},
	Author = {I. S. Gradshteyn and I. M. Ryzhik},
	Publisher = {Academic Press},
	Title = {Table of Integrals, Series, and Products},
	Year = {2000}}

@book{Kushner_Yin03:Stoc_Appro,
	Address = {New York},
	Author = {Harold J. Kushner and G. George Yin},
	Publisher = {Springer},
	Title = {Stochastic Approximation and Recursive Algorithms and Applications},
	Year = {2003}}

@book{Lasdon70:Opt_Large,
	Address = {New York},
	Author = {Leon S. Lasdon},
	Publisher = {Macmillian},
	Title = {Optimization Theory for Large Systems},
	Year = {1970}}

@book{Luenberger73:Intro_LP_NLP,
	Address = {Reading, MA},
	Author = {David G. Luenberger},
	Publisher = {Addison-Wesley},
	Title = {Introduction to Linear and Nonlinear Programming},
	Year = {1973}}

@book{Magnus_Neudecker99:Mtrx_Diff_Calc,
	Address = {New York},
	Author = {J. R. Magnus and H. Neudecker},
	Publisher = {Wiley},
	Title = {Matrix Differential Calculus with Applications in Statistics and Economics},
	Year = {1999}}

@book{Marshall79:Majorization,
	Address = {New York},
	Author = {Albert W. Marshall and Ingram Olkin},
	Publisher = {Academic Press},
	Title = {Inequalities: Theory of Majorization and Its Applications},
	Year = {1979}}

@book{Marti05:Stochastic_Opt,
	Address = {Berlin/Heidelberg/New York},
	Author = {Krut Marti},
	Publisher = {Springer},
	Title = {Stochastic Optimization Methods},
	Year = {2005}}

@book{Metha04:Random_Mtrx,
	Address = {London, UK},
	Author = {Madan Lal Metha},
	Edition = {3},
	Publisher = {Academic Press},
	Title = {Random Matrices},
	Year = {2004}}

@book{Nemhauser99:Integer_Progr,
	Address = {New York},
	Author = {George L. Nemmhauser and Laurence A. Wolsey},
	Edition = {2},
	Publisher = {Wiley-Interscience Publication},
	Title = {Integer and Combinatorial Optimization},
	Year = {1999}}

@book{Nesterov01:InteriorPoint,
	Address = {Philadelphia, PA},
	Author = {Yurii Nesterov and Arkadii Nemirovskii},
	Edition = {3},
	Publisher = {SIAM},
	Title = {Interior-Point Polynomial Algorithms in Convex Programming},
	Year = {2001}}

@phdthesis{Bickson09:GaBP,
	Author = {Danny Bickson},
	School = {Hebrew University of Jerusalem},
	Title = {{G}aussian Belief Propagation: {T}heory and Application},
	Year = {2009}}

@phdthesis{Edelman84:thesis,
	Author = {Alan Edelman},
	School = {Massachusetts Institute of Technology},
	Title = {Eigenvalues and Condition Numbers of Random Matrices},
	Year = {1989}}

@phdthesis{Mitola00:thesis,
	Author = {{Mitola III}, J.},
	School = {KTH Royal Institute of Technology},
	Title = {Cognitive Radio: An Integrated Agent Architecture for Software Defined Radio},
	Year = {2000}}

@book{Proakis90:Digi_Com,
	Address = {New York},
	Author = {John G. Proakis},
	Publisher = {McGraw-Hill},
	Title = {Digital Communications},
	Year = {2000}}

@book{Rappaport02:WCOM,
	Address = {Upper Saddle River, NJ},
	Author = {Theodore S. Rappaport},
	Publisher = {Prentice Hall},
	Title = {Wireless Communications: Principles and Practice},
	Year = {2002}}

@book{Reed02:SDR,
	Address = {Upper Saddle River, NJ},
	Author = {J. H. Reed},
	Publisher = {Prentice Hall},
	Title = {Software Radio: A Modern Approach to Radio Engineering},
	Year = {2002}}

@book{Reed05:UWB,
	Address = {Upper Saddle River, NJ},
	Editor = {Jeffrey Hugh Reed},
	Publisher = {Prentice Hall},
	Title = {An Introduction to Ultra Wideband Communication Systems},
	Year = {2005}}

@book{Rudin76:RealAnalysis,
	Address = {New York, NY},
	Editor = {Walter Rudin},
	Publisher = {McGraw-Hill},
	Title = {Principles of Mathematical Analysis},
	Year = {1976}}

@book{Sherali_Adams99:RLT,
	Address = {Boston, MA},
	Author = {H. D. Sherali and W. P. Adams},
	Publisher = {Kluwer Academic Publishing},
	Title = {A Reformulation-Linearization-Technique for Solving Discrete and Continuous Nonconvex Problems},
	Year = {1999}}

@book{Simon85:SS,
	Address = {Rockvill, MD},
	Author = {Marvin Simon},
	Publisher = {Computer Science Press},
	Title = {Spread Spectrum Communications},
	Year = {1985}}

@book{Roman07:Linear_Algebra,
	Address = {New York, NY},
	Author = {Steven Roman},
	Publisher = {Springer},
	Title = {Advanced Linear Algebra},
	Year = {2007}}

@book{Taylor95:UWB,
	Address = {Boca Raton, FL},
	Editor = {J. D. Taylor},
	Publisher = {CRC Press},
	Title = {Introduction to {U}ltra-{W}ideband Radar Systems},
	Year = {1995}}

@book{Tse_Viswanath05:WCom,
	Address = {Cambridge, UK},
	Author = {David Tse and Pramod Viswanath},
	Publisher = {Cambridge University Press},
	Title = {Fundamentals of Wireless Communication},
	Year = {2005}}

@book{Verdu04:Random_Matrix,
	Address = {Hanover, MA},
	Author = {Antonia M. Tulino and Sergio Verd\'{u}},
	Publisher = {now Publishers Inc.},
	Title = {Random Matrix Theory and Wireless Communications},
	Year = {2004}}

@book{Verdu98:MUD,
	Address = {Cambridge, UK},
	Author = {Sergio Verd\'{u}},
	Publisher = {Cambridge University Press},
	Title = {Multiuser Detection},
	Year = {1998}}

@book{Wilkinson_Reinsch_71:Linear_Algebra,
	Address = {Berlin-Heidelberg-New York},
	Author = {J. H. Wilkinson and C. Reinsch},
	Publisher = {Springer-Verlag},
	Title = {Handbook for Automatic Computation, volumn II: Linear Algebra},
	Year = {1971}}

@book{Wyglinski10:CR_textbook,
	Author = {A. M. Wyglinski and M. Nekovee and Y. Thomas Hou},
	Note = {{ISBN}: 978-0-12-374715-0},
	Publisher = {Academic Press/Elsevier},
	Title = {Cognitive Radio Communications and Networks: Principles and Practices},
	Year = {2010}}

@phdthesis{Zheng02:Thesis,
	Author = {Lizhong Zheng},
	School = {University of California, Berkeley},
	Title = {Diversity-Multiplexing Tradeoff: A Comprehensive View of Multiple Antenna Systems},
	Year = {2002}}

@article{Adickes02:IndoorAP,
	Author = {Martin D. Adickes and Richard E. Billo and Bryan A. Norman and Sujata Banerjee and Bartholomew O. Nnaji and Jayant Rajgopal},
	Journal = {IIE Transactions},
	Number = {9},
	Pages = {823-836},
	Title = {Optimization of Indoor Wireless Communication Network Layout},
	Volume = {34},
	Year = 2002}

@article{Andrews05:IC,
	Author = {Jeffrey G. Andrews},
	Journal = IEEE_M_WC,
	Month = apr,
	Number = {2},
	Pages = {19-29},
	Title = {Interference Cancellation for Cellular Systems: {A} Contemporary Overview},
	Volume = {12},
	Year = 2005}

@article{Alamouti98:STBC,
	Author = {Siavash M. Alamouti},
	Journal = IEEE_J_JSAC,
	Month = Oct,
	Number = {8},
	Pages = {1451-1458},
	Title = {A Simple Transmit Diversity Technique for Wireless Communications},
	Volume = {16},
	Year = 1998}

@article{Alard87:OFDM,
	Author = {M. Alard and R. Lassalle},
	Journal = {EBU Technical Review},
	Month = aug,
	Number = {224},
	Pages = {168-190},
	Title = {Principles of Modulation and Channel Coding for Digital Broadcasting for Mobile Receivers},
	Year = 1987}

@article{Alba07:Multi_Criteria,
	Author = {E. Alba and B. Dorronsoro and F. Luna and A. J. Nebro and P. Bouvry and L. Hogie},
	Journal = {Elsevier Computer Communications Journal},
	Month = feb,
	Number = {4},
	Pages = {685-697},
	Title = {A cellular multi-objective genetic algorithm for optimal broadcasting strategy in metropolitan {MANETs}},
	Volume = {30},
	Year = 2007}

@article{Al-Dhahir01:STCode,
	Author = {N. Al-Dhahir},
	Journal = IEEE_J_COML,
	Month = jul,
	Number = {7},
	Pages = {304-306},
	Title = {Single-Carrier Frequency-Domain Equalization for Space-Time Block-Coded Transmissions Over Frequency-Selective Fading Channels},
	Volume = {5},
	Year = 2001}

@article{Akhtar04:LFB,
	Author = {J. Akhtar and D. Gesbert},
	Journal = IEEE_J_WCOM,
	Month = nov,
	Number = {11},
	Pages = {1959-1962},
	Title = {Extending orthogonal block codes with partial feedback},
	Volume = {3},
	Year = 2004}

@article{Athuraliya00:ScaledSubgradient,
	Author = {Sanjeewa Athuraliya and Steven Low},
	Journal = {Journal of Telecommunication Systems},
	Number = {3-4},
	Pages = {345-358},
	Title = {Optimization Flow Control with {N}ewton-Like Algorithm},
	Volume = {15},
	Year = 2000}

@article{Athuraliya00:DNewton,
	Author = {S},
	Journal = IEEE_J_WCOM,
	Month = nov,
	Number = {11},
	Pages = {1959-1962},
	Title = {Extending orthogonal block codes with partial feedback},
	Volume = {3},
	Year = 2004}

@article{Bahri05:IndoorAP,
	Author = {Abderraouf Bahri and Steven Chamberland},
	Journal = {Computer Networks},
	Pages = {856-866},
	Title = {On the Wireless Local Area Network Design Problem with Performance Guarantees},
	Volume = {48},
	Year = 2005}

@article{Baro00:STCode,
	Author = {S. Baro and G. Bauch and A. Hansmann},
	Journal = IEEE_J_COML,
	Month = jan,
	Number = {1},
	Pages = {20-22},
	Title = {Improved Codes for Space-Time Trellis-Coded Modulation},
	Volume = {4},
	Year = 2000}

@article{Bergman74:Degraded_BC,
	Author = {P. Bergman},
	Journal = IEEE_J_IT,
	Month = mar,
	Number = {3},
	Pages = {279-280},
	Title = {A Simple Converse of Broadcast Channels with Additive {G}aussian Noise},
	Volume = {20},
	Year = {1974}}

@article{Bertsekas83:PNewton,
	Author = {Demitri P. Bertsekas and E. M. Gafni},
	Journal = IEEE_J_AC,
	Month = dec,
	Number = {12},
	Pages = {1090-1096},
	Title = {Projected {N}ewton Methods and Optimization of Multi-commodity Flows},
	Volume = {28},
	Year = {1983}}

@article{Bingham90:OFDM,
	Author = {John A. C. Bingham},
	Journal = IEEE_M_COM,
	Month = may,
	Number = {5},
	Pages = {5-14},
	Title = {Multicarrier Modulation for Data Transmission: {A}n Idea Whose Time Has Come},
	Volume = {28},
	Year = {1990}}

@article{Blum03:MIMO_SH_AdHoc,
	Author = {Rick S. Blum},
	Journal = IEEE_J_JSAC,
	Month = jun,
	Number = {5},
	Pages = {793-801},
	Title = {{MIMO} capacity with interference},
	Volume = {21},
	Year = {2003}}

@article{Boche06:Stability,
	Author = {H. Boche and M. Wiczanowski},
	Journal = {EURASIP Signal Process. J. (Special Issue on Advances in Signal Processing-Assisted Cross-Layer Designs)},
	Month = aug,
	Number = {8},
	Pages = {1815-1833},
	Title = {Stability optimal transmission policy for the multiple antenna multiple access channel in the geometric view},
	Volume = {86},
	Year = {2006}}

@article{Bolcskei03:STCode,
	Author = {H. Bolcskei and M. Borgmann and A. J. Paulraj},
	Journal = IEEE_J_JSAC,
	Month = apr,
	Number = {3},
	Pages = {427-439},
	Title = {Impact of the Propagation Environment on the Performance of Space-Frequency Coded {MIMO-OFDM}},
	Volume = {21},
	Year = {2003}}

@article{Boyd05:Mtrx_Nrns,
	Author = {Stephen Boyd and Lin Xiao},
	Journal = {SIAM Journal on Matrix Analysis and Applications},
	Month = nov,
	Number = {2},
	Pages = {532-546},
	Title = {Least-Squares Covariance Matrix Adjustment},
	Volume = {27},
	Year = 2005}

@article{Bui08:RsrcAlloc,
	Author = {Loc Bui and R. Srikant and Alexander Stolyar},
	Journal = {Philosophical Transactions of the Royal Society},
	Month = mar,
	Number = {1872},
	Pages = {2059-2074},
	Title = {Optimal Resource Allocation for Multicast Sessions in Multi-hop Wireless Networks},
	Volume = {366},
	Year = 2008}

@article{Butterworth00:IndoorAP,
	Author = {Butterworth, K.S. and Sowerby, K.W. and Williamson, A.G.},
	Journal = IEEE_J_COM,
	Month = apr,
	Number = {4},
	Pages = {658-669},
	Title = {Base Station Placement for In-Building Mobile Communication Systems to Yield High Capacity and Efficiency},
	Volume = {48},
	Year = 2000}

@article{Cadambe08:IA,
	Author = {Viveck R. Cadambe and Syed Ali Jafar},
	Journal = IEEE_J_IT,
	Month = aug,
	Number = {8},
	Pages = {3425-3441},
	Title = {Interference Alignment and Degrees of Freedom of the ${K}$ User Interference Channel},
	Volume = {54},
	Year = {2008}}

@article{Caire03:DPC,
	Author = {G. Caire and {Shamai (Shitz)}, Shlomo},
	Journal = IEEE_J_IT,
	Month = jul,
	Number = {7},
	Pages = {1691-1706},
	Title = {On the Achievable Throughput of a Multiantenna {G}aussian Broadcast Channel},
	Volume = {49},
	Year = {2003}}

@article{Catreux00:Interference,
	Author = {S. Cartreux and L. J. Greenstein and P. F. Dressen},
	Journal = IEEE_J_COML,
	Month = nov,
	Number = {11},
	Pages = {334-336},
	Title = {Simulation results for an interference-limited multiple-input multiple-output cellular system},
	Volume = {4},
	Year = {2000}}

@article{Chandrasekhar08:FemtoSurvey,
	Author = {Vikram Chandrasekhar and Jeffrey G. Andrews and Alan Gatherer},
	Journal = IEEE_M_COM,
	Month = sep,
	Number = {9},
	Pages = {59-67},
	Title = {Femtocell Networks: {A} Survey},
	Volume = {46},
	Year = {2008}}

@article{Chandrasekhar09:FemtoCoverage,
	Author = {Vikram Chandrasekhar and Marios Kountouris and Jeffrey G. Andrews},
	Journal = IEEE_J_WCOM,
	Month = oct,
	Number = {10},
	Pages = {5314-5327},
	Title = {Coverage in Multi-Antenna Two-Tier Networks},
	Volume = {8},
	Year = {2009}}

@article{Chandrasekhar09:FemtoIntfManagement,
	Author = {Vikram Chandrasekhar and Jeffrey G. Andrews},
	Journal = IEEE_J_WCOM,
	Month = jul,
	Number = {7},
	Pages = {3498-3509},
	Title = {Uplink Capacity and Interference Avoidance for Two-Tier Femtocell Networks},
	Volume = {8},
	Year = {2009}}

@article{Chen06:MIMO_SH_Asymtoptic,
	Author = {Biao Chen and Michael J. Gans},
	Journal = IEEE_J_SP,
	Month = Jul,
	Number = {7},
	Pages = {2773-2783},
	Title = {{MIMO} Communications in Ad Hoc Networks},
	Volume = {54},
	Year = 2006}

@article{Cimini85:OFDM,
	Author = {{Cimini, Jr}, Leonard J.},
	Journal = IEEE_J_COM,
	Month = Jul,
	Number = {7},
	Pages = {665-675},
	Title = {Analysis and Simulation of a Digital Mobile Channel Using Orthogonal Frequency Division Multiplexing},
	Volume = {33},
	Year = 1985}

@article{Chiang05:NLP_JSAC,
	Journal = IEEE_J_JSAC,
	Month = aug,
	Number = {8},
	Title = {Special Issue on Nonlinear Optimization of Communication Systems},
	Volume = {24},
	Year = 2006}

@article{Choi04:MIMO_ZFBF,
	Author = {Lai-U Choi and Ross D. Murch},
	Journal = IEEE_J_WCOM,
	Month = Jan,
	Number = {1},
	Pages = {20-24},
	Title = {A transmit preprocessing technique for multiuser {MIMO} systems using a decomposition approach},
	Volume = {3},
	Year = 2004}

@article{Cicconetti06:WiMAX_QoS,
	Author = {Claudio Cicconeti and Luciano Lenzini and Enzo Mingozzi and Carl Eklund},
	Journal = IEEE_M_NET,
	Month = mar,
	Number = {2},
	Pages = {503-55},
	Title = {Quality of Service Support in {IEEE} 802.16 Networks},
	Volume = {20},
	Year = 2006}

@article{Chen01:STCode,
	Author = {Z. Chen and J. Yuan and B. Vucetic},
	Journal = IEEE_M_IT,
	Month = mar,
	Number = {7},
	Pages = {440-441},
	Title = {Improved Space-Time Trellis Coded Modulation Scheme on Slow {R}ayleigh Fading Channels},
	Volume = {37},
	Year = 2001}

@article{Chiang06:Decompose,
	Author = {Mung Chiang and S. H. Low and A. R. Calderbank and J. C. Doyle},
	Journal = IEEE_J_PROC,
	Month = jan,
	Number = {1},
	Pages = {255-312},
	Title = {Layering as Optimization Decomposition: A Mathematical Theory of Network Architecture},
	Volume = {95},
	Year = {2007}}

@article{Cohen97:Bilinear,
	Author = {Scott Cohen and Carlo Tomasi},
	Journal = {Technical Report, Deptment of CS, Stanford University},
	Title = {Systems of Bilinear Equations},
	Url = {ftp://reports.stanford.edu/pub/cstr/reports/cs/tr/97/1588/CS-TR-97-1588.pdf},
	Bdsk-Url-1 = {ftp://reports.stanford.edu/pub/cstr/reports/cs/tr/97/1588/CS-TR-97-1588.pdf}}

@article{Cook83:SS,
	Author = {C. Cook and H. Marsh},
	Journal = IEEE_M_COM,
	Month = mar,
	Number = {2},
	Pages = {8-16},
	Title = {An Introduction to Spread Spectrum},
	Volume = {21},
	Year = {1983}}

@article{Costa03:DPC,
	Author = {M. Costa},
	Journal = IEEE_J_IT,
	Month = may,
	Number = {3},
	Pages = {439-441},
	Title = {Writing on Dirty Paper},
	Volume = {29},
	Year = {1983}}

@article{Cover72:BC,
	Author = {Thomas M. Cover},
	Journal = IEEE_J_IT,
	Month = jan,
	Number = {1},
	Pages = {2-14},
	Title = {Broadcast Channels},
	Volume = {18},
	Year = {1972}}

@article{Csiszar78:BC-Wiretap,
	Author = {I. Csisz\'{a}r and J. K\"{o}rner},
	Journal = IEEE_J_IT,
	Month = may,
	Number = {3},
	Pages = {339-348},
	Title = {Broadcast Channels with Confidential Messages},
	Volume = {24},
	Year = {1978}}

@article{Cui04:Energy,
	Author = {Shuguang Cui and Andrea J. Goldsmith and Ahmad Bahai},
	Journal = IEEE_J_JSAC,
	Month = aug,
	Number = {6},
	Pages = {1089-1098},
	Title = {Energy-Efficiency of {MIMO} and Cooperative {MIMO} Techniques in Sensor Networks},
	Volume = {22},
	Year = 2004}

@article{Cui05:Energy,
	Author = {Shuguang Cui and Andrea J. Goldsmith},
	Journal = {EURASIP/Elsevier Signal Processing Journal, Special Issue on Advances in Signal Processing-Based Cross-Layer Designs},
	Month = aug,
	Number = {8},
	Title = {Cross-Layer Design in Energy-Constrained Networks Using Cooperative {MIMO} Techniques},
	Volume = {86},
	Year = 2006}

@article{Dabbagh06:RVQ,
	Author = {Amir D. Dabbagh and David J. Love},
	Journal = IEEE_J_IT,
	Month = may,
	Number = {5},
	Pages = {2190-2202},
	Title = {Feedback Rate-Capacity Loss Tradeoff for Limited Feedback {MIMO} Systems},
	Volume = {52},
	Year = 2006}

@article{Dai_TWCOM04,
	Author = {Huaiyu Dai and Andreas F. Molisch and H. Vicent Poor},
	Journal = IEEE_J_WCOM,
	Month = Mar,
	Number = {2},
	Pages = {442-453},
	Title = {Downlink Capacity of Interference-Limited {MIMO} Systems with Joint Detection},
	Volume = {3},
	Year = 2004}

@article{Damen02:SD_Tradeoff,
	Author = {O. M. Damen and A. Tewfik and J.-C. Belfiore},
	Journal = IEEE_J_IT,
	Month = mar,
	Number = {3},
	Pages = {753-760},
	Title = {A Construction of a Space-Time Code Based on Number Theory},
	Volume = {48},
	Year = 2002}

@article{Damen04:SMPLEX,
	Author = {O. Damen and A. Chkeif and J.-C. Belfiore},
	Journal = IEEE_J_COML,
	Month = may,
	Number = {5},
	Pages = {161-163},
	Title = {Lattice Code Decoder for Space-Time Codes},
	Volume = {4},
	Year = 2000}

@article{Dohler05:MIMO_Approx,
	Author = {Mischa Dohler and Hamid Aghvami},
	Journal = IEEE_J_WCOM,
	Month = jan,
	Number = {1},
	Pages = {30-34},
	Title = {On the Approximation of {MIMO} Capacity},
	Volume = {4},
	Year = 2005}

@article{Eggers93:MIMOChanModel,
	Author = {P. C. F. Eggers and J. Toftgard and A. M. Oprea},
	Journal = IEEE_J_JSAC,
	Month = sep,
	Number = {7},
	Pages = {1046-1057},
	Title = {Antenna Systems for Base Station Diversity in Urban Small and Micro Cells},
	Volume = {11},
	Year = 1993}

@article{Eryilmaz06:Backpressure,
	Author = {Atilla Eryilmaz and R. Srikant},
	Journal = IEEE_J_JSAC,
	Month = aug,
	Number = {8},
	Pages = {1514-1524},
	Title = {Joint Congestion Control, Routing, and {MAC} for Stability and Fairness in Wireless Networks},
	Volume = {24},
	Year = 2006}

@article{Etkin08:IC,
	Author = {R. H. Etkin and David N. C. Tse and Hua Wang},
	Journal = IEEE_J_IT,
	Month = dec,
	Number = {12},
	Pages = {5534-5562},
	Title = {Gaussian Interference Channel Capacity to Within One Bit},
	Volume = {54},
	Year = 2008}

@article{Feng07:StreamProc,
	Author = {H. Feng and Z. Liu and C. Xia and L. Zhang},
	Journal = {Performance Evaluation},
	Month = oct,
	Number = {9-12},
	Pages = {1102-1120},
	Title = {Load Shedding and Distributed Resource Control of Stream Processing Networks},
	Volume = {64},
	Year = 2007}

@article{Feng10:DJacobian,
	Author = {H. Feng and C. Xia and Z. Liu and L. Zhang},
	Journal = {Performance Evaluation},
	Month = aug,
	Number = {10},
	Pages = {1107-1122},
	Title = {Linear-speed interior-path algorithms for distributed control of information networks},
	Volume = {67},
	Year = 2010}

@article{Forsgren02:IPM,
	Author = {Anders Forsgren and Philip E. Gill and Margaret H. Wright},
	Journal = {SIAM Review},
	Month = oct,
	Number = {4},
	Pages = {525-597},
	Title = {Interior Methods for Nonlinear Optimization},
	Volume = {44},
	Year = 2002}

@article{Foschini96:VBLAST,
	Author = {G. J. Foschini},
	Journal = {Bell Labs Tech. J.},
	Number = {2},
	Pages = {41-59},
	Title = {Layered space-time architecture for wireless communication in a fading envorinment when using multi-element antennas},
	Volume = {1},
	Year = 1996}

@article{Foschini98:Limits,
	Author = {G. J. Foschini and M. J. Gans},
	Journal = {Wireless Personal Commun.},
	Month = mar,
	Pages = {311-355},
	Title = {On limits of wireless communications in a fading environment when using multiple antennas},
	Volume = {6},
	Year = {1998}}

@article{Fragouli02:STCode,
	Author = {C. Fragouli and N. Al-Dhahir and S. N. Diggavi and W. Turin},
	Journal = IEEE_J_COM,
	Month = may,
	Number = {5},
	Pages = {742-753},
	Title = {Prefiltered Space-Time {M-BCJR} Equalizer for Frequency-Selective Channels},
	Volume = {50},
	Year = {2002}}

@article{Francheschetti07:ScalingLaw,
	Author = {M. Francheschetti and O. Dousse and D. Tse and P. Thiran},
	Journal = IEEE_J_IT,
	Month = mar,
	Number = {3},
	Pages = {1009-1018},
	Title = {Closing the Gap in the Capacity of Random Wireless Networks via Percolation Theory},
	Volume = {53},
	Year = {2007}}

@article{Gallager74:Degraded_BC,
	Author = {Robert G. Gallager},
	Journal = {Problems of Information Transmission},
	Month = sep,
	Pages = {3-14},
	Title = {Coding for Degraded Broadcast Channels},
	Volume = {X},
	Year = {1974}}

@article{Gallager88:Noisy_BC_INC,
	Author = {Robert G. Gallager},
	Journal = IEEE_J_IT,
	Month = mar,
	Number = {2},
	Pages = {176-180},
	Title = {Finding Parity in Simple Broadcast Networks},
	Volume = {34},
	Year = {1988}}

@article{Ganesan02:STCode,
	Author = {G. Ganesan and P. Stoica},
	Journal = IEEE_J_SP,
	Month = feb,
	Number = {2},
	Pages = {57-60},
	Title = {Differential Modulation Using Space-Time Block Codes},
	Volume = {9},
	Year = {2002}}

@article{Giridhar05:INC,
	Author = {A. Giridhar and P. R. Kumar},
	Journal = IEEE_J_JSAC,
	Month = apr,
	Number = {4},
	Pages = {755-764},
	Title = {Computing and Communicating Functions over Sensor Networks},
	Volume = {23},
	Year = {2005}}

@article{Grossglauser02:ScalingLaw,
	Author = {M. Grossglauser and D. Tse},
	Journal = IEEE_J_NET,
	Month = aug,
	Number = {4},
	Pages = {477-486},
	Title = {Mobility Increases the Capacity of Adhoc Wireless Networks},
	Volume = {10},
	Year = {2002}}

@article{Ge09:MIMO_Pipe,
	Author = {Weiyan Ge and Junshan Zhang and Guoliang Xue},
	Note = {in submission to {\em IEEE Transaction on Wireless Communications}.},
	Title = {{MIMO}-pipe Modeling and Schduling for Interference Management in Multi-hop {MIMO} Networks}}

@article{Gesbert02:MIMOChannelModel,
	Author = {David Gesbert and H. Bolcskei and D. Gore and A. Paulraj},
	Journal = IEEE_J_com,
	Month = dec,
	Number = {12},
	Pages = {1926-1934},
	Title = {Outdoor {MIMO} Wireless Channels: {M}odels and Performance Prediction},
	Volume = {50},
	Year = 2002}

@article{Gesbert03:MIMO_Overview,
	Author = {David Gesbert and Mansoor Shafi and Dashan Shiu and Peter J. Smith and Ayman Naguib},
	Journal = IEEE_J_JSAC,
	Month = Apr,
	Number = {3},
	Pages = {281-302},
	Title = {From Theory to Practice: An Overview of {MIMO} Space-Time Coded Wireless Systems},
	Volume = {21},
	Year = 2003}

@article{Goldsmith03:MIMO_Limit,
	Author = {A. Goldsmith and S. A. Jafar and N. Jindal and S. Vishwanath},
	Journal = IEEE_J_JSAC,
	Month = jun,
	Number = {1},
	Pages = {684-702},
	Title = {Capacity limits of {MIMO} channels},
	Volume = {21},
	Year = {2003}}

@article{Gore02:AntennaSelect,
	Author = {D. Gore and A. Paulraj},
	Journal = IEEE_J_SP,
	Month = oct,
	Number = {10},
	Pages = {2580-2588},
	Title = {{MIMO} Antenna Subset Selection with Space-Time Coding},
	Volume = {50},
	Year = {2002}}

@article{Geoffrion71:Perturbation,
	Author = {A. M. Geoffrion},
	Journal = {SIAM Review},
	Month = jan,
	Number = {1},
	Pages = {1-37},
	Title = {Duality in Nonlinear Programming: A Simplified Applications-Oriented Development},
	Volume = {13},
	Year = {1971}}

@article{Gilhousen90:SS,
	Author = {K. S. Gilhousen and I. M. Jacobs and R. Padovani and {Weaver Jr}, L. A.},
	Journal = IEEE_J_JSAC,
	Month = may,
	Number = {4},
	Pages = {503-514},
	Title = {Increased Capacity Using CDMA for Mobile Satellite Communications},
	Volume = {8},
	Year = {1990}}

@article{Golden99:SMPLEX,
	Author = {G. D. Golden and C. J. Foschini and R. A. Valenzuela and P. W. Wolniansky},
	Journal = {Electronics Letter},
	Month = jan,
	Number = {1},
	Pages = {14-16},
	Title = {Detection Algorithm and Initial Laboratory Results Using {V-BLAST} Space-Time Communication Architecture},
	Volume = {35},
	Year = {1999}}

@article{Gupta00:capacity,
	Author = {P. Gupta and P. R. Kumar},
	Journal = IEEE_J_IT,
	Month = mar,
	Number = {2},
	Pages = {388-404},
	Title = {The Capacity of Wireless Networks},
	Volume = {46},
	Year = {2000}}

@article{Hamdaoui07:MIMO_DOF_JSAC,
	Author = {B. Hamdaoui and P. Ramanathan},
	Journal = IEEE_J_JSAC,
	Month = may,
	Number = {4},
	Pages = {667-677},
	Title = {Cross-layer optimized conditions for {Q}o{S} support in multi-hop wireless networks with {MIMO} links},
	Volume = {25},
	Year = {2007}}

@article{Hamdaoui07:MIMO_DOF_WCOM,
	Author = {B. Hamdaoui and P. Ramanathan},
	Journal = IEEE_J_WCOM,
	Month = nov,
	Number = {11},
	Pages = {4014-4024},
	Title = {A Cross-Layer Admission Control Framework for Wireless Ad-Hoc Networks Using Multiple Antennas},
	Volume = {6},
	Year = {2007}}

@article{Hammons00:STCode,
	Author = {A. R. Hammons and H. E. Gamal},
	Journal = IEEE_J_IT,
	Month = mar,
	Number = {2},
	Pages = {524-542},
	Title = {On the Theory of Space-Time Codes for {PSK} Modulation},
	Volume = {46},
	Year = {2000}}

@article{Han81:IC_Outerbound,
	Author = {T. S. Han and K. Kobayashi},
	Journal = IEEE_J_IT,
	Month = jan,
	Number = {1},
	Pages = {49-60},
	Title = {A New Achievable Rate Region for the Interference Channel},
	Volume = {27},
	Year = {1981}}

@article{Hande07:Inelastic,
	Author = {Prashanth Hande and Shengyu Zhang and Mung Chiang},
	Journal = IEEE_J_NET,
	Month = dec,
	Number = {6},
	Pages = {1240-1253},
	Title = {Distributed Rate Allocation for Inelastic Flows},
	Volume = {15},
	Year = {2007}}

@article{Hassibi02:SD_Tradeoff,
	Author = {Babak Hassibi and B. Hochwald},
	Journal = IEEE_J_IT,
	Month = sep,
	Number = {7},
	Pages = {1804-1824},
	Title = {High Rates Codes That Are Linear in Space and Time},
	Volume = {48},
	Year = {2002}}

@article{Hassibi07:MIMO_BC,
	Author = {Babak Hassibi and Masoud Sharif},
	Journal = IEEE_J_JSAC,
	Month = sep,
	Number = {7},
	Pages = {1333-1344},
	Title = {Fundamental Limits in {MIMO} Broadcast Channels},
	Volume = {25},
	Year = {2007}}

@article{Haykin05:cognitive,
	Author = {S. Haykin},
	Journal = IEEE_J_JSAC,
	Month = feb,
	Number = {2},
	Pages = {201-220},
	Title = {Cognitive Radio: Brain-Empowered Wireless Communications},
	Volume = {23},
	Year = {2005}}

@article{He04:Indoor_Placement,
	Author = {Jian He and Verstak, A.A. and Watson, L.T. and Stinson, C.A. and Ramakrishnan, N. and Shaffer, C.A. and Rappaport, T.S. and Anderson, C.R. and Bae, K.K. and Jing Jiang and Tranter, W.H.},
	Journal = IEEE_J_WCOM,
	Month = jun,
	Number = {6},
	Pages = {1906-1911},
	Title = {Globally Optimal Transmitter Placement for Indoor Wireless Communication Systems},
	Volume = {3},
	Year = {2004}}

@article{Heath01:AntennaSelect,
	Author = {Robert W. Heath and S. Sandhu and A. Paulraj},
	Journal = IEEE_J_COML,
	Month = apr,
	Number = {4},
	Pages = {142-143},
	Title = {Antenna Selection for Spatial Multiplexing Systems with Linear Receivers},
	Volume = {5},
	Year = {2001}}

@article{Hero03:Secure-MIMO,
	Author = {Alfred O. Hero},
	Journal = IEEE_J_IT,
	Month = dec,
	Number = {12},
	Pages = {3235-3249},
	Title = {Secure Space-Time Communication},
	Volume = {49},
	Year = {2003}}

@article{Heun02:GraphEmbedding,
	Author = {V. Heun and E. W. Mayr},
	Journal = {Journal of Algorithms},
	Pages = {51-84},
	Title = {Efficient Dynamic Embeddings of Arbitrary Binary Trees into Hypercubes},
	Volume = {43},
	Year = {2002}}

@article{Hills04:Indoor_Estimate,
	Author = {Hills, A. and Schlegel, J. and Jenkins, B.},
	Journal = IEEE_J_WCOM,
	Month = jan,
	Number = {1},
	Pages = {17-19},
	Title = {Estimating Signal Strengths in the Design of an Indoor Wireless Network},
	Volume = {3},
	Year = {2004}}

@article{Hou07:TVT,
	Author = {Y. T. Hou and Y. Shi and H. D. Sherali and J. E. Wieselthier},
	Journal = IEEE_J_VT,
	Month = may,
	Number = {3},
	Pages = {1333-1344},
	Title = {Multicast Communications in Ad Hoc Networks Using Directional Dntennas: A Lifetime-Centric Approach},
	Volume = {56},
	Year = {2007}}

@article{Hou06:TMC,
	Author = {Y. T. Hou and Y. Shi and J. Pan and S. F. Midkiff},
	Journal = IEEE_J_MC,
	Month = sep,
	Number = {9},
	Pages = {1255-1266},
	Title = {Maximizing the Lifetime of Wireless Sensor Networks Through Optimal Single-Session Flow Routing},
	Volume = {5},
	Year = {2006}}

@article{Hou06:TVT,
	Author = {Y. T. Hou and Y. Shi and H. D. Sherali},
	Journal = IEEE_J_VT,
	Month = may,
	Number = {3},
	Pages = {813-821},
	Title = {Optimal Base Station Selection for Anycast Routing in Wireless Sensor Networks},
	Volume = {55},
	Year = {2006}}

@article{Hou05:MONET,
	Author = {Y. T. Hou and Y. Shi and H. D. Sherali},
	Journal = {ACM/Springer Mobile Networks and Applications ({MONET})},
	Month = dec,
	Number = {6},
	Pages = {865-878},
	Title = {On Node Lifetime Problem for Energy-Constrained Wireless Sensor Networks},
	Volume = {10},
	Year = {2005}}

@article{Hou05:TWC,
	Author = {Y. T. Hou and Y. Shi and H. D. Sherali and S. F. Midkiff},
	Journal = IEEE_J_WCOM,
	Month = sep,
	Number = {5},
	Pages = {2579-2590},
	Title = {On Energy Provisioning and Relay Node Placement for Wireless Sensor Networks},
	Volume = {4},
	Year = {2005}}

@article{Hou06:TVT_anycast,
	Author = {Y. Thomas Hou and Yi Shi and Hanif D. Sherali},
	Journal = IEEE_J_VT,
	Month = may,
	Number = {3},
	Pages = {813-821},
	Title = {Optimal base station selection for anycast routing in wireless sensor networks},
	Volume = {55},
	Year = {2006}}

@article{Hou06:TMC_Single_Session,
	Author = {Y. Thomas Hou and Yi Shi and Jianping Pan and Scott F. Midkiff},
	Journal = IEEE_J_MC,
	Month = sep,
	Number = {9},
	Pages = {1255-1266},
	Title = {Maximizing the lifetime of wireless sensor networks through optimal single-session flow routing},
	Volume = {5},
	Year = {2006}}

@article{Hou07:TVT_Multicast,
	Author = {Y. Thomas Hou and Yi Shi and Hanif D. Sherali and J. E. Wieselthier},
	Journal = IEEE_J_VT,
	Month = may,
	Number = {3},
	Pages = {1333-1344},
	Title = {Multicast communications in ad hoc networks using directional antennas: {A} lifetime-centric approach},
	Volume = {56},
	Year = {2007}}

@article{Hou07:TWC_Variable,
	Author = {Y. Thomas Hou and Yi Shi},
	Journal = IEEE_J_WCOM,
	Month = jun,
	Number = {6},
	Pages = {2140--2148},
	Title = {Variable bit rate flow routing in wireless sensor networks},
	Volume = {6},
	Year = {2007}}

@article{Hou08:ToN_rate,
	Author = {Y. Thomas Hou and Yi Shi and Hanif D. Sherali},
	Journal = IEEE_J_NET,
	Month = apr,
	Number = {2},
	Pages = {321-334},
	Title = {Rate allocation and network lifetime problems for wireless sensor networks},
	Volume = {16},
	Year = {2008}}

@article{Hou08:JSAC_spectrum,
	Author = {Y. Thomas Hou and Yi Shi and Hanif D. Sherali},
	Journal = IEEE_J_JSAC,
	Month = jan,
	Number = {1},
	Pages = {146-155},
	Title = {Spectrum sharing for multi-hop networking with cognitive radios},
	Volume = {26},
	Year = {2008}}

@article{Hu04:MIMO_Hop_Dist,
	Author = {M. Hu and Junshan Zhang},
	Journal = {Special Issue on Mobile Ad Hoc Networks, Journal of Communications and Networks},
	Month = dec,
	Pages = {317-330},
	Title = {{MIMO} Ad Hoc Networks: Medium Access Control, Saturation Throughput, and Optimal Hop Distance},
	Year = 2004}

@article{Hu10:video_over_CRN,
	Author = {D. Hu and Shiwen Mao and Y. Thomas Hou and J. H. Reed},
	Journal = IEEE_J_JSAC,
	Note = {(to appear)},
	Title = {Scalable video multicast in cognitive radio networks}}

@article{Hurley02:CellularPlaning,
	Author = {Stephen Hurley},
	Journal = IEEE_J_VT,
	Month = mar,
	Number = {2},
	Pages = {243-253},
	Title = {Planing Effective Cellular Mobile Radio Networks},
	Volume = {51},
	Year = 2002}

@article{Jafar06:DoF-2U-IC,
	Author = {Syed Ali Jafar and Maralle Jamal Fakhereddin},
	Journal = IEEE_J_IT,
	Month = jul,
	Number = {7},
	Pages = {2637-2642},
	Title = {Degrees of Freedom for the {MIMO} Interference Channel},
	Volume = {53},
	Year = 2007}

@article{Jafar08:X_Chan_DoF,
	Author = {Syed Ali Jafar and {Shamai (Shitz)}, Shlomo},
	Journal = IEEE_J_IT,
	Month = jan,
	Number = {1},
	Pages = {151-170},
	Title = {Degrees of Freedom Region for the {MIMO} {X} Channel},
	Volume = {54},
	Year = 2008}

@article{Jafarkhani01:STCode,
	Author = {H. Jafarkhani},
	Journal = IEEE_J_COM,
	Month = jan,
	Number = {1},
	Pages = {1-4},
	Title = {A quasi orthogonal space time block code},
	Volume = {49},
	Year = 2001}

@article{Jafarkhani03:STCode,
	Author = {H. Jafarkhani and N. Seshadri},
	Journal = IEEE_J_IT,
	Month = apr,
	Number = {4},
	Pages = {937-950},
	Title = {Super-Orthogonal Space-Time Trellis Codes},
	Volume = {49},
	Year = 2003}

@article{Jiang_TCOM06,
	Author = {Jing Jiang and R. Michael Buehrer and William H. Tranter},
	Journal = IEEE_J_COM,
	Month = May,
	Number = {5},
	Pages = {789-793},
	Title = {Greedy Scheduling Performance for a Zero-Forcing Dirty Paper Coded System},
	Volume = {54},
	Year = 2006}

@article{Jiang07:MC_MIMO,
	Author = {Ming Jiang and Lajos Hanzo},
	Journal = IEEE_J_PROC,
	Month = jul,
	Number = {7},
	Pages = {1430-1469},
	Title = {Multiuser {MIMO-OFDM} for Next Generation Wireless Systems},
	Volume = {95},
	Year = {2007}}

@article{Jindal04:MIMO_BC_IWF,
	Author = {Nihar Jindal and Wonjong Rhee and Sriram Vishwanath and Syed Ali Jafar and Andrea Goldsmith},
	Journal = IEEE_J_IT,
	Month = apr,
	Number = {4},
	Pages = {1570-1580},
	Title = {Sum power iterative water-filling for multi-antenna {G}aussian broadcast channels},
	Volume = {51},
	Year = {2005}}

@article{Jindal06:BC_LFB,
	Author = {Nihar Jindal},
	Journal = IEEE_J_IT,
	Month = nov,
	Number = {11},
	Pages = {5045-5059},
	Title = {{MIMO} Broadcast Channels with Finite Rate Feedback},
	Volume = {52},
	Year = {2006}}

@article{Jindal07:BC_High_SNR,
	Author = {Nihar Jindal},
	Journal = IEEE_J_IT,
	Month = dec,
	Number = {12},
	Pages = {4787-4792},
	Title = {High {SNR} Analysis for {MIMO} Broadcast Channels: Dirty Paper Coding Versus Linear Precoding},
	Volume = {53},
	Year = {2007}}

@article{Jo09:FemtoIntfManagement,
	Author = {Han-Shin Jo and Cheol Mun and June Moon and Jong-Gwan Yook},
	Journal = IEEE_J_WCOM,
	Month = oct,
	Number = {10},
	Pages = {4906-4910},
	Title = {Interference Mitigation Using Uplink Power Control for Two-Tier Femtocell Networks},
	Volume = {8},
	Year = {2009}}

@article{Jo10:FemtoCoverage,
	Author = {Han-Shin Jo and Cheol Mun and June Moon and Jong-Gwan Yook},
	Journal = IEEE_J_WCOM,
	Month = oct,
	Number = {10},
	Pages = {2977-2982},
	Title = {Self-Optimized Coverage Coordination in Femtocell Networks},
	Volume = {9},
	Year = {2010}}

@article{Jongren04:LFB,
	Author = {G. J\"{o}ngren and M. Skoglund},
	Journal = IEEE_J_IT,
	Month = oct,
	Number = {10},
	Pages = {2473-2486},
	Title = {Quantized feedback information in orthogonal space-time block coding},
	Volume = {50},
	Year = {2004}}

@article{Jorswieck04:MIMO_SH_AdHoc_WorstCase,
	Author = {E. A. Jorswieck and H. Boche},
	Journal = {EURASIP Journal on Wireless Communications and Networking},
	Month = Feb,
	Number = {2},
	Pages = {273-285},
	Title = {Performance analysis of Capacity of {MIMO} systems under multiuser interference based on worst-case noise behavior},
	Volume = {2004},
	Year = {2004}}

@article{Kelly98:NUM,
	Author = {F. P. Kelly and A. K. Malullo and D. K. H. Tan},
	Journal = {Journal of the Operational Research Society},
	Pages = {237-252},
	Title = {Rate Control in Communications Networks: Shadow Prices, Proportional Fairness and Stability},
	Volume = {49},
	Year = 1998}

@article{Khoshnevis08:MIMO_AF_LFB,
	Author = {Behrouz Khoshnevis and Wei Yu and Raviraj Adve},
	Journal = IEEE_J_JSAC,
	Month = oct,
	Number = {8},
	Pages = {1397-1407},
	Title = {Grassmannian Beamforming for {MIMO} Amplify-and-Forward Relaying},
	Volume = {26},
	Year = {2008}}

@article{Kim07:MIMO_MultiHop,
	Author = {Seung-Jun Kim and Xiaodong Wang and Mohammad Madihian},
	Journal = IEEE_J_MC,
	Month = nov,
	Number = {11},
	Pages = {1259-1269},
	Title = {Cross-Layer Design of Wireless Multihop Backhaul Networks with Multiantenna Beamforming},
	Volume = {6},
	Year = {2007}}

@article{Kim09:WiMAXFemtoSurvey,
	Author = {Ronny Yongho Kim and Jin Sam Kwak and Kamran Etemad},
	Journal = IEEE_M_COM,
	Month = sep,
	Number = {9},
	Pages = {84-91},
	Title = {Wi{MAX} Femtocell: {R}equirements, Challenges, and Solutions},
	Volume = {47},
	Year = {2009}}

@article{Klincewicz83:DNewton,
	Author = {John G. Klincewicz},
	Journal = {Networks},
	Month = mar,
	Number = {3},
	Pages = {427-442},
	Title = {A {N}ewton Method for Convex Separable Network Flow Problems},
	Volume = {13},
	Year = {1983}}

@article{Kobayashi06:MIMO_BC_MWS,
	Author = {Mari Kobayashi and Giuseppe Caire},
	Journal = IEEE_J_JSAC,
	Number = {8},
	Pages = {1640-1646},
	Title = {An Iterative Water-Filling Algorithm for Maximum Weighted Sum-Rate of {G}aussian {MIMO-BC}},
	Volume = {24},
	Year = 2006}

@article{Kozono84:MIMOChanModel,
	Author = {S. Kozono and T. Tsuruhara and M. Sakamoto},
	Journal = IEEE_J_VT,
	Month = nov,
	Number = {4},
	Pages = {301-306},
	Title = {Base Station Polarization Diversity Reception for Mobile Radio},
	Volume = {33},
	Year = 1984}

@article{Laneman04:CC,
	Author = {J. N. Laneman and D. N.C. Tse and G. W. Wornell},
	Journal = IEEE_J_IT,
	Month = dec,
	Number = {12},
	Pages = {3062-3080},
	Title = {Cooperative Diversity in Wireless Networks: Efficient Protocols and Outage Behavior},
	Volume = {50},
	Year = 2004}

@article{Larsson02:LFB,
	Author = {E. G. Larsson and G. Ganesan and P. Stoica and W.\-H. Wong},
	Journal = IEEE_J_COML,
	Month = nov,
	Number = {11},
	Pages = {487-489},
	Title = {On the performance of orthogonal space-time block coding with quantized feedback},
	Volume = {6},
	Year = 2002}

@article{Lau04:LFB1,
	Author = {Vicent Lau and Y. Liu and T.\-A. Chen},
	Journal = IEEE_J_COM,
	Month = jan,
	Number = {1},
	Pages = {62-70},
	Title = {On the design of {MIMO} block-fading channels with feedback-link capacity constraint},
	Volume = {52},
	Year = 2004}

@article{Lau04:LFB2,
	Author = {Vicent Lau and Y. Liu and T.\-A. Chen},
	Journal = IEEE_J_IT,
	Month = sep,
	Number = {9},
	Pages = {2038-2049},
	Title = {Capacity of memoryless channels and block-fading channels with designable cardinality-constrained channel state feedback},
	Volume = {50},
	Year = 2004}

@article{Lee05:Random_Access,
	Author = {Jang-Won Lee and Mung Chiang and A. Robert Calderbank},
	Journal = IEEE_J_COML,
	Month = mar,
	Number = {3},
	Pages = {216-218},
	Title = {Jointly Optimal Congestion and Medium Access Control in Ad Hoc Wireless Networks},
	Volume = {10},
	Year = 2005}

@article{Lee11:MapReduce,
	Author = {Kyong-Ha Lee and Yoon-Joon Lee and Hyunsik Choi and Yon Dohn Chung and Bongki Moon},
	Journal = {ACM SIGMOD Record},
	Month = dec,
	Number = {4},
	Pages = {11-20},
	Title = {Parallel Data Processing with {M}ap{R}educe: A Survey},
	Volume = {40},
	Year = 2011}

@article{Lee06:Rate_Reliability,
	Author = {Jang-Won Lee and Mung Chiang and A. Robert Calderbank},
	Journal = IEEE_J_JSAC,
	Month = may,
	Number = {5},
	Pages = {962-976},
	Title = {Price-Based Distributed Algorithms for Rate-Reliability Tradeoff in Network Utility Maximization},
	Volume = {24},
	Year = 2006}

@article{Leung-Yan-Cheong78:Gaussian-Wiretap,
	Author = {S. K. Leung-Yan-Cheong and M. E. Hellman},
	Journal = IEEE_J_IT,
	Month = jul,
	Number = {4},
	Pages = {451-456},
	Title = {The {G}aussian Wire-Tap Channel},
	Volume = {24},
	Year = 1978}

@article{Leighton92:GraphEmbedding,
	Author = {F. T. Leighton and M. J. Newman and A. G. Ranade and E. J. Schwabe},
	Journal = {SIAM Journal of Computing},
	Month = aug,
	Number = {4},
	Pages = {639-654},
	Title = {Dynamic Tree Embeddings in Butterflies and Hypercubes},
	Volume = {21},
	Year = 1992}

@article{Liang08:MAC-Secrecy,
	Author = {Yingbing Liang and H. Vicent Poor},
	Journal = IEEE_J_IT,
	Month = mar,
	Number = {3},
	Pages = {976-1002},
	Title = {Multiple Access Channels with Confidential Messages},
	Volume = {54},
	Year = 2008}

@article{Lim06:Trust_Region,
	Author = {C. Lim, H.D. Sherali},
	Journal = {Mathematical Methods of Operations Research},
	Month = aug,
	Number = {1},
	Pages = {33-53},
	Title = {A Trust Region Target Value Method for Optimizing Nondifferentiable Lagrangian Duals of Linear Programs},
	Volume = {64},
	Year = 2006}

@article{Lin06:ImperfectSch,
	Author = {Xiaojun Lin and Ness B. Shroff},
	Journal = IEEE_J_NET,
	Month = apr,
	Number = {2},
	Pages = {302-315},
	Title = {The Impact of Imperfect Scheduling on Cross-Layer Congestion Control in Wireless Networks},
	Volume = {14},
	Year = 2006}

@article{Lin06:CrossLayer_Wireless,
	Author = {Xiaojun Lin and Ness B. Shroff and R. Srikant},
	Journal = IEEE_J_JSAC,
	Month = aug,
	Number = {8},
	Pages = {1452-1463},
	Title = {A Tutorial on Cross-Layer Optimization in Wireless Networks},
	Volume = {24},
	Year = 2006}

@article{Ling06:AP_Placement,
	Author = {Xiang Ling and Kwan Lawrence Yeung},
	Journal = IEEE_J_WCOM,
	Month = oct,
	Number = {10},
	Pages = {2705-2711},
	Title = {Joint Access Point Placement and Channel Assignment for 802.11 wireless {LAN}},
	Volume = {5},
	Year = 2006}

@article{Liu01:STCode,
	Author = {Z. Liu and G. B. Giannakis and S. Barbarossa and A. Scaglione},
	Journal = IEEE_J_JSAC,
	Month = jul,
	Number = {7},
	Pages = {1352-1364},
	Title = {Transmit Antenna Space-Time Block Coding for Generalized {OFDM} in the Presence of Unknown Multipath},
	Volume = {19},
	Year = 2001}

@article{Liu02:STCode,
	Author = {Youjian Liu and M. P. Fitz and O. Y. Takeshita},
	Journal = IEEE_J_IT,
	Month = dec,
	Number = {12},
	Pages = {3062-3079},
	Title = {A Rank Criterion for {QAM} Space-Time Codes},
	Volume = {48},
	Year = 2002}

@article{Liu06:MIMO_SH_AdHoc_BBRLT,
	Author = {Jia Liu and Y. Thomas Hou and Yi Shi and Hanif D. Sherali},
	Journal = IEEE_J_WCOM,
	Month = feb,
	Number = {2},
	Pages = {488-494},
	Title = {On the Capacity of Multiuser {MIMO} Networks with Interference},
	Volume = {7},
	Year = 2008}

@article{Liu06:MIMO_MH_TR,
	Author = {Jia Liu and Y. Thomas Hou and Yi Shi and Hanif D. Sherali},
	Journal = {Technical Report, Deptment of ECE, Virginia Tech},
	Month = sep,
	Title = {Cross-Layer Optimization on Routing and Power Control of {MIMO}-based mesh Networks},
	Url = {http://www.ece.vt.edu/thou/Publications/publications.html},
	Year = 2006,
	Bdsk-Url-1 = {http://www.ece.vt.edu/thou/Publications/publications.html}}

@article{Liu07:MIMO_MH_JSAC,
	Author = {Jia Liu and Y. Thomas Hou and Yi Shi and Hanif D. Sherali},
	Journal = IEEE_J_JSAC,
	Month = aug,
	Number = {6},
	Pages = {913-926},
	Title = {Cross-Layer Optimization for {MIMO}-Based Wireless Ad Hoc Networks: Routing, Power Allocation, and Bandwidth Allocation},
	Volume = {26},
	Year = 2008}

@article{Liu07_CL_MIMO_BC_TR,
	Author = {Jia Liu and Y. Thomas Hou},
	Journal = {Technical Report, Deptment of ECE, Virginia Tech},
	Month = mar,
	Title = {Cross-Layer Optimization of {MIMO}-Based Mesh Networks with Gaussian Vector Broadcast Channels},
	Url = {http://filebox.vt.edu/users/kevinlau/publications/},
	Year = 2007,
	Bdsk-Url-1 = {http://filebox.vt.edu/users/kevinlau/publications/}}

@article{Liu07:MIMO_BC_GP,
	Author = {Jia Liu and Y. Thomas Hou and Hanif D. Sherali},
	Journal = {Technical Report, Deptment of ECE, Virginia Tech},
	Month = nov,
	Title = {Conjugate Gradient Projection Method for Multi-Antenna {G}aussian Broadcast Channels},
	Url = {http://www.ece.vt.edu/thou/Publications/publications.html},
	Year = 2006,
	Bdsk-Url-1 = {http://www.ece.vt.edu/thou/Publications/publications.html}}

@article{Liu07:MIMO_BC_MWS,
	Author = {Jia Liu and Y. Thomas Hou and Hanif D. Sherali},
	Journal = IEEE_J_JSAC,
	Title = {Maximum Weighted Sum Rate of {MIMO} Gaussian Broadcast Channels},
	Year = {submitted for publication}}

@article{Liu07:MIMO_BC_MWS_TR,
	Author = {Jia Liu and Y. Thomas Hou and Hanif D. Sherali},
	Journal = {Technical Report, Deptment of ECE, Virginia Tech},
	Month = nov,
	Title = {Maximum Weighted Sum Rate of {MIMO} Gaussian Broadcast Channels},
	Url = {http://www.ece.vt.edu/thou/Publications/publications.html},
	Year = 2006,
	Bdsk-Url-1 = {http://www.ece.vt.edu/thou/Publications/publications.html}}

@article{Liu07:MIMO_BC_LFB_TR,
	Author = {Jia Liu and Y. Thomas Hou},
	Journal = {Technical Report, Deptment of ECE, Virginia Tech},
	Month = dec,
	Title = {Stochastic Approximation Approach for {MIMO} Gaussian Broadcast Channels with Imperfect {CSI}},
	Url = {http://filebox.vt.edu/users/kevinlau/publications},
	Year = 2007,
	Bdsk-Url-1 = {http://filebox.vt.edu/users/kevinlau/publications}}

@article{Liu07:MIMO_BC_WPF_TR,
	Author = {Jia Liu and Y. Thomas Hou},
	Journal = {Technical Report, Dept. of ECE, Virginia Tech},
	Month = jun,
	Title = {Weighted Proportional Fairness Capacity for {G}aussian {MIMO} Broadcast Channels},
	Url = {http://filebox.vt.edu/users/kevinlau/publications},
	Year = 2007,
	Bdsk-Url-1 = {http://filebox.vt.edu/users/kevinlau/publications}}

@article{Liu07:CL_MIMO_LFB_TR,
	Author = {Jia Liu and Y. Thomas Hou},
	Journal = {Technical Report, Deptment of ECE, Virginia Tech},
	Month = nov,
	Title = {Limited Feedback-Based Cross-Layer Optimization for {MIMO}-Based Mesh Networks},
	Url = {http://filebox.vt.edu/users/kevinlau/publications},
	Year = 2007,
	Bdsk-Url-1 = {http://filebox.vt.edu/users/kevinlau/publications}}

@article{Liu08:MIMO_OFDM_WIMAX_TR,
	Author = {Jia Liu and Y. Thomas Hou},
	Journal = {Technical Report, Department of ECE, Virginia Tech},
	Month = mar,
	Title = {Optimal Downlink Power Allocation and Scheduling for {MIMO}-Based {W}i{MAX} Access Networks},
	Url = {http://filebox.vt.edu/users/kevinlau/publications},
	Year = 2008,
	Bdsk-Url-1 = {http://filebox.vt.edu/users/kevinlau/publications}}

@article{Liu08:MIMO_OFDM_WIMAX,
	Author = {Jia Liu and Y. Thomas Hou},
	Journal = IEEE_J_JSAC,
	Title = {Optimal Downlink Power Allocation and Scheduling for {MIMO}-Based {W}i{MAX} Access Networks},
	Url = {http://filebox.vt.edu/users/kevinlau/publications},
	Year = {submitted for publication, [Online]. Available: http://filebox.vt.edu/users/kevinlau/publications},
	Bdsk-Url-1 = {http://filebox.vt.edu/users/kevinlau/publications}}

@article{Liu08:MIMO_OFDM_AdHoc,
	Author = {Jia Liu and Y. Thomas Hou},
	Journal = {Technical Report, Department of ECE, Virginia Tech},
	Month = aug,
	Title = {Power Spectrum Optimization of Multi-Carrier MIMO Ad Hoc Networks: A Dual Stochastic Approximation Approach},
	Url = {http://filebox.vt.edu/users/kevinlau/publications},
	Year = 2008,
	Bdsk-Url-1 = {http://filebox.vt.edu/users/kevinlau/publications}}

@article{Liu09:MIMO_SEC_SL_TR,
	Author = {Jia Liu and Y. Thomas Hou and Hanif D. Sherali},
	Journal = {Technical Report, Department of ECE, Virginia Tech},
	Month = jan,
	Title = {Maximum Achievable Rates in {G}aussian {MIMO} Wire-Tap Channels with Gaussian Signaling},
	Url = {http://filebox.vt.edu/users/kevinlau/publications},
	Year = 2009,
	Bdsk-Url-1 = {http://filebox.vt.edu/users/kevinlau/publications}}

@article{Liu09:CL_MIMO_DOF_TR,
	Author = {Jia Liu and Yi Shi and Y. Thomas Hou},
	Journal = {Technical Report, Department of ECE, Virginia Tech},
	Month = jul,
	Title = {A Tractable and Accurate Cross-Layer Model for Multi-Hop {MIMO} Ad Hoc Networks},
	Url = {http://filebox.vt.edu/users/kevinlau/publications},
	Year = 2009,
	Bdsk-Url-1 = {http://filebox.vt.edu/users/kevinlau/publications}}

@article{Liu12:FBS_JSAC,
	Author = {Jia Liu and Tianyou Kou and Qian Chen and Hanif D. Sherali},
	Journal = IEEE_J_JSAC,
	Month = apr,
	Number = {3},
	Pages = {652-663},
	Title = {Femtocell Base Station Deployment in Commercial Buildings: A Global Optimization Approach},
	Volume = {30},
	Year = 2012}

@article{Liu11:DNewton_TR,
	Author = {Jia Liu and Hanif D. Sherali},
	Journal = {Technical Report, Dept. of ECE, Ohio State University},
	Month = jul,
	Title = {A Distributed {N}ewton's Method for Joint Multi-Hop Routing and Flow Control: Theory and Algorithm},
	Url = {http://www2.ece.ohio-state.edu/~liu/publications/DNewton.pdf},
	Year = 2011,
	Bdsk-Url-1 = {http://www2.ece.ohio-state.edu/~liu/publications/DNewton.pdf}}

@article{Liu12:DNewton_INC_TR,
	Author = {Jia Liu and Cathy H. Xia and Ness B. Shroff},
	Journal = {Technical Report, Dept. of ECE, Ohio State University},
	Month = jul,
	Title = {Optimal Network Flow for In-Network Computing: A Distributed Second-Order Approach},
	Url = {http://www2.ece.ohio-state.edu/~liu/publications/DNewton_INC.pdf},
	Year = 2012,
	Bdsk-Url-1 = {http://www2.ece.ohio-state.edu/~liu/publications/DNewton_INC.pdf}}

@article{Liu12:DNewton_Wireless_TR,
	Author = {Jia Liu and Cathy H. Xia and Ness B. Shroff and Hanif D. Sherali},
	Journal = {Technical Report, Dept. of ECE, Ohio State University},
	Month = jul,
	Title = {Distributed Cross-Layer Optimization in Wireless Networks: A Second-Order Approach},
	Url = {http://www2.ece.ohio-state.edu/~liu/publications/DNewton_Wireless.pdf},
	Year = 2012,
	Bdsk-Url-1 = {http://www2.ece.ohio-state.edu/~liu/publications/DNewton_Wireless.pdf}}

@article{Liu12:CPS_Bldg_TR,
	Author = {Jia Liu and Qian Chen and Tianyou Kou and Hanif D. Sherali},
	Journal = {Technical Report, Dept. of ECE, Ohio State University},
	Month = jul,
	Title = {On Wireless Network Infrastructure Optimization for Cyber-Physical Systems in Future Smart Buildings},
	Url = {http://www2.ece.ohio-state.edu/~liu/publications/CPS_Bldg_TR.pdf},
	Year = 2011,
	Bdsk-Url-1 = {http://www2.ece.ohio-state.edu/~liu/publications/CPS_Bldg_TR.pdf}}

@article{Liu06:WiMAX,
	Author = {Qingwen Liu and Xin Wang and Georgios B. Giannakis},
	Journal = IEEE_J_VT,
	Month = may,
	Number = {3},
	Pages = {839-847},
	Title = {A Cross-Layer Scheduling Algorithm with {Q}o{S} Support in Wireless Networks},
	Volume = {55},
	Year = 2006}

@article{LiuZ09:StreamProc,
	Author = {Zhen Liu and Ao Tang and Cathy H. Xia and Li Zhang},
	Journal = {Annals of Operations Research},
	Month = sep,
	Number = {1},
	Pages = {161-182},
	Title = {A Decentralized Control Mechanism for Stream Processing Networks},
	Volume = {170},
	Year = 2009}

@article{Love08:LFB,
	Author = {David J. Love and Robert W. Heath and V. K. N. Lau and D. Gesbert and B. D. Rao and M. Andrews},
	Journal = IEEE_J_JSAC,
	Month = oct,
	Number = {8},
	Pages = {1341-1365},
	Title = {An Overview of Limited Feedback in Wireless Communication Systems},
	Volume = {26},
	Year = 2008}

@article{Love03:LFB,
	Author = {David J. Love and Robert W. Heath and T. Strohmer},
	Journal = IEEE_J_IT,
	Month = oct,
	Number = {10},
	Pages = {2735-2747},
	Title = {Grassmannian beamforming for multiple-input multiple-output wireless systems},
	Volume = {49},
	Year = 2003}

@article{Love05:LFB,
	Author = {David J. Love and Robert W. Heath},
	Journal = IEEE_J_SP,
	Month = jan,
	Number = {1},
	Pages = {64-73},
	Title = {Limited feedback unitary precoding for orthogonal space-time block codes},
	Volume = {53},
	Year = 2005}

@article{Love05:LFB2,
	Author = {David J. Love and Robert W. Heath},
	Journal = IEEE_J_IT,
	Month = aug,
	Number = {8},
	Pages = {2967-2976},
	Title = {Limited feedback unitary precoding for spatial multiplexing systems},
	Volume = {51},
	Year = 2005}

@article{Love05:LFB3,
	Author = {David J. Love and Robert W. Heath},
	Journal = IEEE_J_WCOM,
	Month = jan,
	Number = {1},
	Pages = {20-23},
	Title = {Necessary and sufficient conditions for full diversity order in correlated {R}ayleigh fading beamforming and combining systems},
	Volume = {4},
	Year = 2005}

@article{Love04:LFB4,
	Author = {David J. Love and Robert W. Heath},
	Journal = IEEE_J_COML,
	Month = may,
	Number = {5},
	Pages = {305-307},
	Title = {Diversity performance of precoded orthogonal space-time block codes using limited feedback},
	Volume = {8},
	Year = 2004}

@article{Love04:LFB_ComMag,
	Author = {David J. Love and Robert W. Heath and Wiroonsak Santipach and Michael Honig},
	Journal = IEEE_M_COM,
	Month = oct,
	Number = {10},
	Pages = {54-59},
	Title = {What is the value of limited feedback for {MIMO} channels?},
	Volume = {42},
	Year = 2004}

@article{Low04:Distr_Algs,
	Author = {Stven Low and R. Srikant},
	Journal = {Network and Spatial Economics},
	Number = {1},
	Pages = {75-101},
	Title = {A Mathematical framework for Designing a Low-Loss Low-Delay Internet},
	Volume = {4},
	Year = 2004}

@article{MacKenzie09:ProcIEEE,
	Author = {A.B.~MacKenzie and J. H. Reed and P. Athanas and C. W. Bostian and R.M.~Buehrer and L. A. DaSilva and S. W. Ellingson and Y. Thomas Hou and M. Hsiao and J.-M. Park and C. Patterson and S. Raman and C. R. C. M. da Silva},
	Journal = IEEE_J_PROC,
	Month = apr,
	Number = {4},
	Pages = {660-688},
	Title = {Cognitive radio and networking research at {V}irginia {T}ech},
	Volume = {97},
	Year = 2009}

@article{Madan06:Energy,
	Author = {Ritesh Madan and Shuguang Cui and Sanjay Lall and Andrea Goldsmith},
	Journal = IEEE_J_WCOM,
	Month = nov,
	Number = {11},
	Pages = {3142-3152},
	Title = {Cross-Layer Design for Lifetime Maximization in Interference-Limited Wireless Sensor Networks},
	Volume = {5},
	Year = 2006}

@article{Mao07:TMC_BeamStar,
	Author = {Shiwen Mao and Y. Thomas Hou},
	Journal = IEEE_J_MC,
	Month = nov,
	Number = {11},
	Pages = {1284-1296},
	Title = {BeamStar: An edge-based approach to routing in wireless sensor networks},
	Volume = {6},
	Year = 2007}

@article{Mao08:WirelessMag,
	Author = {Shiwen Mao and Y. Thomas Hou and M.-Y. Wu},
	Journal = IEEE_M_WC,
	Month = aug,
	Number = {4},
	Pages = {67-73},
	Title = {Exploiting edge capabilities for wireless sensor networking},
	Volume = {15},
	Year = 2008}

@article{Marcenko67:Wishart,
	Author = {V. A. Mar\u{c}enko and L. A. Pastur},
	Journal = {Math. USSR-Sbornik},
	Pages = {457-483},
	Title = {Distributions of eigenvalues for some sets of random matrices},
	Volume = {1},
	Year = {1967}}

@article{Malick05:Mtrx_Nrns,
	Author = {J. Malick},
	Journal = {SIAM Journal on Matrix Analysis and Applications},
	Month = sep,
	Number = {1},
	Pages = {272-284},
	Title = {A Dual Approach to Semidefinite Least-Squares Problems},
	Volume = {26},
	Year = 2005}

@article{Mathar00:CellularPlanning,
	Author = {R. Mathar and Thomas Niessen},
	Journal = {Wireless Networks},
	Pages = {421-428},
	Title = {Optimum Positioning of Base Stations for Cellular Radio Networks},
	Volume = {6},
	Year = 2000}

@article{May98:OFDM,
	Author = {Thomas May and Hermann Rohling and Volker Engels},
	Journal = IEEE_J_COM,
	Month = sep,
	Number = {1},
	Pages = {182-190},
	Title = {Performance Analysis of {V}iterbi Decoding for 64-{DAPSK} and 64-{QAM} Modulated {OFDM} Signals},
	Volume = {46},
	Year = 1998}

@article{Mehlfuhrer08:MIMO_WiMAX,
	Author = {Christian Mehlf\"{u}hrer and Sebastian Caban and Markus Rupp},
	Journal = {EURASIP Journal on Advances in Signal Processing},
	Title = {Experimental Evaluation of Adaptive Modulation and Coding in {MIMO} {W}i{MAX} with Limited Feedback},
	Url = {http://hindawi.com/RecentlyAcceptedArticlePDF.aspx?journal=ASP\&number=837102},
	Year = {accepted for publication},
	Bdsk-Url-1 = {http://hindawi.com/RecentlyAcceptedArticlePDF.aspx?journal=ASP%5C&number=837102}}

@article{Molisch05:AntennSelect,
	Author = {A. F. Molisch and M. Z. Win and Yang-Seok Choi and J. H. Winters},
	Journal = IEEE_J_WCOM,
	Month = jul,
	Number = {4},
	Pages = {1759-1772},
	Title = {Capacity of {MIMO} Systems with Antenna Selection},
	Volume = {4},
	Year = 2005}

@article{Molisch05:UWB,
	Author = {Andreas F. Molisch},
	Journal = IEEE_J_VT,
	Month = sep,
	Number = {5},
	Pages = {1528-1545},
	Title = {Ultrawideband Propagation Channels -- {T}heory, Measurement, and Modeling},
	Volume = {54},
	Year = 2005}

@article{Moose94:OFDM,
	Author = {Paul H. Moose},
	Journal = IEEE_J_COM,
	Month = oct,
	Number = {10},
	Pages = {2908-2914},
	Title = {A Technique for Orthogonal Frequency Division Multiplexing Frequency Offset Correction},
	Volume = {42},
	Year = 1994}

@article{Muller97:OFDM,
	Author = {Stefan H. M\"{u}ller and Robert W. B\"{a}uml and Robert F. H. Fischer and Johannes B. Huber},
	Journal = {Annual of Telecommunications},
	Month = feb,
	Number = {1-2},
	Pages = {58-67},
	Title = {{OFDM} with Reduced Peak-to-Average Power Ratio by Multiple Signal Representation},
	Volume = {52},
	Year = 1997}

@article{Munoz-Medina07:MIMO_Relay,
	Author = {Olga Mu$\tilde{\mbox{n}}$oz-Medina and Josep Vidal and Adri\'{a}n Agust\'{i}n},
	Journal = IEEE_J_SP,
	Month = jun,
	Number = {6},
	Pages = {2593-2604},
	Title = {Linear Transceiver Design in Nonregenerative Relays with Channel State Information},
	Volume = {55},
	Year = 2007}

@article{Mukkavilli03:LFB,
	Author = {K. K. Mukkavilli and A. Sabharwal and E. Erkip and B. Aazhang},
	Journal = IEEE_J_IT,
	Month = oct,
	Number = {10},
	Pages = {2562-2579},
	Title = {On beamforming with finite rate feedback in multiple antenna systems},
	Volume = {49},
	Year = 2003}

@article{Naguib00:STCode,
	Author = {A. Naguib and N. Seshadri and R. Calderbank},
	Journal = IEEE_M_SP,
	Month = may,
	Number = {3},
	Pages = {76-92},
	Title = {Increasing Data Rate Over Wireless channels},
	Volume = {17},
	Year = 2000}

@article{Nabar02:MIMOChanModel,
	Author = {R. Nabar and H. Bolcskei and V. Erceg and D. Gesbert and A. Paulraj},
	Journal = IEEE_J_COM,
	Month = oct,
	Number = {10},
	Pages = {2553-2562},
	Title = {Performance of Multiantenna Signaling Techniques in the Precense of Polarization Diversity},
	Volume = {50},
	Year = 2002}

@article{Narasimhan03:SMPLEX,
	Author = {Ravi Naraimhan},
	Journal = IEEE_J_IT,
	Month = nov,
	Number = {11},
	Pages = {2829-2838},
	Title = {Spatial Multiplexing With Transmit Antenna and Constellation Selection for Correlated {MIMO} Fading Channels},
	Volume = {51},
	Year = 2003}

@article{Narula98:LFB,
	Author = {A. Narula and M. J. Lopez and M. D. Trott and G. W. Wornell},
	Journal = IEEE_J_JSAC,
	Month = oct,
	Number = {8},
	Pages = {1423-1436},
	Title = {Efficient use of side informationin multiple-antenna data transmission over fading channels},
	Volume = {16},
	Year = 1998}

@article{Neely03:PA_Routing,
	Author = {Michael J. Neely and Eytan Modiano and C. E. Rohrs},
	Journal = IEEE_J_NET,
	Month = feb,
	Number = {2},
	Pages = {138-152},
	Title = {Power Allocation and Routing in Multibeam Satellites with Time-Varying Channels},
	Volume = {11},
	Year = 2003}

@article{Neely05:BackPressure,
	Author = {Michael J. Neely and Eytan Modiano and C. E. Rohrs},
	Journal = IEEE_J_JSAC,
	Month = jan,
	Number = {1},
	Pages = {89-103},
	Title = {Dynamic Power Allocation and Routing for Time Varying Wireless Networks},
	Volume = {23},
	Year = 2005}

@article{Neely08:BackPressure,
	Author = {Michael J. Neely and Eytan Modiano and Chih-Ping Li},
	Journal = IEEE_J_NET,
	Month = apr,
	Number = {2},
	Pages = {396-409},
	Title = {Faireness and Optimal Stochastic Control for Heterogeneous Networks},
	Volume = {16},
	Year = 2008}

@article{Goel08:Screcy-MIMO,
	Author = {Satashu Goel and Rohit Negi},
	Journal = IEEE_J_WCOM,
	Month = jun,
	Number = {6},
	Pages = {2180-2189},
	Title = {Guaranteeing Secrecy Using Artificial Noise},
	Volume = {7},
	Year = 2008}

@article{Nosratinia04:Coop_Com,
	Author = {Aria Nosratinia and Todd E. Hunter and Ahmadreza Hedayat},
	Journal = IEEE_M_COM,
	Month = oct,
	Number = {10},
	Pages = {74-80},
	Title = {Cooperative Communication in Wireless Networks},
	Volume = {42},
	Year = 2004}

@article{Novakovic00:PC_Overview,
	Author = {Dejan M. Novakovic and Miroslav L. Dukic},
	Journal = {IEEE Communications Surveys and Tutorials},
	Month = oct,
	Number = {4},
	Pages = {2-15},
	Title = {Evolution of the Power Control Techniques for {DS-CDMA} toward 3{G} Wireless Communication Systems},
	Volume = {3},
	Year = 2000}

@article{Oggier11:MIMO-Wiretap,
	Author = {Frederique Oggier and Babak Hassibi},
	Journal = IEEE_J_IT,
	Month = aug,
	Number = {8},
	Pages = {4961-4972},
	Title = {The Secrecy Capacity of the {MIMO} Wiretap Channe},
	Volume = {57},
	Year = 2011}

@article{Ozgur07:ScalingLaw,
	Author = {A. Ozgur and O. Leveque and D. Tse},
	Journal = IEEE_J_IT,
	Month = oct,
	Number = {10},
	Pages = {3549-3572},
	Title = {Hierarchical Cooperation Achieves Optimal Capacity Scaling in Ad Hoc Networks},
	Volume = {53},
	Year = 2007}

@article{Paloma03:MC-MIMO,
	Author = {Daniel P\'{e}rez Paloma and John M. Cioffi and Miguel Angel Lagunas},
	Journal = IEEE_J_SP,
	Month = sep,
	Number = {9},
	Pages = {2381-2401},
	Title = {Joint {T}x-{R}x Beamforming Design for Multicarrier {MIMO} Channels: {A} Unified Framework for Convex Optimization},
	Volume = {51},
	Year = 2003}

@article{Paloma05:MC-MIMO,
	Author = {Daniel P\'{e}rez Paloma},
	Journal = IEEE_J_SP,
	Month = dec,
	Number = {12},
	Pages = {4661-4674},
	Title = {Convex Primal Decomposition for Multicarrier Linear {MIMO} Transceivers},
	Volume = {53},
	Year = 2005}

@article{Pan05:TMC,
	Author = {J. Pan and L. Cai and Y. T. Hou and Y. Shi and S. X. Shen},
	Journal = IEEE_J_MC,
	Month = sep # {/} # oct,
	Number = {5},
	Pages = {458-473},
	Title = {Optimal Base-Station Locations in Two-Tiered Wireless Sensor Networks},
	Volume = {4},
	Year = 2005}

@article{Pande07:MIMO_OFDM_LFB_AS,
	Author = {Tarkesh Pande and David J. Love and James V. Krogmeier},
	Journal = IEEE_J_SP,
	Month = may,
	Number = {5},
	Pages = {2284-2293},
	Title = {Reduced Feedback {MIMO-OFDM} Precoding and Antenna Selection},
	Volume = {55},
	Year = 2007}

@article{Pantazis07:WSN_Lifetime,
	Author = {N. A. Pantazis and D. D. Vergados},
	Journal = {IEEE Communications Surveys and Tutorials},
	Month = oct,
	Number = {4},
	Pages = {86-107},
	Title = {A Survey on Power Control Issues in Wireless Sensor Networks},
	Volume = {9},
	Year = 2007}

@article{Patel10:FemtoSurvey,
	Author = {Chirag Patel and Mehmet Yavuz and Sanjiv Nanda},
	Journal = IEEE_M_WC,
	Month = oct,
	Number = {5},
	Pages = {6-7},
	Title = {Femtocells: Industry Perspectives},
	Volume = {17},
	Year = 2010}

@article{Paulraj04:overview,
	Author = {A. J. Paulraj and D. A. Gore and R. U. Nabar and H. Bolcskei},
	Journal = IEEE_J_PROC,
	Month = feb,
	Number = {2},
	Pages = {198-218},
	Title = {An Overview of {MIMO} Communications --- A Key to Gigabit Wireless},
	Volume = {92},
	Year = 2004}

@article{Peel05:DPC_Impl,
	Author = {Chris B. Peel and B. Hochwald and A. L. Swindlehurst},
	Journal = IEEE_J_COM,
	Month = jan,
	Number = {1},
	Pages = {195-202},
	Title = {A Vector Perturbation Technique for Near Capacity Multi-Antenna Multi-User Communication -- {P}art {I}: {C}hannel Inversion and Regularization},
	Volume = {53},
	Year = 2005}

@article{Pickholtz82:SS,
	Author = {R. Pickholtz and D. Schilling and L. Milstein},
	Journal = IEEE_J_COM,
	Month = may,
	Number = {5},
	Pages = {855-884},
	Title = {Theory of Spread-Sprectrum Communications -- {A} Tutorial},
	Volume = {30},
	Year = 1982}

@article{Pursley77:SS_MAC_1,
	Author = {Michael B. Pursley},
	Journal = IEEE_J_COM,
	Month = aug,
	Number = {8},
	Pages = {795-799},
	Title = {Performance Evaluation for Phase-Coded Spread Sprectrum Multiple-Access Communication -- {P}art {I}: {S}ystem Analysis},
	Volume = {25},
	Year = 1977}

@article{Pursley77:SS_MAC_2,
	Author = {Michael B. Pursley and D. V. Sarwate},
	Journal = IEEE_J_COM,
	Month = aug,
	Number = {8},
	Pages = {800-803},
	Title = {Performance Evaluation for Phase-Coded Spread Sprectrum Multiple-Access Communication -- {P}art {II}: {C}ode Sequence Analysis},
	Volume = {25},
	Year = 1977}

@article{Pursley87:SS_PRN,
	Author = {Michael B. Pursley},
	Journal = IEEE_J_PROC,
	Month = jan,
	Number = {1},
	Pages = {116-134},
	Title = {The Role of Spread Sprectrum in Packet Radio Networks},
	Volume = {75},
	Year = 1987}

@article{Raghunathan06:WSN_Lifetime,
	Author = {V. Raghunathan and S. Ganeriwal and M. Srivastava},
	Journal = IEEE_M_COM,
	Month = apr,
	Number = {4},
	Pages = {108-114},
	Title = {Emerging Techniques for Long Lived Wireless Sensor Networks},
	Volume = {44},
	Year = 2006}

@article{Rentel08:Synchronization,
	Author = {Carlos H. Rentel and Thomas Kunz},
	Journal = IEEE_J_MC,
	Month = may,
	Number = {5},
	Pages = {633-646},
	Title = {A Mutual Network Synchronization Method for Wireless Ad Hoc and Sensor Networks},
	Volume = {7},
	Year = 2008}

@article{Robbins51:SA,
	Author = {H. Robbins and S. Monro},
	Journal = {Annals of Mathematical Statistics},
	Pages = {400-407},
	Title = {A Stochastic Approximation Method},
	Volume = {22},
	Year = 1951}

@article{Rohling99:OFDM,
	Author = {Hermann Rohling and Thomas May and Karsten Br\"{u}ninghaus and Rainer Gr\"{u}nheid},
	Journal = IEEE_J_PROC,
	Month = oct,
	Number = {10},
	Pages = {1778-1789},
	Title = {Broad-Band {OFDM} Radio Transmission for Multimedia Applications},
	Volume = {87},
	Year = 1999}

@article{Rong07:WiMAX,
	Author = {Bo Rong and Yi Qian and Kejie Lu},
	Journal = {IEEE Trans. Mobile Comput.},
	Month = jun,
	Number = {6},
	Pages = {621-632},
	Title = {Integrated Downlink Resource Management for Multiservice {W}i{MAX} Networks},
	Volume = {6},
	Year = 2007}

@article{Rosen60:GrdPrj,
	Author = {J. B. Rosen},
	Journal = {SIAM J. Applied Mathematics},
	Pages = {181-217},
	Title = {The Gradient Projection Method for Nonlinear Programming, {P}art {I}, {L}inear Constraints},
	Volume = {8},
	Year = {1960}}

@article{Santipach05:RVQ,
	Author = {Wiroonsak Santipach and Michael Honig},
	Journal = IEEE_J_IT,
	Month = oct,
	Number = {10},
	Pages = {3475-3492},
	Title = {Signature optimiation for {CDMA} with limited feedback},
	Volume = {51},
	Year = 2005}

@article{Salvekar04:MIMO_WiMAX,
	Author = {Atul Salvekar and Sumeet Sandhu and Qinghua Li and Minh-Anh Vuong and Xiaoshu Qian},
	Journal = {Intel Technology Journal},
	Month = aug,
	Number = {3},
	Pages = {229-239},
	Title = {Multi-Antenna Technology in {W}i{MAX} Systems},
	Volume = {8},
	Year = {2004}}

@article{Santipach06:RVQ_JIT,
	Author = {Wiroonsak Santipach and Michael Honig},
	Journal = IEEE_J_IT,
	Title = {Capacity of a Multi-Antenna Fading Channel with a Quantized Precoding Matrix},
	Url = {http://arxiv.org/abs/0704.0217v1},
	Year = {submitted for publication},
	Bdsk-Url-1 = {http://arxiv.org/abs/0704.0217v1}}

@article{Scaglione06:Coop_Com,
	Author = {Anna Scaglione and Dennis L. Goeckel and J. Nicholas Laneman},
	Journal = IEEE_M_SP,
	Month = sep,
	Number = {5},
	Pages = {18-29},
	Title = {Cooperative Communications in Mobile Ad Hoc Networks},
	Volume = {23},
	Year = 2006}

@article{Seidel92:FAF,
	Author = {S. Y. Seidel and T. S. Rappaport},
	Journal = IEEE_J_AP,
	Month = feb,
	Number = {2},
	Pages = {207-217},
	Title = {914 {MH}z Path Loss Prediction Models for Indoor Wireless Communications in Multifloored Buildings},
	Volume = {40},
	Year = 1992}

@article{Shannon49:secrecy,
	Author = {Claude Shannon},
	Journal = {Bell Syst. Tech. J.},
	Pages = {656-715},
	Title = {Communication Theory of Secrecy Systems},
	Volume = {29},
	Year = 1949}

@article{Sharif05:BC_LFB,
	Author = {M. Sharif and B. Hassibi},
	Journal = IEEE_J_IT,
	Month = feb,
	Number = {2},
	Pages = {506-522},
	Title = {On the Capacity of {MIMO} Broadcast Channels with Partial Side Information},
	Volume = {51},
	Year = 2005}

@article{Sherali92:BBRLT,
	Author = {Hanif D. Sherali and Cihan H. Tuncbilek},
	Journal = {Journal of Global Optimization},
	Number = {1},
	Pages = {101-112},
	Title = {A Global Optimization Algorithm for Polynomial Programming Problems Using a Reformulation-Linearization Technique},
	Volume = {2},
	Year = 1992}

@article{Sherali_ORL96,
	Author = {Hanif D. Sherali and Gyunghyun Choi},
	Journal = {Operations Research Letters},
	Month = sep,
	Number = {3},
	Pages = {105-113},
	Title = {Recovery of Primal Solutions When Using Subgradient Optimization Methods to Solve {L}agrangian Duals of Linear Programs},
	Volume = {19},
	Year = 1996}

@article{Sherali96:OptTxLocation,
	Author = {Sherali, H.D. and Pendyala, C.M. and Rappaport, T.S.},
	Journal = IEEE_J_JSAC,
	Month = may,
	Number = {4},
	Pages = {662-673},
	Title = {Optimal Location of Transmitters for Micro-cellular Radio Communication System Design},
	Volume = {14},
	Year = 1996}

@article{Sherali_Wang_ORL01,
	Author = {H. D. Sherali and H. Wang},
	Journal = {Mathematical Programming},
	Number = {3},
	Pages = {459-478},
	Title = {Global optimization of nonconvex factorable programming problems},
	Volume = {89},
	Year = {2001}}

@article{Sherali_OR98,
	Author = {Hanif D. Sherali and Warren P. Adams and Patrick J. Driscoll},
	Journal = {Operations Research},
	Month = may,
	Number = {3},
	Pages = {396-405},
	Title = {Exploiting Special Structures in Constructing a Hierarchy of Relaxations for 0-1 Mixed Integer Problems},
	Volume = {46},
	Year = 1998}

@article{Sherali_Fraticelli_JGO00,
	Author = {H. D. Sherali and B. M. Fraticelli},
	Journal = {Journal of Global Optimization},
	Number = {1-4},
	Pages = {223-261},
	Title = {Enhancing {RLT} relaxations via a new class of {Semidefinit Cuts}},
	Volume = {22},
	Year = {2000}}

@article{Sherali_Smith01:Lexico,
	Author = {Hanif D. Sherali and J. Cole Smith},
	Journal = {Management Sciences},
	Number = {10},
	Pages = {1396-1407},
	Title = {Improving Discrete Model Representations via Symmetry Considerations},
	Volume = {47},
	Year = {2001}}

@article{Shi06:JSAC_UWB,
	Author = {Yi Shi and Y. Thomas Hou and Hanif D. Sherali and Scott F. Midkiff},
	Journal = IEEE_J_JSAC,
	Month = apr,
	Number = {4},
	Pages = {857--863},
	Title = {Optimal routing for {UWB}-based sensor networks},
	Volume = {24},
	Year = {2006}}

@article{Shi08:Comp_Net_UWB,
	Author = {Yi Shi and Y. Thomas Hou},
	Journal = {Computer Networks (Elsevier)},
	Month = oct,
	Number = {14},
	Pages = {2797--2804},
	Title = {On the capacity of {UWB}-based wireless sensor networks},
	Volume = {52},
	Year = {2008}}

@article{Shi08:TMC_UWB,
	Author = {Yi Shi and Y. Thomas Hou and Hanif D. Sherali},
	Journal = IEEE_J_MC,
	Month = jun,
	Number = {6},
	Pages = {764--777},
	Title = {Cross-layer optimization for data rate utility problem in {UWB}-based ad hoc networks},
	Volume = {7},
	Year = {2008}}

@article{Shi09:WINET,
	Author = {Y. Shi and Y. T. Hou and A. Efrat},
	Journal = {ACM/Springer Wireless Networks (WINET)},
	Number = {1},
	Pages = {21-38},
	Title = {Algorithm Design for A Class of Base Station Location Problems in Sensor Networks},
	Volume = {15},
	Year = {2009}}

@article{Shi09:ToSN,
	Author = {Yi Shi and Y. Thomas Hou},
	Journal = {ACM Transactions on Sensor Networks},
	Month = nov,
	Number = {4},
	Title = {Optimal Base Station Placement in Wireless Sensor Networks},
	Volume = {5},
	Year = {2009}}

@article{Shi10:power_control,
	Author = {Y. Shi and Y. Thomas Hou and H. Zhou},
	Journal = IEEE_J_WCOM,
	Note = {(to appear)},
	Title = {Per-node based optimal power control for multi-hop cognitive radio networks}}

@article{Shin07:MIMO_OFDM_CHAN_ESTI,
	Author = {Changyong Shin and Robert W. Heath and Edward J. Powers},
	Journal = IEEE_J_VT,
	Month = mar,
	Number = {2},
	Pages = {670-685},
	Title = {Blind Channel Estimation for {MIMO-OFDM} Systems},
	Volume = {56},
	Year = {2007}}

@article{Simeone10:FemtoIntfManagement,
	Author = {O. Simeone and E. Erkip and {Shamai (Shitz)}, Shlomo},
	Journal = IEEE_J_JSAC,
	Month = dec,
	Number = {9},
	Pages = {1469-1478},
	Title = {Robust Transmission and Interference Management for Femtocells with Unreliable Network Access},
	Volume = {28},
	Year = {2010}}

@article{Siwamogsatham02:STCode,
	Author = {S. Siwamogsatham and M. P. Fitz},
	Journal = IEEE_J_SP,
	Month = oct,
	Number = {10},
	Pages = {2408-2416},
	Title = {Robust Space-Time Codes for Correlated Rayleigh Fading Channels},
	Volume = {50},
	Year = {2002}}

@article{So07:WiMAX,
	Author = {Jae-Woo So},
	Journal = IEEE_J_COML,
	Month = feb,
	Number = {2},
	Pages = {155-157},
	Title = {A Downlink Performance Analysis of {V}o{IP} Services over an {IEEE} 802.16e {OFDMA} System},
	Volume = {11},
	Year = {2007}}

@article{Snow07:WiMAX_ErrorRate,
	Author = {Chris Snow and Lutz Lampe and Robert Schober},
	Journal = IEEE_J_COM,
	Month = sep,
	Number = {9},
	Pages = {1736-1746},
	Title = {Error Rate Analysis for Coded Multicarrier Systems Over Quasi-Static Fading Channels},
	Volume = {55},
	Year = {2007}}

@article{So08:AP_Placement,
	Author = {Aaron So and Ben Liang},
	Journal = IEEE_J_VT,
	Month = jan,
	Number = {1},
	Pages = {532-547},
	Title = {Efficient Wireless Extension Point Placement Algorithm in Urban REctilineal {WLAN}s},
	Volume = {57},
	Year = {2008}}

@article{Song05:OFDM1,
	Author = {Guocong Song and Ye Li},
	Journal = IEEE_J_WCOM,
	Month = mar,
	Number = {2},
	Pages = {614-624},
	Title = {Cross-Layer Optimization for {OFDM} Wireless Networks -- {P}art {I}: {T}heoretical Framework},
	Volume = {4},
	Year = {2005}}

@article{Song05:OFDM2,
	Author = {Guocong Song and Ye Li},
	Journal = IEEE_J_WCOM,
	Month = mar,
	Number = {2},
	Pages = {625-634},
	Title = {Cross-Layer Optimization for {OFDM} Wireless Networks -- {P}art {II}: {A}lgorithm Development},
	Volume = {4},
	Year = {2005}}

@article{Specer04:MIMO_BC,
	Author = {Quentin H. Spencer and A. Lee Swindlehurst and Martin Haardt},
	Journal = IEEE_J_SP,
	Month = feb,
	Number = {2},
	Pages = {461-471},
	Title = {Zero-Forcing Methods for Downlink Spatial Multiplexing in Multiuser {MIMO} Channels},
	Volume = {52},
	Year = {2004}}

@article{Stamatelos96:IndoorBS,
	Author = {Stamatelos, D. and Ephremides, A.},
	Journal = IEEE_J_JSAC,
	Month = may,
	Number = {4},
	Pages = {651-661},
	Title = {Spectral Efficiency and Optimal Base Placement for Indoor Wireless Networks},
	Volume = {14},
	Year = {1996}}

@article{Sundaresan03:MIMO_SH_MAC,
	Author = {Karthikeyan Sundaresan and Raghupathy Sivakumar and Mary Ann Ingram and Tae-Young Chang},
	Journal = IEEE_J_MC,
	Month = oct,
	Number = {4},
	Pages = {350-365},
	Title = {Medium Access Control in Ad Hoc Networks with {MIMO} Links: Optimization Considerations and Algorithms},
	Volume = {3},
	Year = 2004}

@article{Szalay11:BigData,
	Author = {Alexander S. Szalay},
	Journal = {IEEE Computing in Science and Engineering},
	Month = nov,
	Number = {6},
	Pages = {34-41},
	Title = {Extreme Data-Intensive Scientific Computing},
	Volume = {13},
	Year = 2011}

@article{Tang07:MIMO_Relay,
	Author = {Xiaojun Tang and Yingbo Hua},
	Journal = IEEE_J_WCOM,
	Month = apr,
	Number = {4},
	Pages = {1398-1407},
	Title = {Optimal Design of Non-Regenerative {MIMO} Wireless Relays},
	Volume = {6},
	Year = 2007}

@article{Tarokh98:STTC,
	Author = {V. Tarokh and N. Seshadri and A. R. Calderbank},
	Journal = IEEE_J_IT,
	Month = Mar,
	Number = {2},
	Pages = {744-765},
	Title = {Space-time Codes for High Data Rate Wireless Communication: Performance Criterion and Code Construction},
	Volume = {44},
	Year = 1998}

@article{Tassiulas92:BackPressure,
	Author = {L. Tassiulas and A. Ephremides},
	Journal = IEEE_J_AC,
	Month = dec,
	Number = {12},
	Pages = {1936-1948},
	Title = {Stability Properties of constrained Queuing Systems and Scheduling Policies for Maximum Throughput in Multihop Radio Networks},
	Volume = {37},
	Year = 1992}

@article{Tarokh99:STBC,
	Author = {V. Tarokh and H. Jafarkhani and A. R. Calderbank},
	Journal = IEEE_J_IT,
	Month = Mar,
	Number = {3},
	Pages = {451-460},
	Title = {Space-Time Block Codes for Wireless Communications: Performance Results},
	Volume = {17},
	Year = 1999}

@article{Telatar99:Limits,
	Author = {I. E. Telatar},
	Journal = {European Trans. Telecomm.},
	Month = nov,
	Number = {6},
	Pages = {585-596},
	Title = {Capacity of multi-antenna {G}aussian channels},
	Volume = {10},
	Year = {1999}}

@article{Tse98:MAC_Struct,
	Author = {David Tse and S. Hanly},
	Journal = IEEE_J_IT,
	Month = nov,
	Number = {7},
	Pages = {2796-2815},
	Title = {Multi-Access Fading Channels: Part {I}: Polymatroid Structure, Optimal Resource Allocation and Throughput Capacities},
	Volume = {44},
	Year = {1998}}

@article{Tse04:MAC_Tradeoff,
	Author = {David Tse and Pramod Viswanath and Lizhong Zheng},
	Journal = IEEE_J_IT,
	Month = sep,
	Number = {9},
	Pages = {1859-1874},
	Title = {Diversity-Multiplexing Tradeoff in Multiple Access Channels},
	Volume = {50},
	Year = {2004}}

@article{Turkmani95:MIMOChanModel,
	Author = {A. M. D. Turkmani and A. A. Arowojolu and P. A. Jefford and C. J. Kellett},
	Journal = IEEE_J_VT,
	Month = may,
	Number = {2},
	Pages = {318-326},
	Title = {An Experimental Evaluation of the Performance of Two-Branch Space and Polarization of the Performance of Two-Branch Space and Polization Diversity Schemes at 1800 {MH}z},
	Volume = {44},
	Year = {1995}}

@article{Vishwanath03:duality,
	Author = {S. Vishwanath and N. Jindal and A. Goldsmith},
	Journal = IEEE_J_IT,
	Month = oct,
	Number = {10},
	Pages = {2658-2668},
	Title = {Duality, Achievable Rates, and Sum-Rate Capacity of {MIMO} Broadcast Channels},
	Volume = {49},
	Year = {2003}}

@article{Viswanathan03:MIMO_BC_SD,
	Author = {Harish Viswanathan and Sivarama Venkatesan and Howard Huang},
	Journal = IEEE_J_JSAC,
	Month = jun,
	Number = {5},
	Pages = {802-811},
	Title = {Downlink Capacity Evaluation of Cellular Networks with Known-Interference Cancellation},
	Volume = {21},
	Year = {2003}}

@article{Visotsky01:LFB,
	Author = {E. Visotsky and U. Madhow},
	Journal = IEEE_J_IT,
	Month = sep,
	Number = {6},
	Pages = {2632-2639},
	Title = {Space-Time Transmit Precoding with Imperfect Feedback},
	Volume = {47},
	Year = {2001}}

@article{Viswanath02:Opportunistic,
	Author = {P. Viswanath and D. N. C. Tse and R. Laroia},
	Journal = IEEE_J_IT,
	Month = jun,
	Number = {6},
	Pages = {1277-1294},
	Title = {Opportunistic Beamforming using Dumb Antennas},
	Volume = {48},
	Year = {2002}}

@article{Viswanath03:duality,
	Author = {P. Viswanath and D. N. C. Tse},
	Journal = IEEE_J_IT,
	Month = aug,
	Number = {8},
	Pages = {1912-1921},
	Title = {Sum Capacity of the Vector {G}aussian Broadcast Channel and Uplink-Downlink Duality},
	Volume = {49},
	Year = {2003}}

@article{Visuri06:MC_MIMO_MAC,
	Author = {Samuli Visuri and Helmut B\"{o}lcskei},
	Journal = IEEE_J_IT,
	Month = sep,
	Number = {9},
	Pages = {3980-3993},
	Title = {Multiple-Access Strategies for Frequency-Selective {MIMO} Channels},
	Volume = {52},
	Year = {2006}}

@article{Verdu_TIT89,
	Author = {Verd\'{u}},
	Journal = IEEE_J_IT,
	Month = may,
	Number = {1},
	Pages = {85-96},
	Title = {Minimum probability of error for asynchronous {G}aussian multiple-access channels},
	Volume = {32},
	Year = {1986}}

@article{Wang07:WiMAX,
	Author = {Jianfeng Wang and Muthaiah Venkatachalam and Yuguang Fang},
	Journal = IEEE_J_JSAC,
	Month = may,
	Number = {4},
	Pages = {712-721},
	Title = {System Architecture and Cross-Layer Optimization of Video Broadcast over {W}i{MAX}},
	Volume = {25},
	Year = {2007}}

@article{Wang07:SA,
	Author = {Xin Wang and Georgios B. Giannakis and Antonio G. Marques},
	Journal = IEEE_J_PROC,
	Month = dec,
	Number = {12},
	Pages = {2410-2431},
	Title = {A Unified Approach to {Q}o{S}-Guaranteed Scheduling for Channel-Adaptive Wireless Networks},
	Volume = {95},
	Year = {2007}}

@article{Weingarten06:MIMO_BC,
	Author = {Hanan Weingarten and Yossef Steinberg and {Shamai (Shitz)}, Shlomo},
	Journal = IEEE_J_IT,
	Month = sep,
	Number = {9},
	Pages = {3936-3964},
	Title = {The Capacity Region of the {G}aussian Multiple-Input Multiple-Output Broadcast Channel},
	Volume = {52},
	Year = {2006}}

@article{Weinstein71:OFDM,
	Author = {S. B. Weinstein and Paul M. Ebert},
	Journal = IEEE_J_COM,
	Month = oct,
	Number = {5},
	Pages = {628-634},
	Title = {Data Transmission by Frequency-Division Multiplexing Using the Discrete Fourier Transform},
	Volume = {19},
	Year = {1971}}

@article{Win98:UWB,
	Author = {Moe Z. Win and Robert A. Scholtz},
	Journal = IEEE_J_COML,
	Month = feb,
	Number = {2},
	Pages = {36-38},
	Title = {Impulse radio: {H}ow it works},
	Volume = {2},
	Year = {1998}}

@article{Winters87:Limits,
	Author = {Jack Winters},
	Journal = IEEE_J_JSAC,
	Month = Jun,
	Number = {5},
	Pages = {871-878},
	Title = {On the Capacity of Radio Communication Systems with Diversity in a {R}ayleigh Fading Environment},
	Volume = {5},
	Year = 1987}

@article{Winters06:MIMO-AdHoc,
	Author = {Jack H. Winters},
	Journal = IEEE_M_WC,
	Month = aug,
	Number = {4},
	Pages = {77-83},
	Title = {Smart Antenna Techniques and Their Application to Wireless Ad Hoc Networks},
	Volume = {13},
	Year = 2006}

@article{Wong99:Multiuser_OFDM,
	Author = {Cheong Yui Wong and Roger S. Cheng and Khaled Ben Letaief and Ross D. Murch},
	Journal = IEEE_J_JSAC,
	Month = oct,
	Number = {10},
	Pages = {1747-1758},
	Title = {Multiuser {OFDM} with Adaptive Subcarrier, Bit, and Power Allocation},
	Volume = {17},
	Year = {1999}}

@article{Wong06:IndoorAP,
	Author = {Wong, J.K.L. and Mason, A.J. and Neve, M.J. and Sowerby, K.W.},
	Journal = {IEE Proceedings -- Communications},
	Month = oct,
	Number = {5},
	Pages = {771-778},
	Title = {Base Station Placement in Indoor Wireless Systems Using Binary Integer Programming},
	Volume = {153},
	Year = {2006}}

@article{Woznicki01:Mtrx_Split,
	Author = {Zbigniew I. Woznicki},
	Journal = {International Journal of Mathematics and Mathematical Sciences},
	Month = may,
	Number = {5},
	Pages = {251-284},
	Title = {Matrix Splitting Principles},
	Volume = {28},
	Year = {2001}}

@article{Wyner75:wire-tap,
	Author = {A. D. Wyner},
	Journal = {Bell Syst. Tech. J.},
	Number = {8},
	Pages = {1355-1387},
	Title = {The Wire-Tap Channel},
	Volume = {54},
	Year = {1975}}

@article{Xia10:FemtoAccCon,
	Author = {Ping Xia and Vikram Chandrasekhar and Jeffrey G. Andrews},
	Journal = IEEE_J_WCOM,
	Month = dec,
	Number = {12},
	Pages = {3798-3809},
	Title = {Open vs. Closed Access Femtocells in the Uplink},
	Volume = {9},
	Year = 2010}

@article{Xiao04:SRRA,
	Author = {Lin Xiao and Mikael Johansson and Stephen P. Boyd},
	Journal = IEEE_J_COM,
	Month = jul,
	Number = {7},
	Pages = {1136-1144},
	Title = {Simultaneous Routing and Resource Allocation via Dual Decomposition},
	Volume = {52},
	Year = 2004}

@article{Yan02:STCode,
	Author = {Qing Yan and R. S. Blum},
	Journal = IEEE_J_COM,
	Month = jul,
	Number = {7},
	Pages = {1136-1144},
	Title = {Improved Space-Time Convolutional Codes for Quasi-Static Slow Fading Channels},
	Volume = {52},
	Year = 2004}

@article{Ye03:MIMO_SH_AdHoc,
	Author = {S. Ye and R. S. Blum},
	Journal = IEEE_J_SP,
	Month = nov,
	Number = {11},
	Pages = {2839-2848},
	Title = {Optimized signaling for {MIMO} interference systems with feedback},
	Volume = {51},
	Year = {2003}}

@article{Yoo06:BC_LFB,
	Author = {T. Yoo and Andrea Goldsmith},
	Journal = IEEE_J_JSAC,
	Month = mar,
	Number = {3},
	Pages = {528-541},
	Title = {On the Optimality of Multiantenna Broadcast Scheduling Using Zero-Forcing Beamforming},
	Volume = {24},
	Year = {2006}}

@article{Yoo07:BC_LFB,
	Author = {T. Yoo and Nihar Jindal and Andrea Goldsmith},
	Journal = IEEE_J_JSAC,
	Month = sep,
	Number = {7},
	Pages = {1478-1491},
	Title = {Multi-Antenna Downlink Channels with Limited Feedback and User Selection},
	Volume = {25},
	Year = {2007}}

@article{Yu04:MIMO_BC_SUMRATE_DPC,
	Author = {Wei Yu and John M. Cioffi},
	Journal = IEEE_J_IT,
	Month = sep,
	Number = {9},
	Pages = {1875-1892},
	Title = {Sum Capacity of {G}aussian Vector Broadcast Channels},
	Volume = {50},
	Year = {2004}}

@article{Yu04:MIMO_MAC_IWF,
	Author = {Wei Yu and Wonjong Rhee and Stephen Boyd and John M. Cioffi},
	Journal = IEEE_J_IT,
	Month = jan,
	Number = {1},
	Pages = {145-152},
	Title = {Iterative Water-Filling for {G}aussian Vector Multiple-Access Channels},
	Volume = {50},
	Year = {2004}}

@article{Yu05:DPC_Impl,
	Author = {Wei Yu and David Varodayan and John M. Cioffi},
	Journal = IEEE_J_IT,
	Month = jul,
	Number = {7},
	Pages = {1220-1230},
	Title = {Trellis and Convolutional Precoding for Transmitter-Based Interference Presubtraction},
	Volume = {53},
	Year = {2005}}

@article{Yu06:duality,
	Author = {Wei Yu},
	Journal = IEEE_J_IT,
	Month = feb,
	Number = {2},
	Pages = {361-374},
	Title = {Uplink-Downlink Duality via Minimax Duality},
	Volume = {52},
	Year = {2006}}

@article{Yu06:MC_Nonconvex,
	Author = {Wei Yu and Raymond Lui},
	Journal = IEEE_J_COM,
	Month = jul,
	Number = {7},
	Pages = {1310-1322},
	Title = {Dual Methods for Nonconvex Spectrum Optimization of Multicarrier Systems},
	Volume = {54},
	Year = {2006}}

@article{Yu06:MIMO_BC_DD,
	Author = {Wei Yu},
	Journal = IEEE_J_IT,
	Month = feb,
	Number = {2},
	Pages = {754-759},
	Title = {Sum-Capacity Computation for the Gaussian Vector Broadcast Channel via Dual Decomposition},
	Volume = {52},
	Year = {2006}}

@article{Zamir02:DPC_Impl,
	Author = {R. Zamir and {Shamai (Shitz)}, Shlomo and U. Erez},
	Journal = IEEE_J_IT,
	Month = jun,
	Number = {6},
	Pages = {1250-1277},
	Title = {Nested Linear/Lattice codes for structured Multiterminal Binning},
	Volume = {48},
	Year = {2002}}

@article{Zeng07:MIMO_OFDM_CHAN_EST,
	Author = {Yonghong Zeng and A. Rahim Leyman and Tung-Sang Ng},
	Journal = IEEE_J_COM,
	Month = dec,
	Number = {12},
	Pages = {2270-2278},
	Title = {Joint Semiblind Frequency Offset and Channel Estimation for Multiuser {MIMO-OFDM} Uplink},
	Volume = {55},
	Year = {2006}}

@article{Zhang08:SA,
	Author = {Junshan Zhang and Dong Zheng and Mung Chiang},
	Journal = IEEE_J_IT,
	Month = feb,
	Number = {2},
	Pages = {645-665},
	Title = {The Impact of Stochastic Noisy Feedback on Distributed Network Utility Maximization},
	Volume = {54},
	Year = 2008}

@article{Zheng03:Tradeoff,
	Author = {Lizhong Zheng and David N. C. Tse},
	Journal = IEEE_J_IT,
	Month = May,
	Number = {5},
	Pages = {1073-1096},
	Title = {Diversity and Multiplexing: A Fundamental Tradeoff in Multiple-Antenna Channels},
	Volume = {49},
	Year = 2003}

@article{Zhou05:LFB,
	Author = {S. Zhou and W. Wang and G. B. Giannakis},
	Journal = IEEE_J_WCOM,
	Month = jul,
	Number = {7},
	Pages = {1948-1957},
	Title = {Quantifying the power loss when transmit beamforming relies on finite-rate feedback},
	Volume = {4},
	Year = 2005}

@article{Zorzi06:MIMO_MAC_Issues,
	Author = {Michele Zorzi and James Zeidler and Adam Anderson and Bhaskar Rao and John Proakis and A. Lee Swindlehurst and Michael Jensen and Srikanth Krishnamurthy},
	Journal = IEEE_M_WC,
	Month = aug,
	Number = {4},
	Pages = {62-76},
	Title = {Cross-Layer Issues in {MAC} Protocol Design for {MIMO} Ad Hoc Networks},
	Volume = {13},
	Year = 2006}

@article{Geier05:SNR_Cutoff,
	Author = {J. Geier},
	Journal = {WiFi Planet Online Tutorial},
	Title = {{SNR} Cutoff Recommendations},
	Url = {http://www.wi-fiplanet.com/tutorials/article.php/3468771},
	Year = {2005},
	Bdsk-Url-1 = {http://www.wi-fiplanet.com/tutorials/article.php/3468771}}

@inproceedings{Aguayo04:80211_Measurement,
	Address = {Portland, OR},
	Author = {D. Aguayo and J. Bicket and S. Biswas and G. Judd and R. Morris},
	Booktitle = {Proc. ACM SIGCOMM},
	Month = aug # {30--} # sep # {3,},
	Pages = {121-131},
	Title = {Link-Level Measurements from an 802.11b Mesh Network},
	Year = {2004}}

@inproceedings{Andrew07:MC_Scheduling,
	Address = {Montr\'{e}al, Qu\'{e}bec, Canada},
	Author = {Mathew Andrews and Lisa Zhang},
	Booktitle = {Proc. ACM Mobicom},
	Month = sep # { 9-14,},
	Pages = {3-14},
	Title = {Scheduling Algorithms for Multi-Carrier Wireless Data Systems},
	Year = {2007}}

@inproceedings{Amraoui03:MIMO_BC,
	Address = {Yokohama, Japan},
	Author = {A. Amraoui and G. Kramer and {Shamai (Shitz)}, Shlomo},
	Booktitle = {Proc.~IEEE ISIT},
	Month = jun # {29--} # jul # {4,},
	Pages = {296},
	Title = {Coding for the {MIMO} broadcast channel},
	Year = {2003}}

@inproceedings{Adya04:multiradio,
	Address = {San Jose, CA},
	Author = {A. Adya and P. Bahl and J. Padhye and A. Wolman and L. Zhou},
	Booktitle = {Proc.~IEEE International Conference on Broadband Networks},
	Month = oct # { 5-9,},
	Pages = {344-354},
	Title = {A Multi-radio Unification Protocol for {IEEE} 802.11 Wireless Networks},
	Year = {2004}}

@inproceedings{Alicherry05:MCMR,
	Address = {Cologne, Germany},
	Author = {M. Alicherry and R. Bhatia and L. Li},
	Booktitle = {Proc. ACM Mobicom},
	Month = aug,
	Pages = {58-72},
	Title = {Joint Channel Assignment and Routing for Throughput Optimization in Multi-radio Wireless Mesh Networks},
	Year = {2005}}

@inproceedings{Ashraf10:FemtoCoverage,
	Address = {Cape Town, South Africa},
	Author = {Imran Ashraf and Holger Claussen and Lester T.W. Ho},
	Booktitle = {Proc. IEEE International Conference on Communications (ICC)},
	Month = may # { 23-27,},
	Pages = {1-6},
	Title = {Distributed Radio Coverage Optimization in Enterprise Femtocell Networks},
	Year = {2010}}

@inproceedings{Bejerano04:IndoorAP,
	Address = {Philadelphia, PA},
	Author = {Yigan Bejerano and Seung-Jae han and Li (Erran) Li},
	Booktitle = {Proc. ACM MobiCom},
	Month = sep # {26--} # oct # {1,},
	Pages = {2326-2330},
	Title = {Fairness and Load Balancing in Wireless {LAN} Using Association Control},
	Year = {2004}}

@inproceedings{Bhatia07:MIMO_MeshNetwork,
	Address = {Anchorage, AK},
	Author = {Randeep Bhatia and Li Li},
	Booktitle = {Proc. IEEE INFOCOM},
	Month = may # { 6-12,},
	Pages = {2326-2330},
	Title = {Throughput Optimization of Wireless Mesh Networks with {MIMO} Links},
	Year = {2007}}

@inproceedings{Bickson08:DNewton,
	Address = {Monticello, IL},
	Author = {Danny Bickson and Yoav Tock and Ori Shental and Danny Dolev},
	Booktitle = {Proc. Allerton Conference on Communication, Control, and Computing},
	Month = sep # { 23-26,},
	Pages = {895-901},
	Title = {Polynomial Linear Programming with {G}aussian Belief Propagation},
	Year = {2008}}

@inproceedings{Bickson09:DNewton,
	Address = {Seoul, Korea},
	Author = {Danny Bickson and Yoav Tock and Argyris Zymnis and Stephen Boyd and Danny Dolev},
	Booktitle = {Proc. IEEE International Symposium on Information Theory (ISIT)},
	Month = jun # {28--} # jul # {3,},
	Pages = {829-833},
	Title = {Distributed Large Scale Network Utility Maximization},
	Year = {2009}}

@inproceedings{Biglieri07:MIMO_WiMAX,
	Address = {Orlando, FL},
	Author = {E. Biglieri and Andrea Goldsmith and B. Muquet and H. Sari},
	Booktitle = {Proc. IEEE Mobile WiMAX Symposium},
	Month = mar # { 25-29,},
	Pages = {74-79},
	Title = {Diversity Interference Cancellation and Spatial Multiplexing in {MIMO} Mobile {W}i{MAX} Systems},
	Year = {2007}}

@inproceedings{Blum03:LFB,
	Author = {Rick S. Blum},
	Booktitle = {Proc. IEEE Int. Conf. Acoust., Speech and Sig. Proc.},
	Month = apr,
	Pages = {89-92},
	Title = {{MIMO} with limited feedback of channel state information},
	Volume = {4},
	Year = {2003}}

@inproceedings{Boche_IZS04,
	Address = {Zurich, Switzerland},
	Author = {H. Boche and M. Wiczanowski},
	Booktitle = {Proc. IZS},
	Month = feb,
	Pages = {18-21},
	Title = {Stability Region of Arrival Rates and Optimal Scheduling for {MIMO-MAC} -- {A} Cross-Layer Approach},
	Year = {2004}}

@inproceedings{Bogdanov04:BS_Placement,
	Address = {Hong Kong SAR, China},
	Author = {Bogdanov, A. and Maneva, E. and Riesenfeld, S.},
	Booktitle = {Proc. IEEE INFOCOM},
	Month = mar # { 7-11,},
	Pages = {575-585},
	Title = {Power-aware Base Station Positioning for Sensor Networks},
	Year = {2004}}

@inproceedings{Bolognani10:DQNewton,
	Address = {Annecy, France},
	Author = {Saverio Bolognani and Sandro Zampieri},
	Booktitle = {Proc. 2nd IFAC Workshop on Distributed Estimation and Control in Networked Systems},
	Month = sep # { 13-14,},
	Pages = {305-310},
	Title = {Distributed Quasi-{N}ewton Method and Its Applications to the Optimal Reactive Power Flow Problem},
	Year = {2010}}

@inproceedings{Buehrer01:MIMOChanModel,
	Author = {R. M. Buehrer and S. Arunachalam and K. Wu and A. Tonello},
	Booktitle = {Proc. IEEE VTC},
	Month = may,
	Pages = {342-346},
	Title = {Spatial Channel Models and Measurements for {IMT}-2000 Systems},
	Year = {2001}}

@inproceedings{Carroll10:SmartPhoneEnergy,
	Author = {A. Carroll and G. Heiser},
	Booktitle = {Proc. of USENIX ATC},
	Month = jun,
	Title = {An Analysis of Power Consumption in a Smartphone},
	Year = {2010}}

@inproceedings{Chan07:MIMO_WiMAX,
	Address = {Orlando, FL},
	Author = {Tsz Ho Chan and Chui Ying Cheung and Mounir Hamdi and Maode Ma},
	Booktitle = {Proc. IEEE Mobile WiMAX Symposium},
	Month = mar # { 25-29,},
	Pages = {98-103},
	Title = {Overview of Rate Adaptation Algorithms Based on {MIMO} Technology in {W}i{MAX} Networks},
	Year = {2007}}

@inproceedings{Choi08:FemtoAccess,
	Address = {New Orleans, LA},
	Author = {David Choi and Pooya Monajemi and Shinjae Kang and John Willasenor},
	Booktitle = {Proc. IEEE Globecom},
	Month = nov # {30--} # dec # {4,},
	Pages = {1-5},
	Title = {Dealing with Loud Neighbors: The Benefits and Tradeoffs of Adaptive Femtocell Access},
	Year = {2008}}

@inproceedings{Chu08:MIMO_Coop,
	Address = {Hong Kong SAR, China},
	Author = {Shan Chu and Xin Wang},
	Booktitle = {Proc. ACM Mobihoc},
	Month = may # { 26-30,},
	Pages = {63-72},
	Title = {Opportunistic and Cooperative Spatial Multiplexing in {MIMO} Ad Hoc Networks},
	Year = {2008}}

@inproceedings{Dean04:MapReduce,
	Address = {San Francisco, CA},
	Author = {Jeffrey Dean and Sanjay Ghemawat},
	Booktitle = {Proc. USENIX OSDI},
	Month = dec # { 6-8,},
	Pages = {137-149},
	Title = {Map{R}educe: Simplified Data Processing on Large Clusters},
	Year = {2004}}

@inproceedings{Demirkol01:Stream_Control,
	Address = {Atlantic City, NJ, U.S.A.},
	Author = {M. F. Demirkol and M. A. Ingram},
	Booktitle = {Proc. IEEE Veh. Technol. Conf.},
	Month = oct,
	Pages = {187-191},
	Title = {Power-controlled capacity for interfering {MIMO} links},
	Year = {2001}}

@inproceedings{Demirkol03:Stream_Control,
	Address = {New Orleans, LA, U.S.A.},
	Author = {M. F. Demirkol and M. A. Ingram},
	Booktitle = {Proc. IEEE Wireless Commun. and Networking Conf.},
	Month = mar,
	Pages = {343-348},
	Title = {Stream Control in Network with Interfereing {MIMO} Links},
	Year = {2003}}

@inproceedings{Dhillon03:SN_Placement,
	Address = {New Orleans, LA},
	Author = {S. S. Dhillon and K. Chakrabarty},
	Booktitle = {Proc. IEEE Wireless Communications \& Networking Conference (WCNC)},
	Month = mar # { 16-20,},
	Pages = {1609-1614},
	Title = {Sensor Placement for Effective Coverage and Surveillance in Distributed Sensor Networks},
	Year = {2003}}

@inproceedings{Draves04:mesh,
	Address = {Philadelphia, PA},
	Author = {R. Draves and J. Padhye and B. Zill},
	Booktitle = {Proc. ACM Mobicom},
	Month = sep,
	Pages = {114-128},
	Title = {Routing in Multi-Radio, Multi-Hop Wireless Mesh Networks},
	Year = {2004}}

@inproceedings{Efrat05:BS_Placement,
	Address = {Boston, MA},
	Author = {Efrat, A. and Har-Peled, S. and Mitchell, J.S.B.},
	Booktitle = {Proc. IEEE Conference on Broadband Networks (BroadNets)},
	Month = oct # { 3-7,},
	Pages = {767-776},
	Title = {Approximation Algorithms for Two Optimal Location Problems in Sensor Networks},
	Year = {2005}}

@inproceedings{Eryilmaz05:WirelessCLO,
	Address = {Miami, FL},
	Author = {A. Eryilmaz and R. Srikant},
	Booktitle = {Proc. IEEE INFOCOM},
	Month = mar,
	Pages = {1804-1814},
	Title = {Fair Resource Allocation in Wireless Networks Using Queue-Length-Based Scheduling and Congestion Control},
	Year = {2005}}

@inproceedings{Fitz98:STCode,
	Author = {M. P. Fitz and J. V. Krogmeier},
	Booktitle = {Proc. Allerton Conf. on Commun., Control, and Comput.},
	Month = sep,
	Pages = {391-400},
	Title = {Further Results on Space-Time Codes for {R}ayleigh Fading},
	Year = {1998}}

@inproceedings{Hamdaoui07:MIMO_Network,
	Address = {Montr\'{e}al, Qu\'{e}bec, Canada},
	Author = {Bechir Hamdaoui and Kang G. Shin},
	Booktitle = {Proc. ACM Mobihoc},
	Month = sep,
	Pages = {120-129},
	Title = {Characterization and Analysis of Multi-Hop Wireless {MIMO} Network Throughput},
	Year = {2007}}

@inproceedings{Hou07:SDR,
	Address = {Anchorage, Alaska, USA},
	Author = {Y. T. Hou and Y. Shi, and H. D. Sherali},
	Booktitle = {Proc. IEEE INFOCOM},
	Month = may # { 6-12,},
	Pages = {1-9},
	Title = {Optimal Spectrum Sharing for Multi-Hop Software Defined Radio Networks},
	Year = {2007}}

@inproceedings{Hou03:Milcom_lifetime,
	Address = {Boston, MA},
	Author = {Y. Thomas Hou and Yi Shi and Jianping Pan},
	Booktitle = {Proc. IEEE Military Communications Conference (MILCOM)},
	Month = oct # { 13-16,},
	Pages = {603-608},
	Title = {A Lifetime-aware flow routing algorithm for energy-constrained wireless sensor networks},
	Year = {2003}}

@inproceedings{Hou05:Infocom,
	Address = {Miami, FL},
	Author = {Y. T. Hou and Y. Shi and H. D. Sherali and J.E. Wieselthier},
	Booktitle = {Proc. IEEE INFOCOM},
	Month = mar # { 13-17,},
	Pages = {761-772},
	Title = {Online Lifetime-Centric Multicast Routing for Ad Hoc Networks with Directional Antennas},
	Year = {2005}}

@inproceedings{Hou05:SECON_Prolong,
	Address = {Santa Clara, CA},
	Author = {Y. Thomas Hou and Yi Shi and Hanif D. Sherali and Scott F. Midkiff},
	Booktitle = {Proc. IEEE Communications Society Conference on Sensor and Ad Hoc Communications and Networks (SECON)},
	Month = sep # { 26-29,},
	Pages = {295-304},
	Title = {Prolonging sensor network lifetime with energy provisioning and relay node placement},
	Year = {2005}}

@inproceedings{Hou04:MobiHoc,
	Address = {Tokyo, Japan},
	Author = {Y. T. Hou and Y. Shi and H. D. Sherali},
	Booktitle = {Proc. ACM MobiHoc},
	Month = may # { 24-26,},
	Pages = {67-77},
	Title = {Rate Allocation in Wireless Sensor Networks with Network Lifetime Requirement},
	Year = {2004}}

@inproceedings{Huai11:DOT,
	Address = {Cascais, Portugal},
	Author = {Yin Huai and Rubao Lee and Simon Zhang and Cathy H. Xia and Xiaodong Zhang},
	Booktitle = {Proc. ACM SOCC},
	Month = oct # { 27-28,},
	Title = {{DOT}: A Matrix Model for Analyzing, Optimizing and Deploying Software for Big Data Analytics in Distributed Systems},
	Year = {2011}}

@inproceedings{Isard07:Dryad,
	Address = {Lisboa, Portugal},
	Author = {Michael Isard and Mihai Budiu and Yuan Yu and Andrew Birrell and Dennis Fetterly},
	Booktitle = {Proc. ACM SIGOPS/Eurosys},
	Month = mar # { 21-23,},
	Pages = {59-72},
	Title = {Dryad: Distributed Data-Parallel Programs from Sequential Building Blocks},
	Year = {2007}}

@inproceedings{Jadbabaie09:DNewton,
	Address = {Shanghai, China},
	Author = {Ali Jadbabaie and Asuman Ozdaglar and Michael Zargham},
	Booktitle = {Proc. IEEE Conference on Decision and Control (CDC)},
	Month = dec # { 16-18,},
	Title = {A Distributed {N}ewton Method for Network Optimization},
	Year = {2009}}

@inproceedings{Jafar01:LFB_CoAdap,
	Author = {Syed Ali Jafar and Sriram Vishwanath and Andrea Goldsmith},
	Booktitle = {Proc. IEEE ICC},
	Month = jun # { 11-14,},
	Pages = {2266-2270},
	Title = {Greedy Scheduling Performance for a Zero-Forcing Dirty Paper Coded System},
	Year = 2001}

@inproceedings{Jaffres-Runser06:AP_Placement,
	Address = {Montreal, QC},
	Author = {Katia Jaffr\`{e}s-Runser and Jean-Marie Gorce and St\'{e}phane Ub\'{e}da},
	Booktitle = {Proc.~IEEE VTC Fall},
	Month = sep # { 25-28,},
	Pages = {1-5},
	Title = {Multiobjective {Q}o{S}-Oriented Planning for Indoor Wireless {LAN}},
	Year = {2006}}

@inproceedings{Jain01:multichannel,
	Address = {Scottsdale, AZ},
	Author = {N. Jain and S. R. Das and A. Nasipuri},
	Booktitle = {Proc.~IEEE ICCCN},
	Month = oct # { 15-17,},
	Pages = {432-439},
	Title = {A Multichannel {CSMA MAC} Protocol with Receiver-Based Channel Selection for Multihop Wireless Networks},
	Year = {2001}}

@inproceedings{Jiang03:AP_Placement,
	Address = {Beijing, China},
	Author = {T. Jiang and G. Zhu},
	Booktitle = {Proc.~IEEE PIMRC},
	Month = sep # { 7-10,},
	Pages = {2302-2306},
	Title = {Uniform Design Simulated Annealing for Optimal Access Point Placement of High Data Rate Indoor Wireless {LAN} Using {OFDM}},
	Year = {2003}}

@inproceedings{Kang07:OFSDMA,
	Address = {Orlando, FL},
	Author = {Mingyu Kang and Duho Rhee and Hae Gwang Hwang and Kwang Soon Kim},
	Booktitle = {Proc.~IEEE MILCOM},
	Month = oct # { 29-31,},
	Title = {Opportunistic Scheduling with Partial {CQRI} Feedback in {OFSDMA} Systems},
	Year = {2007}}

@inproceedings{Kamath08:INC,
	Address = {Toronto, Canada},
	Author = {S. Kamath and D. Manjunath},
	Booktitle = {Proc.~IEEE International Symposium on Information Theory (ISIT)},
	Month = jul # { 6-11,},
	Pages = {647-651},
	Title = {On Distributed Function Computation in Structure-Free Random Networks},
	Year = {2008}}

@inproceedings{Kanoria07:INC,
	Address = {Nice, France},
	Author = {Y. Kanoria and D. Manjunath},
	Booktitle = {Proc.~IEEE International Symposium on Information Theory (ISIT)},
	Month = jun # { 24-29,},
	Pages = {626-630},
	Title = {On Distributed Computation in Noisy Random Planar Networks},
	Year = {2007}}

@inproceedings{Kappes04:AP_Placement,
	Address = {Dallas, TX},
	Author = {M. Kappes and A. S. Krishnakumar and P. Krishnan},
	Booktitle = {Proc.~IEEE GLOBECOM},
	Month = nov # {29--} # dec # {3,},
	Pages = {3264-3269},
	Title = {Estimating Signal Strength Coverage for a Wireless Access Point},
	Year = {2004}}

@inproceedings{Kobayashi00:AP_Placement,
	Address = {London, UK},
	Author = {M. Kobayashi and S. Haruyama and R. Kohno and M. Nakagawa},
	Booktitle = {Proc.~IEEE PIMRC},
	Month = sep # { 18-21,},
	Pages = {200-204},
	Title = {Optimal Access Point Placement in Simultaneous Broadcast System Using {OFDM} for Indoor Wireless {LAN}},
	Year = {2000}}

@inproceedings{Kodialam05:MCMR,
	Address = {Cologne, Germany},
	Author = {M. Kodialam and T. Nandagopal},
	Booktitle = {Proc. ACM Mobicom},
	Month = aug,
	Pages = {73-87},
	Title = {Characterizing the Capacity Region in Multi-Radio Multi-Channel Wireless Mesh Networks},
	Year = {2005}}

@inproceedings{Komolafe05:Multi_Criteria,
	Address = {Miami, FL},
	Author = {O. Komolafe and J. Sventek},
	Booktitle = {Proc.~IEEE INFOCOM},
	Month = mar # { 13-17,},
	Pages = {2447-2457},
	Title = {{RSVP} performance evaluation using multi-objective evolutionary optimisation},
	Year = {2005}}

@inproceedings{Kyasanur05:multichannel-2,
	Address = {Cologne, Germany},
	Author = {P. Kyasanur and N. H. Vaidya},
	Booktitle = {Proc. ACM Mobicom},
	Month = aug # {28--} # sep # {2,},
	Pages = {43-57},
	Title = {Capacity of Multi-Channel Wireless Networks: Impact of Number of Channels and Interfaces},
	Year = {2005}}

@inproceedings{Lan_Yu_GLOBECOM04,
	Address = {Dallas, TX, U.S.A.},
	Author = {T. Lan and Wei Yu},
	Booktitle = {Proc. IEEE GLOBECOM},
	Month = nov,
	Pages = {420-424},
	Title = {Input Optimization for Multi-Antenna Broadcast Channels and Per-Antenna Power Constraints},
	Volume = {1},
	Year = {2004}}

@inproceedings{Lee02:AP_Placement,
	Address = {Tempa, FL},
	Author = {Youngseok Lee and Kyoungae Kim and Yanghee Choi},
	Booktitle = {Proc. IEEE Local Computer Networks (LCN)},
	Month = nov # { 6-8,},
	Pages = {831-836},
	Title = {Optimization of {AP} Placement and Channel Assignment in wireless {LAN}s},
	Year = {2002}}

@inproceedings{Lin04:WirelessCLO,
	Address = {Atlantis, Paradise Island, Bahamas},
	Author = {Xiaojun Lin and Ness B. Shroff},
	Booktitle = {Proc. IEEE CDC},
	Month = dec,
	Pages = {1484-1489},
	Title = {Joint Rate Control and Scheduling in Multihop Wireless Networks},
	Year = {2006}}

@inproceedings{Liu06:MIMO_BBRLT,
	Address = {San Francisco, CA},
	Author = {Jia Liu and Y. Thomas Hou and Yi Shi and Hanif D. Sherali},
	Booktitle = {Proc. IEEE GLOBECOM},
	Month = nov,
	Title = {Optimization of Multiuser {MIMO} Networks with Interference},
	Year = {2006}}

@inproceedings{Liu07:MIMO_LAGGP,
	Address = {Hong Kong, P. R. China},
	Author = {Jia Liu and Y. Thomas Hou and Yi Shi and Hanif D. Sherali},
	Booktitle = {Proc. IEEE WCNC},
	Month = mar,
	Title = {Cross-Layer Optimization on Routing and Power Control of {MIMO} Ad Hoc Networks},
	Year = {2007}}

@inproceedings{Liu07:MIMO_BC_CGP,
	Address = {Nice, France},
	Author = {Jia Liu and Y. Thomas Hou and Hanif D. Sherali},
	Booktitle = {Proc. IEEE International Symposium on Information Theory (ISIT)},
	Month = jun # { 24-29,},
	Pages = {781-785},
	Title = {Conjugate Gradient Projection Approach for {MIMO} {G}aussian Broadcast Channels},
	Year = {2007}}

@inproceedings{Liu08:CL_MIMO_BC,
	Address = {Beijing, China},
	Author = {Jia Liu and Y. Thomas Hou and Hanif D. Sherali},
	Booktitle = {Proc. IEEE ICC},
	Month = may # { 19-23,},
	Note = {({\bf Best Paper Award})},
	Pages = {2859-2864},
	Title = {Routing and Power Allocation for {MIMO}-based Ad Hoc Networks with Dirty Paper Coding},
	Year = {2008}}

@inproceedings{Liu08:MIMO_BC_MWSR,
	Address = {Beijing, China},
	Author = {Jia Liu and Y. Thomas Hou and Hanif D. Sherali},
	Booktitle = {Proc. IEEE ICC},
	Month = may # { 19-23,},
	Pages = {3664-3668},
	Title = {On the Maximum Weighted Sum-Rate of {MIMO} {G}aussian Broadcast Channels},
	Year = {2008}}

@inproceedings{Liu08:MIMO_BC_WPF,
	Address = {Phoenix, AZ},
	Author = {Jia Liu and Y. Thomas Hou},
	Booktitle = {Proc. IEEE INFOCOM},
	Month = apr # { 13-18,},
	Pages = {385-393},
	Title = {Weighted Proportional Fairness Capacity of {G}aussian {MIMO} Broadcast Channels},
	Year = {2008}}

@inproceedings{Liu08:MIMO_MH_CSI_Milcom,
	Address = {San Diego, CA},
	Author = {Jia Liu and Y. Thomas Hou},
	Booktitle = {Proc. IEEE MILCOM},
	Month = nov,
	Pages = {1-7},
	Title = {On the Performance of {MIMO}-Based Ad Hoc Networks under Imperfect {CSI}},
	Year = {2008}}

@inproceedings{Liu09:Secrecy,
	Address = {Baltimore, MD},
	Author = {Jia Liu and Y. Thomas Hou and Hanif D. Sherali},
	Booktitle = {Proc. IEEE CISS},
	Month = mar,
	Title = {Optimal Power Allocation for Achieving Perfect Secrecy Capacity in {MIMO} Wire-Tap Channels},
	Year = 2009}

@inproceedings{Liu09:MC_MIMO_Mobihoc,
	Address = {New Orleans, LA},
	Author = {Jia Liu and Y. Thomas Hou and Yi Shi and Hanif D. Sherali},
	Booktitle = {Proc. ACM Mobihoc},
	Month = may # { 18-21,},
	Pages = {43-54},
	Title = {On Performance Optimization for Multi-Carrier {MIMO} Ad Hoc Networks},
	Year = 2009}

@inproceedings{Liu12:DNewton_INFOCOM,
	Address = {Orlando, FL},
	Author = {Jia Liu and Hanif D. Sherali},
	Booktitle = {Proc. IEEE INFOCOM},
	Month = mar # { 25-30,},
	Pages = {2489-2497},
	Title = {A Distributed {N}ewton's Method for Joint Multi-Hop Routing and Flow Control: Theory and Algorithm},
	Year = 2012}

@inproceedings{Liu13:DNewton_INFOCOM,
	Address = {Turin, Italy},
	Author = {Jia Liu and Cathy H. Xia and Ness B. Shroff and Hanif D. Sherali},
	Booktitle = {Proc. IEEE INFOCOM},
	Month = apr # { 14-19,},
	Title = {Distributed Cross-Layer Optimization in Wireless Networks: A Second-Order Approach},
	Year = 2013}

@inproceedings{LiuR06:MAC-Secrecy,
	Address = {Seatle, WA},
	Author = {Ruoheng Liu and Ivana Mari\'{c} and Roy D. Yates and Predrag Spasojevi\'{c}},
	Booktitle = {Proc. IEEE ISIT},
	Month = jul,
	Pages = {957-961},
	Title = {The Discrete Memoryless Multiple Access Channel with Confidential Messages},
	Year = 2006}

@inproceedings{Liu09:MH_MIMO_DOF_Infocom,
	Address = {San Diego, CA},
	Author = {Jia Liu and Yi Shi and Y. Thomas Hou},
	Booktitle = {Proc. IEEE INFOCOM},
	Month = mar # { 15-19},
	Note = {2010 (to appear). Technical report available online at http://filebox.vt.edu/users/kevinlau/publications/},
	Title = {A Tractable and Accurate Cross-Layer Model for Multi-Hop {MIMO} Ad Hoc Networks}}

@inproceedings{Liu12:FBS_Infocom,
	Address = {Orlando, FL},
	Author = {Jia Liu and Qian Chen and Hanif D. Sherali},
	Booktitle = {Proc. IEEE INFOCOM},
	Month = mar # { 25-30},
	Pages = {3233-3237},
	Title = {Algorithm Design for Femtocell Base Station Placement in Commercial Building Environments},
	Year = {2012}}

@inproceedings{Lu06:AP_Placement,
	Address = {Montreal, QC},
	Author = {Jia-Liang Lu and Katia Jaffr\`{e}s-Runser and Jean-Marie Gorce and Fabrice Valois},
	Booktitle = {Proc. IEEE International Conference on Wireless and Mobile Computing, Networking and Communications},
	Month = jun # { 19-21,},
	Pages = {152-158},
	Title = {Indoor w{LAN} Planning with a {Q}o{S} Constraint Based on a {M}arkovian Performance Evaluation Model},
	Year = {2006}}

@inproceedings{Malik08:Synchronization,
	Address = {Bangalore, India},
	Author = {Asad Waqar Malik and Alfred Park and Richard M. Fujimoto},
	Booktitle = {Proc. IEEE International Conference on Cloud Computing},
	Month = sep # { 21-25,},
	Pages = {49-56},
	Title = {Optimistic Synchronization of Parallel Simulations in Cloud Computing Environements},
	Year = {2009}}

@inproceedings{Mo08:Synchronization,
	Address = {San Diego, CA},
	Author = {Shaomin Mo and J. Hsu and J. Gu and Ming Luo and R. Ghanadan},
	Booktitle = {Proc. IEEE MILCOM},
	Month = nov # { 17-19,},
	Pages = {1-7},
	Title = {Network Synchronization for distributed {MANET}},
	Year = {2008}}

@inproceedings{Mundarath04:MIMO_NULLHOC,
	Address = {Dallas, TX},
	Author = {J. Mundarath and P. Ramanathan and B.D. van Veen},
	Booktitle = {Proc. IEEE GLOBECOM},
	Month = nov # {29--} # dec # {3,},
	Pages = {2765-2769},
	Title = {{NULLHOC}: {A} {MAC} protocol for adaptive antenna array based wireless ad hoc networks in multipath environments},
	Year = {2004}}

@inproceedings{Muquet07:MIMO_WiMAX,
	Address = {Hong Kong, China},
	Author = {Bertrand Muquet and Ezio Biglieri and Hikmet Sari},
	Booktitle = {Proc. IEEE WCNC},
	Month = mar # { 11-15,},
	Pages = {1812-1815},
	Title = {{MIMO} Link Adaptation in Mobile {W}i{MAX} Systems},
	Year = {2007}}

@inproceedings{Nasipuri99:multichannel,
	Address = {Hong Kong, P. R. China},
	Author = {A. Nasipuri and J. Zhuang and S. R. Das},
	Booktitle = {Proc. IEEE WCNC},
	Month = sep # { 21-24,},
	Pages = {1402-1406},
	Title = {A Multichannel {CSMA MAC} Protocol for Multihop Wireless Networks},
	Year = {1999}}

@inproceedings{Oh07:MIMO_TCP,
	Address = {Orlando, FL},
	Author = {Soon Y. Oh and Mario Gerla and Joon-Sang Park},
	Booktitle = {Proc. IEEE MILCOM},
	Month = oct # { 29-31,},
	Title = {{MIMO} and {TCP}: {A} Case for Cross-Layer Design},
	Year = {2007}}

@inproceedings{Oh07:MIMO_CAST,
	Address = {Orlando, FL},
	Author = {Soon Y. Oh and Mario Gerla and Pengkai Zhao and Babak Daneshrad and Guangyu Pei and Jae H. Kim},
	Booktitle = {Proc. IEEE MILCOM},
	Month = oct # { 29-31,},
	Title = {{MIMO-CAST}: {A} Cross-Layer Ad Hoc Multicast Protocol Using {MIMO} Radios},
	Year = {2007}}

@inproceedings{Pan03:Mobicom_topology,
	Address = {San Diego, CA},
	Author = {Jianping Pan and Y. Thomas Hou and Lin Cai and Yi Shi and S. Xueming Shen},
	Booktitle = {Proc. ACM International Conference on Mobile Computing and Networking (MobiCom)},
	Month = sep # { 14-19,},
	Pages = {286-299},
	Title = {Topology control for wireless sensor networks},
	Year = {2003}}

@inproceedings{Pan85:MatrixInverse,
	Address = {New York, NY},
	Author = {V. Pan and J. Reif},
	Booktitle = {Proc. ACM symposium on Theory of computing},
	Pages = {143-152},
	Title = {Efficient Parallel Solution for Linear Systems},
	Year = {1985}}

@inproceedings{Park04:MIMO_sel_diversity,
	Address = {Dallas, TX},
	Author = {M. Park and R. Heath and S. Nettles},
	Booktitle = {Proc. IEEE GLOBECOM},
	Month = nov # {29--} # dec # {3,},
	Pages = {3363-3367},
	Title = {Improving throughput and fairness of {MIMO} ad hoc networks using antenna selection diversity},
	Year = {2004}}

@inproceedings{Park05:MIMO_CL_MAC_Design,
	Address = {St. Louis, MO},
	Author = {M. Park and S-H. Choi and S. Nettles},
	Booktitle = {Proc. IEEE GLOBECOM},
	Month = nov # {28--} # dec # {2,},
	Pages = {2870-2874},
	Title = {Cross-layer {MAC} design for wireless networks using {MIMO}},
	Year = {2005}}

@inproceedings{Park05:MIMO_MAC_Design,
	Address = {Seoul, Korea},
	Author = {Joon-Sang Park and Alok Nandan and Mario Gerla and Heechoon Lee},
	Booktitle = {Proc. IEEE ICC},
	Month = may # { 16-20,},
	Pages = {3642-3646},
	Title = {{SPACE-MAC}: {E}nabling Spatial Reuse Using {MIMO} Channel-Aware {MAC}},
	Year = {2005}}

@inproceedings{Radunovic_LeBoudec_INFOCOMC04,
	Address = {Hong Kong, China},
	Author = {B. Radunovic and J.-Y. Le Boudec},
	Booktitle = {Proc. IEEE INFOCOM},
	Month = {Mar},
	Pages = {1916-1927},
	Title = {Rate performance objectives of multi-hop wireless networks},
	Year = {2004}}

@inproceedings{Raniwala05:mesh,
	Address = {Miami, FL},
	Author = {A. Raniwala and T. Chiueh},
	Booktitle = {Proc. IEEE INFOCOM},
	Month = mar # { 13-17,},
	Pages = {2223-2234},
	Title = {Architecture and Algorithms for An {IEEE} 802.11-Based Multi-Channel Wireless Mesh Network},
	Year = {2005}}

@inproceedings{Roh04:LFB,
	Author = {J. C. Roh and B. D. Rao},
	Booktitle = {Proc. IEEE WCNC},
	Month = mar,
	Pages = {760-764},
	Title = {An efficient feedback method for {MIMO} systems with slowly time-varying channels},
	Year = {2004}}

@inproceedings{Roh04:LFB2,
	Author = {J. C. Roh and B. D. Rao},
	Booktitle = {Proc. IEEE PIMRC},
	Month = sep,
	Pages = {805-809},
	Title = {Channel feedback quantization methods for {MISO} and {MIMO} systems},
	Year = {2004}}

@inproceedings{Sandhu01:SD_Tradeoff,
	Author = {S. Sandhu and A. Paulraj},
	Booktitle = {Proc. IEEE GLOBECOM},
	Pages = {1073-1077},
	Title = {Unified Design of Linear Space-Time Block Codes},
	Year = {2001}}

@inproceedings{Santipach03:LFB,
	Author = {Wiroonsak Santipach and M. L. Honig},
	Booktitle = {Proc. IEEE Mil. Commun. Conf.},
	Month = oct,
	Pages = {141-146},
	Title = {Asymptotic performance of {MIMO} wireless channels with limited feedback},
	Volume = {1},
	Year = {2003}}

@inproceedings{Santipach06:RVQ,
	Author = {Wiroonsak Santipach and M. L. Honig},
	Booktitle = {Proc. IEEE ISIT},
	Month = jul,
	Pages = {290},
	Title = {Asymptotic capacity of beamforming with limited feedback},
	Year = {2004}}

@inproceedings{Sahin07:MIMO_WiMAX,
	Address = {Baltimore, MD},
	Author = {Mustafa E. Sahin and Huseyin Arslan and Daljeet Singh},
	Booktitle = {Proc. IEEE VTC},
	Month = sep # {30--} # oct # {3,},
	Pages = {666-670},
	Title = {Reception and Measurement of {MIMO-OFDM} Signals with a Single Receiver},
	Year = {2007}}

@inproceedings{Sang09:FemtoSelfOrg,
	Address = {Honolulu, HI},
	Author = {Young Jin Sang and Hae Gwang Hwang and Kwang Soon Kim},
	Booktitle = {Proc. IEEE Globecom},
	Month = nov # {30--} # dec # {04,},
	Pages = {1-5},
	Title = {A Self-Organized Femtocell for {IEEE} 802.16e System},
	Year = {2009}}

@inproceedings{Seidel92:FAF_VTC,
	Address = {Denver, CO},
	Author = {Seidel, S.Y. and Rappaport, T.S. and Feuerstein, M.J. and Blackard, K.L. and Grindstaff, L.},
	Booktitle = {Proc. IEEE VTC},
	Month = may # { 10--13},
	Pages = {814-815},
	Title = {The Impact of Surrounding Buildings on Propagation for Wireless In-Building Personal Communications System Design},
	Year = {1992}}

@inproceedings{Shi05:UWB,
	Address = {Cologne, Germany},
	Author = {Y. Shi and Y. T. Hou and H. D. Sherali and S. F. Midkiff},
	Booktitle = {Proc. ACM Mobicom},
	Month = aug # {28--} # sep # {2,},
	Pages = {299-312},
	Title = {Cross-Layer Optimization for Routing Data Traffic in {UWB}-Based Sensor Networks},
	Year = {2005}}

@inproceedings{Shi06:Milcom_UWB,
	Address = {Washington, DC},
	Author = {Yi Shi and Y. Thomas Hou and Hanif D. Sherali and Sastry Kompella},
	Booktitle = {Proc. IEEE Military Communications Conference (MILCOM)},
	Month = oct # { 23--25},
	Pages = {3589-3595},
	Title = {Cross-layer optimization for {UWB}-based ad hoc networks},
	Year = {2006}}

@inproceedings{Shi06:SDR,
	Address = {Anchorage, Alaska, USA},
	Author = {Y. Shi and Y. T. Hou},
	Booktitle = {Proc. IEEE INFOCOM},
	Month = may # { 6-12,},
	Pages = {1694-1702},
	Title = {Optimal Power Control for Programmable Radio Networks},
	Year = {2007}}

@inproceedings{Shi07:SECON_placement,
	Address = {San Diego, CA},
	Author = {Y. Shi and Y. Thomas Hou},
	Booktitle = {Proc. IEEE Communications Society Conference on Sensor, Mesh and Ad Hoc Communications and Networks (SECON)},
	Month = jun # { 18-21,},
	Pages = {512--519},
	Title = {Approximation algorithm for base station placement in wireless sensor networks},
	Year = {2007}}

@inproceedings{Shi08:BPA,
	Address = {Phoenix, AZ},
	Author = {Yi Shi and Y. Thomas Hou},
	Booktitle = {Proc. IEEE INFOCOM},
	Month = apr # { 13-18,},
	Pages = {376-384},
	Title = {Theoretical results on base station movement problem for sensor network},
	Year = {2008}}

@inproceedings{Shi08:Infocom_distributed,
	Address = {Phoenix, AZ},
	Author = {Yi Shi and Y. Thomas Hou},
	Booktitle = {Proc. IEEE INFOCOM},
	Month = apr # { 13-18,},
	Pages = {1966-1974},
	Title = {A distributed optimization algorithm for multi-hop cognitive radio networks},
	Year = {2008}}

@inproceedings{Shi08:Milcom,
	Address = {San Diego, CA},
	Author = {Yi Shi and Y. Thomas Hou and Sastry Kompella},
	Booktitle = {Proc. IEEE Military Communications Conference (MILCOM)},
	Month = nov,
	Pages = {3723--3729},
	Title = {A cross-layer approach to multi-hop networking with cognitive radios},
	Year = {2008}}

@inproceedings{Shi08:MobiHoc,
	Address = {Hong Kong SAR, China},
	Author = {Yi Shi and Sushant Sharma and Y. Thomas Hou and Sastry Kompella},
	Booktitle = {Proc. ACM MobiHoc},
	Month = may,
	Pages = {3--12},
	Title = {Optimal relay assignment for cooperative communications},
	Year = {2008}}

@inproceedings{Shi09:MobiHoc_protocol,
	Address = {New Orleans, LA},
	Author = {Y. Shi and Y. Thomas Hou and Jia Liu and Sastry Kompella},
	Booktitle = {Proc. ACM Mobihoc},
	Month = may # { 18-21,},
	Pages = {239-248},
	Title = {How to correctly use the protocol interference model for multi-hop wireless networks},
	Year = {2009}}

@inproceedings{Shi09:Smartphone_App,
	Address = {Beijing, China},
	Author = {Weiwei Shi},
	Booktitle = {Proc. IEEE International Conference on Electronics Commerce and Business Intelligence},
	Month = jun # { 6-7,},
	Pages = {106-110},
	Title = {An Empirical Research on Users' Acceptance of Smart Phone Online Application Software},
	Year = {2009}}

@inproceedings{Vedantham04:Channl_Assignment,
	Address = {Las Angeles, CA, USA},
	Author = {Ramanuja Vedantham and Sandeep Kakumanu and Sriram Lakshmanan and Raghupathy Sivakumar},
	Booktitle = {Proc. ACM Mobicom},
	Month = sep # { 23-26,},
	Pages = {378-389},
	Title = {Component Based Channel Assignment in Single Radio, Multi-Channel Ad Hoc Networks},
	Year = {2006}}

@inproceedings{Shah11:NetworkFlowFunc,
	Address = {St. Petersburg, Russia},
	Author = {Virag Shah and Bikash Kumar Dey and D. Manjunath},
	Booktitle = {Proc. IEEE International Symposium on Information Theory (ISIT)},
	Month = jul # {31--} # aug # {5,},
	Pages = {234-238},
	Title = {Network Flows for Functions},
	Year = {2011}}

@inproceedings{Sundaresan05:MIMO_Routing,
	Address = {Boston, MA, U.S.A.},
	Author = {K. Sundaresan and R. Sivakumar},
	Booktitle = {Proc. IEEE International Conf. on Network Protocols},
	Month = Nov,
	Pages = {85-98},
	Title = {Routing in Ad Hoc Networks with {MIMO} Links},
	Year = {2005}}

@inproceedings{Swannack05:BC_LFB,
	Author = {C. Swannack and E. Uysal-Biyikoglu and G. Wornell},
	Booktitle = {Proc. Allerton Conf. on Commun., Control, and Comput.},
	Month = Oct,
	Title = {{MIMO} Broadcast Scheduling with Limited Channel State Information},
	Year = {2005}}

@inproceedings{Tekin06:MAC-Wiretap,
	Author = {E. Tekin and A. Yener},
	Booktitle = {44th Annual Allerton Conference on Communication, Control, and Computing},
	Month = sep,
	Title = {Achievable Rates for the General Gaussian Multiple Access Wire-tap Channel with Collective Secrecy},
	Year = {2006}}

@inproceedings{Vedantham04:Channl_Assignment,
	Address = {Las Angeles, CA, USA},
	Author = {Ramanuja Vedantham and Sandeep Kakumanu and Sriram Lakshmanan and Raghupathy Sivakumar},
	Booktitle = {Proc. ACM Mobicom},
	Month = sep # { 23-26,},
	Pages = {378-389},
	Title = {Component Based Channel Assignment in Single Radio, Multi-Channel Ad Hoc Networks},
	Year = {2006}}

@inproceedings{Wang04:SN_Placement,
	Address = {Hong Kong SAR, China},
	Author = {Guiling Wang and Guohong Cao and {La Porta}, T.},
	Booktitle = {Proc. IEEE INFOCOM},
	Month = mar # { 7-11,},
	Pages = {2469-2479},
	Title = {Movement-Assisted Sensor Deployment},
	Year = {2004}}

@inproceedings{Wang_MOBIHOC05,
	Address = {Urbana-Champaign, Illinois, USA},
	Author = {Xin Wang and Koushik Kar},
	Booktitle = {Proc. MOBIHOC},
	Title = {Cross-Layer Rate Control for End-to-End Proportional Fairness in Wireless Networks with Random Access},
	Year = {2005}}

@inproceedings{Wei10:DNewton,
	Address = {Atlanta, GA},
	Author = {Ermin Wei and Asuman Ozdaglar and Ali Jadbabaie},
	Booktitle = {Proc. IEEE Conference on Decision and Control (CDC)},
	Month = dec # { 15-17,},
	Title = {A Distributed {N}ewton Method for Network Utitlity Maximization},
	Year = {2010}}

@inproceedings{Wohlmuth98:GraphEmbedding,
	Author = {O. Wohlmuth and F. Mayer-Lindenberg},
	Booktitle = {Proc. ACM Symposium on Applied Computing},
	Pages = {569-574},
	Title = {A Method for the Embedding of Arbitrary Trees into Hypercubes},
	Year = {1998}}

@inproceedings{Wu05:SN_Placement,
	Address = {Miami, FL},
	Author = {J. Wu and S. Yang},
	Booktitle = {Proc. IEEE INFOCOM},
	Month = mar # { 13-17,},
	Pages = {2313-2324},
	Title = {Smart: {A} Scan-Based Movement-Assisted Sensor Deployment Method in Wireless Sensor Networks},
	Year = {2005}}

@inproceedings{Xu05:RL_Placement,
	Address = {Santa Clara, CA},
	Author = {Xu, K. and Hassanein, H. and Takahara, G.},
	Booktitle = {Proc. IEEE International Conference on Sensor and Ad Hoc Communications and Networks (SECON)},
	Month = sep # { 26-29,},
	Pages = {575-585},
	Title = {Relay Node Deployment Strategies in Heterogeneous Wireless Sensor Networks: {M}ultiple-Hop Communication Case},
	Year = {2005}}

@inproceedings{Yeh_CISS05,
	Address = {Princton, NJ},
	Author = {Edmund M. Yeh and A. S. Cohen},
	Booktitle = {Proc. Conf. Inf. Sci. Syst.},
	Month = mar,
	Title = {Information Theory, Queueing, and Resource Allocation in Multi-user Fading Communications},
	Year = {2004}}

@inproceedings{Ying06:INC,
	Address = {Boston, MA},
	Author = {Lei Ying and R. Srikant and G. Dullerud},
	Booktitle = {Proc. 4th International Symposium on Modeling and Optimization in Mobile, Ad-Hoc and Wireless Networks (WiOpt)},
	Month = apr # { 3-7,},
	Pages = {1-9},
	Title = {Distributed Symmetric Function Computation in Noisy Wireless Sensor Networks with Binary Data},
	Year = {2006}}

@inproceedings{Yu_CISS03,
	Address = {Baltimore, MD, U.S.A.},
	Author = {Wei Yu},
	Booktitle = {Proc. Conf. Information Sciences and Systems (CISS)},
	Month = mar,
	Title = {A Dual Decomposition Approach to the Sum Power {G}aussian Vector Multiple-Access Channel Sum Capacity Problem},
	Year = {2003}}

@inproceedings{Zhang06:SA,
	Address = {Anchorage, Alaska, USA},
	Author = {Junshan Zhang and Dong Zheng and Mung Chiang},
	Booktitle = {Proc. IEEE INFOCOM},
	Month = may,
	Title = {The Impact of Stochastic Noisy Feedback on Distributed Network Utility Maximization},
	Year = {2007}}

@inproceedings{Zhao10:ForkJoin,
	Address = {New York, NY},
	Author = {Haiquan Zhao and Cathy H. Xia and Zhen Liu and Don Towsley},
	Booktitle = {Proc. ACM Sigmetrics},
	Month = jun # { 14-18,},
	Title = {A Unified Modeling Framework for Distributed Resource Allocation of General Fork and Join Processing Networks},
	Year = {2010}}

@inproceedings{Zou03:BS_Placement,
	Address = {San Francisco, CA},
	Author = {Zou, Y. and Chakrabarty, K.},
	Booktitle = {Proc. IEEE INFOCOM},
	Month = mar # {30--} # apr # {3,},
	Pages = {1293-1303},
	Title = {Sensor Deployment and Target Localization Based on Virtual Forces},
	Year = {2003}}

@inproceedings{Zymnis07:DNewton,
	Address = {Monticello, IL},
	Author = {Argyrios Zymnis and Nikolaos Trichakis and Stephen Boyd and Dan O{\'\i}Neill},
	Booktitle = {Proc. Allerton Conference on Communication, Control, and Computing},
	Month = sep # { 26-28,},
	Title = {An Interior-Point Method for Large Scale Network Utility Maximization},
	Year = {2007}}

@misc{TopTenReview:Smartphone2011,
	Organization = {TopTen Reviews},
	Title = {Smartphone Review},
	Url = {http://cell-phones.toptenreviews.com/smartphones/},
	Year = {2011},
	Bdsk-Url-1 = {http://cell-phones.toptenreviews.com/smartphones/}}

@misc{CNET10Testing-2010,
	Organization = {CNET},
	Title = {Cell phone battery chart},
	Url = {http://reviews.cnet.com/cell-phone-battery-life-charts/?tag=leftNav.0},
	Year = {2010},
	Bdsk-Url-1 = {http://reviews.cnet.com/cell-phone-battery-life-charts/?tag=leftNav.0}}

@misc{IEEE802.11n-2009,
	Month = oct,
	Organization = {IEEE Std 802.11n-2009},
	Title = {{IEEE} Standard for Information Technology--Telecommunications and information exchange between systems--Local and metropolitan area networks--Specific requirements {P}art 11: Wireless LAN Medium Access Control ({MAC}) and Physical Layer ({PHY}) Specifications Amendment 5: Enhancements for Higher Throughput},
	Year = {2009}}

@misc{IEEE802.16-2004,
	Month = oct,
	Organization = {IEEE Std 802.16-2004},
	Title = {Air Interface for Fixed Broadband Wireless Access Systems},
	Year = {2004}}

@misc{IEEE802.16e,
	Month = feb,
	Organization = {IEEE Std 802.16e},
	Title = {Air interface for Fixed and Mobile Broadband Wireless Access Systems},
	Year = {2006}}

@misc{3GPP_LTE,
	Organization = {3GPP. TS-36 Series},
	Title = {Feasibility Study of Evolved {UTRA} and {UTRAN}}}

@misc{Johnson04:DSR,
	Author = {D. B. Johnson and D. A. Maltz and Y.-C. Hu},
	Howpublished = {Internet Draft draft-ieft-manet-dsr-10.txt},
	Month = jul,
	Organization = {IETF MANET Working Group},
	Title = {The dynamic source routing protocol for mobile ad hoc networks ({DSR})},
	Year = {2004}}

@misc{Perkins03:AODV,
	Author = {C. Perkins and E. Belding-Royer and S. Das},
	Howpublished = {RFC 3561},
	Month = jul,
	Organization = {Internet Engineering Task Force},
	Title = {Ad Hoc On-Demand Distance Vector ({AODV}) Routing},
	Year = {2003}}

@misc{Clausen03:OLSR,
	Author = {T. Clausen and P. Jacquet and A. Laouiti and P. Muhlethaler and A. Qayyum and L. Viennot},
	Howpublished = {RFC 3626},
	Month = oct,
	Organization = {Internet Engineering Task Force},
	Title = {Optimized Link State Routing Protocol ({OLSR})},
	Year = {2003}}

@misc{ABIPresentation07:Indoor,
	Organization = {2nd Int'l. Conf. Home Access Points and Femtocells},
	Title = {Presentations by {ABI} {R}esearch, {P}icochip, {A}irvana, {IP}.access, {G}artner, {T}elefonica {E}spana},
	Url = {http://www.avrenevents.com/dallas-femto2007/purchase_presentations.htm},
	Year = {2007},
	Bdsk-Url-1 = {http://www.avrenevents.com/dallas-femto2007/purchase_presentations.htm}}

@misc{BigData10:Economist,
	Organization = {Economist},
	Title = {Data, Data Everywhere},
	Url = {http://www.economist.com},
	Year = {2010},
	Bdsk-Url-1 = {http://www.economist.com}}

@misc{BigData12:IBM,
	Organization = {IBM},
	Title = {What is Big Data?},
	Url = {http://www-01.ibm.com/software/data/bigdata/},
	Year = {2012},
	Bdsk-Url-1 = {http://www-01.ibm.com/software/data/bigdata/}}

@misc{Hadoop,
	Author = {{Hadoop. {\tt http://hadoop.apache.org}}}}

@misc{LEED,
	Author = {{USGBC. {\tt http://new.usgbc.org/leed}}}}

%% file: Supporting_Preambles/symbols_commands.tex
\renewcommand{\u}{\mathbf{u}}

\newcommand{\x}{\mathbf{x}}

\newtheorem{thm}{Theorem}

\newtheorem{lem}[thm]{Lemma}

\algtext*{EndWhile}% Remove "end while" text
\algtext*{EndIf}% Remove "end if" text
\algtext*{EndFor}% Remove "end while" text
\algtext*{EndFunction}% Remove "end if" text
\algtext*{EndProcedure}% Remove "end while" text

%% file: Abstract/Abstract.tex
% !TEX root = ../ML_Networking_INFOCOM20.tex

\begin{abstract}
%Thanks to the rise of machine learning (ML) and its wide range of applications, 
In recent years, to sustain the resource-intensive computational needs for training deep neural networks (DNNs), it is widely accepted that exploiting the parallelism in large-scale computing clusters is critical for the efficient deployments of DNN training jobs.
However, existing resource schedulers for traditional computing clusters are not well suited for DNN training, which results in unsatisfactory job completion time performance.
The limitations of these resource scheduling schemes motivate us to propose a new computing cluster resource scheduling framework that is able to leverage the special layered structure of DNN jobs and significantly improve their job completion times.
Our contributions in this paper are three-fold:
i) We develop a new resource scheduling analytical model by considering DNN's layered structure, which enables us to analytically formulate the resource scheduling optimization problem for DNN training in computing clusters;
ii) Based on the proposed performance analytical model, we then develop an efficient resource scheduling algorithm based on the widely adopted parameter-server architecture using a sum-of-ratios multi-dimensional-knapsack decomposition (SMD) method to offer strong performance guarantee;
iii) We conduct extensive numerical experiments to demonstrate the effectiveness of the proposed schedule algorithm and its superior performance over the state of the art.
\end{abstract}

%% file: Sec1_Intro/Sec1_Intro.tex
% !TEX root = ../ML_Networking_INFOCOM20.tex

\section{Introduction} \label{sec:intro}

In recent years, deep-learning-based applications are quickly finding their ways into our everyday life,
including healthcare, automobile, retail, smart homes, just to name a few.
However, these applications also generate and inject a large volume of resource-intensive computing jobs for training deep neural networks (DNNs), which are used in various systems for computer vision, natural language processing, online recommendation, etc.
In order to sustain such a rapidly growing need for DNN training in recent years, it is widely accepted that a viable solution is to exploit the vast {\em parallelism} in distributed computing architectures to schedule deep learning jobs.
%Indeed, the recent success of machine learning research and applications is due in a large part to the advances in multi-core CPU/GPU technologies (on micro-chip scale) and networked cloud computing (on macro-computing-cluster scale), which enable the deployment of highly parallel DNN frameworks (e.g., MXNet, PyTorch, TensorFlow, etc.).
%
To date, however, most traditional resource scheduling schemes for computing clusters are not designed for DNN training (e.g., Google's Borg System~\cite{Verma15:Eurosys}
%Borg ~\cite{Verma15:Eurosys} by Google 
employs static resource allocation specified by the users upon job submissions).
%In addition, existing resource schedulers do not leverage the special layered structural properties of deep learning frameworks (i.e., the parameter server architecture) and the iterative characteristics of DNN training to maximize system efficiency\footnote{To speed up the training process, the idea of exploiting the special layered structure of DNNs to overlap communication and computation has been explored recently in ~\cite{Hashemi19:SysML,Jayarajan19:SysML,Peng19:SOSP}, which showed that the training throughput of the MXNet framework could be improved by 25\%-70\%.}.
Also, most of the recently proposed scheduling schemes designed for DNN jobs (e.g., Gandiva~\cite{Xiao18:Gandiva} and Tiresias~\cite{Gu19:Tiresias}) are heuristic approaches, which provide no performance guarantee.
In light of the increasing importance of DNN-based applications, there is a pressing need for developing provably efficient resource schedulers tailored for DNN training in computing clusters.
%with theoretical performance guarantee.

However, developing such resource scheduling algorithms for DNN training clusters is highly non-trivial.
In a computing cluster, DNN training jobs are submitted over time with various competing resource requirements (numbers of CPUs and GPUs, size of memory, etc.), and the training process is both resource-intensive and time-consuming.
% (e.g., numbers of parameter servers and workers, size of memory, etc.).
%Different job admission and resource scheduling decisions could lead to drastically different DNN training performances in distributed computing clusters.
% An improperly designed scheduling algorithm could result in excessive delay in DNN training. 
For example, researchers showed that it could take 115 minutes to train a model with ResNet50 dataset~\cite{He16:ResNet50} on a DGX-1 machine with 8 V100 GPUs~\cite{MLPerf19}, and even 3--5 days to train the DeepSpeech2 model~\cite{Amodei16:ICML} on the LibriSpeech dataset~\cite{LibriSpeech} using 16 GPUs~\cite{Amodei16:ICML}.
%It is time-consuming to train a DNN model due to its complexity and large dataset, e.g., It takes .
Also, to date, there is a lack of a tractable and accurate analytical model that takes different mechanisms of communication-computation overlapping into consideration based on the layered structure of DNN.\footnote{To speed up the training process, the idea of exploiting the special layered structure of DNNs to overlap communication and computation has been explored recently in\cite{Hashemi19:SysML,Jayarajan19:SysML,Peng19:SOSP}, which showed that the training throughput of the MXNet framework could be improved by 25\%--70\%.}
Furthermore, similar to the design of most scheduling algorithms for large-scale distributed computing clusters, the computing resource limitation for DNN computing jobs naturally leads to integer bin-packing-like constraints in the scheduling problem, which makes the problem NP-Hard. 
Also, the objective function of the DNN resource scheduling problem has a sum-of-ratios structure due to the computational speed characterization in DNN training.
%As will be shown later
As a result, the scheduling problem is {\em non-convex} even with continuous relaxation, which introduces yet another layer of difficulty to the already-challenging problem.

%Paragraph3: Contributions
In this paper, we overcome the above challenges and develop a suite of scheduling algorithmic techniques for efficient DNN training in computing clusters
%\footnote{{\color{red} We consider the homogenous setting, where straggling issue is not critical.}}.
Our main results and technical contributions are summarized as follows:

\begin{list}{\labelitemi}{\leftmargin=1em \itemindent=-0.0em \itemsep=.2em}

\item We develop a new performance analysis framework for scheduling DNN training jobs.
Specifically, we first propose a {\em unified} analytical model to characterize the DNN training, which captures a variety of ways to overlap communication and computation by exploiting the layered structure of DNNs.%forward and backward propagation times in
We then formulate the job admission and resource scheduling problem for DNN training by associating the above unified analytical model with both synchronous and asynchronous stochastic gradient descent (SGD) algorithms. 
% to maximize the overall system utility in terms of job completion time.
Each job is associated with a utility function, which is non-increasing with respect to its job completion time.
The objective of the analytical model is to maximize the overall utility (i.e., minimize the overall training completion time).
    
\item Based on the above analytical framework, we formulate the resource scheduling problem as a mixed-integer nonlinear programming problem (MINLP), and prove its NP-hardness.
To overcome this fundamental hardness, we propose a divide-and-conquer approach called SMD (\underline{s}um-of-ratio \underline{m}ulti-dimensional-knapsack \underline{d}ecomposition).
Specifically, based on a keen observation of the physical reality of resource requests in most distributed cloud computing systems in practice (e.g., Amazon's EC2), we show that our resource scheduling problem has a {\em decomposition} structure.
Under this decomposition, the inner subproblem is a mixed-integer sum-of-ratios problem with packing constraints.
Thanks to the lower dimensionality of the inner subproblem, we are able to develop an efficient $\epsilon$-approximation algorithm
%\footnote{For a maximization problem in optimization, given $\epsilon\in(0,1)$, we say that an algorithm is an $\epsilon$-approximation algorithm if it guarantees to produce a solution $\tilde{\mathbf{x}}$ with an objective value $\tilde{z}$ that is no less than $(1-\epsilon)z^*$ (i.e., $\tilde{z}\geq(1-\epsilon)z^*$), where $z^*$ represents the optimal objective value.} 
based on grid-searching coupled with randomized rounding
%\footnote{The randomized rounding technique and the proof techniques for its approximation ratio analysis are general and could be of independent interest.} 
for solving this subproblem.

\item Upon solving the inner subproblem, we show that the outer subproblem reduces to a multi-dimensional knapsack problem (MKP), which also admits an efficient $\epsilon$-approximation algorithm. 
By combining both steps, we establish the overall approximation ratio of the proposed SMD approach.
To verify the efficacy of our proposed algorithm, we conduct numerical experiments based on Google cluster traces~\cite{Googletrace15:Github}.
Our results show that the proposed SMD approach significantly outperforms the equal server-worker allocation scheme (widely used in practice) and a state-of-the-art approach called Optimus~\cite{Peng18:ML_EuroSys}.
\end{list}

%Collectively, our results contribute to a comprehensive and fundamental understanding of resource scheduling for DNN training in distributed computing clusters.
The remainder of this paper is organized as follows. In Section~\ref{sec:relatedwork}, we review the literature to put our work in a comparative perspective.
%Section~\ref{sec:prelim} familiarizes readers with necessary DNN background.
In Section~\ref{sec:problemformulation}, we introduce the system model and problem formulation.
Section~\ref{sec:algorithm} presents our algorithms and their performance analysis.
Section~\ref{sec:numerical} shows numerical results, and Section~\ref{sec:conclusion} concludes this paper.

%% file: Sec2_Related/Sec2_Related.tex
% !TEX root = ../ML_Networking_INFOCOM20.tex

\section{Related Work} \label{sec:relatedwork}

Due to the rise of machine learning (ML) applications and their high computational workload, optimizing resource scheduling to facilitate distributed ML frameworks have attracted a great amount of interest in recent years (see, e.g.,~\cite{Ghodsi11:DRF,Jyothi16:Morpheus,Tumanov16:TetriSched,Sun17:Dorm} and the references therein).
%many distributed ML frameworks have been proposed to leverage modern large-scale computing clusters.
% Parameter Server (PS) architecture is widely adopted in these distributed ML frameworks~\cite{Li14:OSDI,Chilimbi14:OSDI}.
DNN training jobs have unique characteristics (e.g., iterativeness and layered structure), which could be leveraged to overlap computation and communication time between iterations to reduce the training time. 
However, these existing works were designed for resource allocation to support {\em general} ML jobs in computing clusters, which %may or 
 may not be tailored for DNN training jobs.
As a result, when being applied in DNN training, their performance is suboptimal in general since they do not leverage the aforementioned characteristics of DNN training.
%thus developing customized efficient resource allocation scheduler on DL jobs attracts increasingly interests.

To date, research on computing cluster scheduling optimization tailored for DNN training remains relatively new with limited results.
Most of the early attempts in this area (e.g.,~\cite{Chilimbi14:OSDI} and references therein) only considered static allocation of workers and parameter servers (PS).
To our knowledge, Yan \textit{et al.}~\cite{Yan15:KDD} was the first to investigate the performance of distributed ML frameworks, and developed a DNN performance model at a layer-level granularity (e.g., the model considers the computation time of each operator on a specific CPU and the NN structures).
Subsequently, Gandiva~\cite{Xiao18:Gandiva} exploited intra-job predictability to time-slice GPUs efficiently across multiple jobs to better-fit GPUs.
% , but they did not consider resource adjustments.
 %To our knowledge, 
Tiresias~\cite{Gu19:Tiresias} aimed to reduce the job completion times when the jobs' execution times are unpredictable due to non-smooth loss curves during a trial-and-error exploration.
The most recent work~\cite{Mahajan20:Themis} focused on making machine learning workloads complete in a finish-time fair manner.
However, these schedulers are based on heuristic approaches.
Also, the training completion time of a job is significantly affected by resource allocation.
Studies in~\cite{Peng18:Eurosys} showed that increasing resources did not contribute to a linear increase of the training speed and could even slow down the training process.
Since static numbers of workers and PSs specified by users are suboptimal in general,
researchers have also started to consider dynamic scheduling algorithms to determine  optimal numbers of workers and PSs to optimize the training speed.
To our knowledge, the first dynamic scheduling algorithm with performance guarantee was reported by Bao \textit{et al.}~\cite{Bao18:ML_INFOCOM}, where they designed an online scheduling algorithm for deep learning jobs.
However, their studies relied on strong assumptions and simplified modeling of deep learning jobs.
%In Optimus~\cite{Peng18:Eurosys}, Peng et al.~proposed a customized scheduler for deep learning jobs based on parameter server architecture with no performance guarantee.

The most related work to ours is Optimus~\cite{Peng18:Eurosys}, where Peng {\em et al.} developed a heuristic resource allocation algorithm for the distributed deep learning jobs.
Our work differs from~\cite{Peng18:Eurosys} in the following key aspects: 
1) In~\cite{Peng18:Eurosys}, the authors built the performance model without taking the DNN layered structure into consideration.
As will be shown later, this yields suboptimal scheduling decisions in general.
By contrast, we develop an analytical model that considers the layered characteristics of DNN training, which captures communication-computation overlapping in state-of-the-art DNN training systems~\cite{Jayarajan19:SysML};
and 2) Optimus proposed a dynamic resource scheduler, which is only a heuristic with no performance guarantee based on their online-fitted resource-performance models.
In comparison, we propose a resource scheduler that leverages an analytical model to offer strong performance guarantees.
% \textbf{Communication and computation overlap: }The state-of-the-art DNN frameworks such as TensorFlow and MXNet support overlap communication with BP.
% Recent work P3 [cite] further proposed a better parameter synchronization by allowing overlap between computation and FP using layer slicing.
% It mainly focused on the MXNet parameter server architecture.
% TicTac[cite] proposed the similar idea to P3, but it was based on TensorFlow parameter server TCP architecture.
% These previous work showed their speedup compared to existing parameter synchronization strategies.
% However, their models were not transferable among different frameworks due to the underlying global barrier.
% The most recent work ByteScheduler [cite] thus proposed a generic model which can work across multiple DNN frameworks and demonstrated its higher speedup.

% In our work, we formulate the DNN training speed model at the finer granularity by considering the special DNN layered structure where different parameter synchronization approaches are studied, then map the model to the high-level resource allocation, where the global optimal scheduling is analyzed.

% \begin{figure}[t]
%     \centering
%     \includegraphics[width=0.4\textwidth]{Figures/fig_cluster.eps}
%     \caption{Overview of the system model.}
%     \label{fig:cluster}
% \end{figure}

%% file: Sec3_Model/Sec3_Model.tex
% !TEX root = ../ML_Networking_INFOCOM20.tex

\section{System Model and Problem Formulation} \label{sec:problemformulation}
% In this section, we first introduce our analytical model for ML jobs both under synchronous training and asynchronous training, and then we formulate our integer nonlinear sum-of-ratios problems.
In this section, we first present our general problem formulation for DNN training in distributed computing clusters in Section~\ref{sec:overall_formulation}.
We then specialize the problem formulation under both synchronous and asynchronous SGD with the basic sequential training model~\cite{Zhang17:Poseidon} in Section~\ref{sec:sequential_model}. 
Lastly, in Section~\ref{sec:advanced_models}, we further generalize and refine our analytical model to include two more advanced DNN training models.
%For example, in Amazon EC2, upon the submission of a training job, the user specifies his/her resource demands (e.g., the numbers of workers and PSs, the numbers of CPUs and GPUs, etc.).

%Clearly, the training completion time of a job is significantly affected by resource allocation.
%For example, studies in~\cite{Peng18:Eurosys} showed that the fastest training speed was achieved when eight workers and 12 PSs were allocated to synchronously train a ResNet-50 model~\cite{He16:ResNet50} on the ImageNet dataset~\cite{ImageNet} (the total number of workers and PSs is fixed to 20) in their cluster.
%It was also shown in~\cite{Peng18:Eurosys} that increasing resources did not contribute to a linear increase of the training speed and could even slow down the training process.
%Clearly, the fixed numbers of workers and PSs specified by users are often suboptimal.
%This motivates us to design resource scheduling algorithm to determine optimal numbers of workers and PSs to achieve the fastest training speed.

\subsection{General Problem Formulation} \label{sec:overall_formulation}
\begin{figure}[t]
\centering
\begin{minipage}{.23\textwidth}
   \centering
  \includegraphics[height=0.11\textheight]{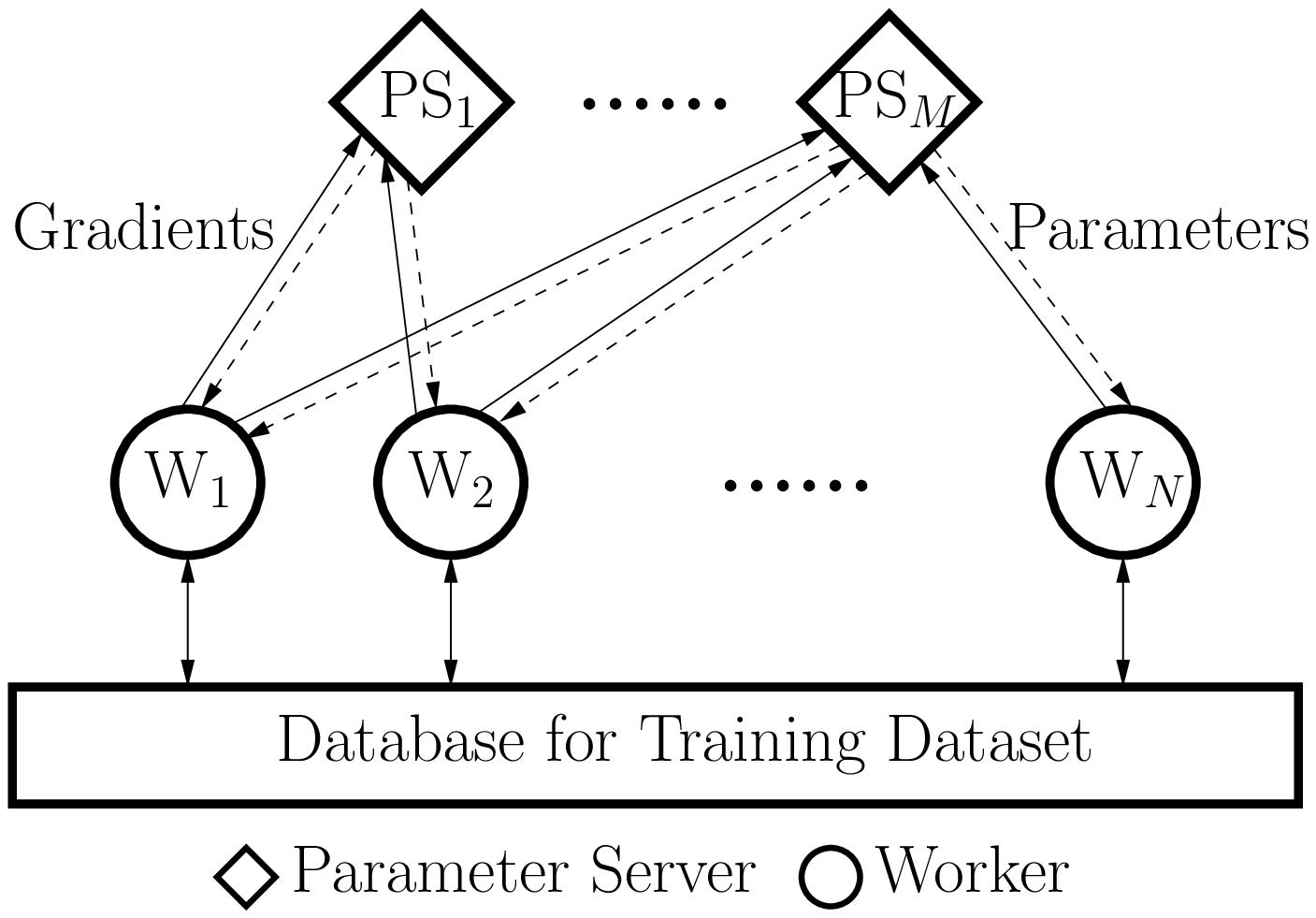}
  \vspace{-.06in}
  \caption{PS-based architecture.}
  \label{fig:architecture}
\end{minipage}%
\hspace{0.001\columnwidth}%
\begin{minipage}{.25\textwidth}
   \centering
  \includegraphics[height=0.116\textheight]{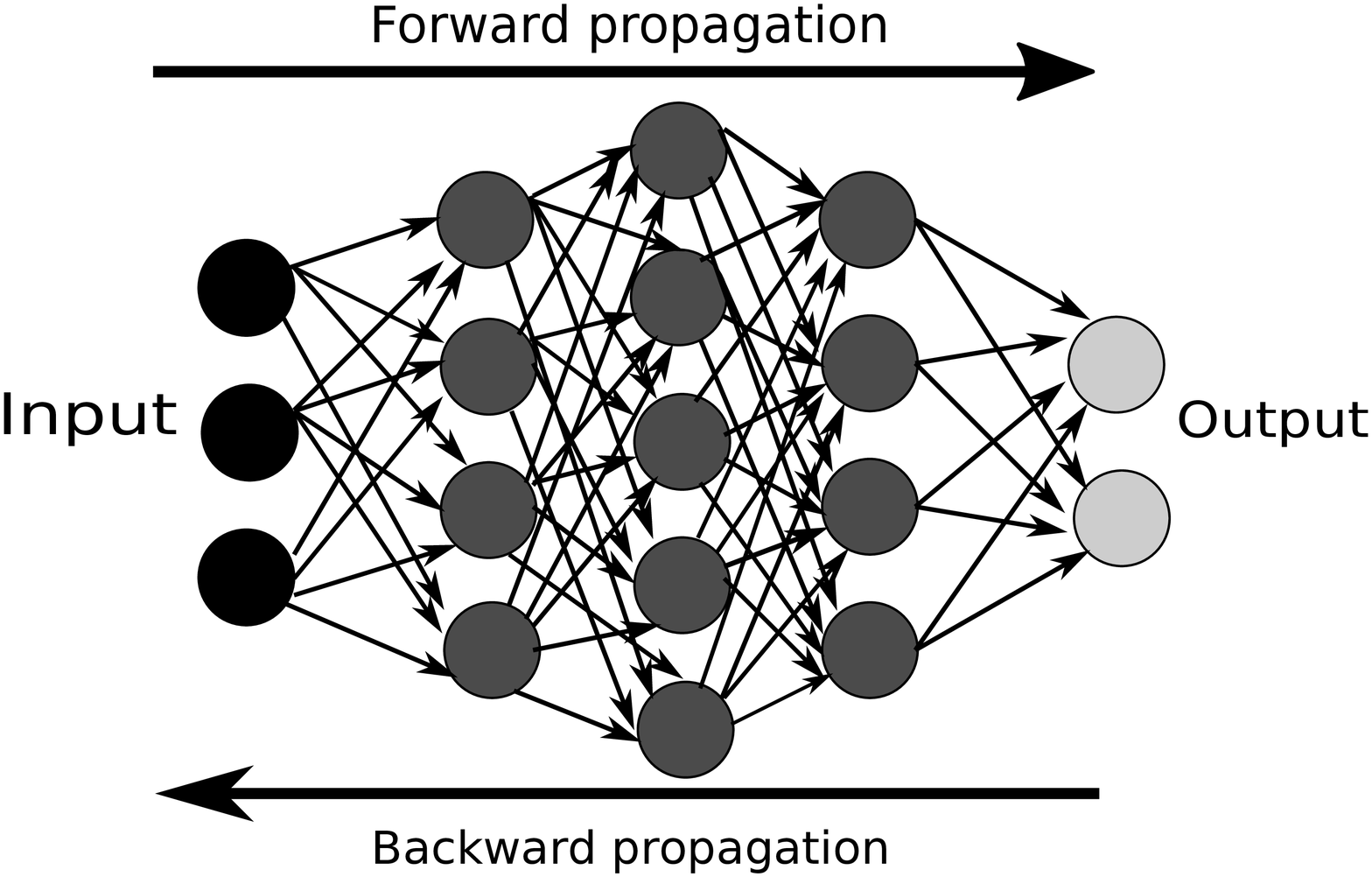}
  \vspace{-.06in}
  \caption{DNN training.}
  \label{fig:DNN}
\end{minipage}
\vspace{-.2in}
\end{figure}

In this paper, we focus on data parallelism in data centers with the PS architecture~\cite{Dean:2012:LSD,Li14:OSDI}\footnote{Due to the homogeneous processing speed in data centers, we assume that the straggling issue is minor in this paper.}.
As shown in Fig.~\ref{fig:architecture}, 
%this architecture includes workers, PSs, and the training dataset.
in each iteration, each worker fetches a mini-batch of samples from its local dataset and computes a stochastic gradient.
Upon finishing computation, each worker sends the gradient to the PSs to update the model parameters following an optimization algorithm, e.g., the stochastic gradient descent (SGD) method.
The workers then will pull the updated parameters from the PSs, fetch the next mini-batch of samples and proceed to the next iteration.
This process repeats until some convergence criterion of the optimization algorithm is met.
%In the training process, an ``epoch'' is referred to as the duration that the workers exhaust all the samples in the training dataset.

The gradient computation at each worker is based on the specific DNN  % model. , which is a deep neural network in essence. 
as illustrated in Fig.~\ref{fig:DNN}.
%a typical DNN includes the input layer, the hidden layer(s), and the output layer.
In each iteration, one mini-batch of samples is used by a worker to compute the loss from the initial layer to the last layer, which is referred to as {\em forward propagation} (FP).
After FP, a stochastic gradient of the DNN model parameters will be computed in the reverse order of layers in a {\em backward propagation} (BP) fashion.
The computed gradients are used for updates at the PSs.

We consider the setting where DNN training jobs are submitted by users to the computing cluster to be processed periodically.
These submitted jobs are trained using workers and PSs implemented via virtual machines or containers, which share resources in the underlying servers.
Upon submission, the user will specify the resource needed for the job, based on which our algorithm will determine the numbers of workers and PSs if the job is admitted in the current scheduling interval.
%The scheduling interval could be some sufficiently large time range (e.g., a day). 
The jobs that are submitted during the scheduling interval are called \textit{active jobs}.
We assume that the scheduling interval is sufficiently large (e.g., hours) such that all the scheduled active jobs can be completed within a scheduling interval.

% In each scheduling interval, which can be some fixed time range such as one day, to collect all the submitted jobs.
We use $E[i]$ to denote the total number of iterations to train for job $i$, which is specified by the user.\footnote{In practice, to prevent spending excessively long time waiting for the training process of a DNN job to converge, a maximum number of training iterations is usually set by the user.} 
We use $f(p[i],w[i])$ to denote the current training speed of job $i$ (i.e., the number of iterations completed per unit time), which is a function of the number of PSs ($p[i]$) and the number of workers ($w[i]$).
Then, the completion time of job $i$ can be estimated as $\frac{E[i]}{f(p[i],w[i])}$.
% $\pi_i\triangleq\frac{E[i]}{f(p[i],w[i])}$.
Let $\mathcal{I}$ denote the set of active jobs submitted in the scheduling interval with $|\mathcal{I}|=I$, and $\mathcal{R}$ denote the set of computing resources (e.g., CPU, GPU and memory). 
Let $O^r[i]$ and $G^r[i]$ be the amount of type-$r$ resources demanded by a worker and a PS for job $i$, respectively.
Let $v^r[i]$ be the resource limit specified by the user for job $i$, i.e., the maximum amount of resource $r$ the user may need.
Let $C^r$ denote the capacity of type-$r$ resource in the computing cluster.
Due to resource limits in the cluster, the system may not be able to process all jobs in $\mathcal{I}$ in the current scheduling interval.
Thus, we use a binary variable $x_i$ to indicate whether job $i$ is admitted in this interval ($x_i=1$) or not ($x_i = 0$). 
%If job $i$ is admitted, then $x_i=1$; otherwise, $x_i = 0$.
Let $\mu_i(\cdot)\geq0$ be the utility function associated with job $i$, which is {\em non-increasing} with respect to the completion time $\frac{E[i]}{f(p[i],w[i])}$.
% We set $\mu_i(\frac{E[i]}{f(p[i],w[i])})=\omega_i\frac{E[i]}{f(p[i],w[i])}$, where $\omega_i>0$.
In this paper, our goal is to optimize the admission decision $\mathbf{x} \triangleq \{x_{i}, i\in\mathcal{I}\}$ and resource allocation decisions $\mathbf{p} \triangleq \{p[i], i\in\mathcal{I}\}$ and $\mathbf{w} \triangleq \{ w[i], i\in\mathcal{I}\}$ to maximize the overall utility for the submitted jobs in each scheduling interval.
This optimization problem can be formulated as:
\begin{align}
    \label{general_formulation}\underset{\mathbf{w,p,x}}{\text{Maximize }} &\sum_{i\in\mathcal{I}}\mu_i \bigg( \frac{E[i]}{f(p[i],w[i])} \bigg)x_i \\
    \label{selet}\text{subject to }&\sum_{i\in\mathcal{I}}v^r[i]x_i\leq C^r, \quad\forall r\in\mathcal{R}, \\
    \label{resr_constr} &(O^r[i] w[i] \!+\!G^r[i] p[i])x_i \!\leq\! v^r[i], \forall i\!\in\!\mathcal{I},r\!\in\!\mathcal{R}, \\
    &p[i]\in \mathbb{Z}^{++}, w[i]\in \mathbb{Z}^{++}, x_i \in \{0,1\}, \quad\forall i\in\mathcal{I}. \nonumber
\end{align}

% where $f(p[i],w[i])$ can be replaced with Eqn.~(\ref{sync_speed_fn}) and Eqn.~(\ref{async_speed_fn}) under synchronous training and asynchronous training, respectively.
% And $I'$ is the set of jobs with $x_i=1$.
%In the above formulation,
Constraint~(\ref{selet}) ensures that the sum of maximum resource demands from admitted % user-specified 
active jobs does not exceed the cluster's resource capacity.
Constraint~(\ref{resr_constr}) ensures that the allocated resources to run workers and PSs in each job $i$ do not exceed the resource limits specified by the job owner.

\subsection{Training Speed Modeling} \label{sec:sequential_model}

In the general problem formulation above, the training speed function $f(p[i],w[i])$ remains to be defined.
Next, we will first derive the training speed $f(p[i],w[i])$ per iteration per worker based on the basic sequential computation-communication model~\cite{Zhang17:Poseidon}, which will serve as a baseline for two more advanced computation-communication models in Sec.~\ref{sec:advanced_models}.
%Next, we will first introduce the key notations.

% In this section, we will model the training speed function $f(p[i],w[i])$ based on computation and communication patterns in the parameter server architecture.
% We use the sequential based model as our analytical baseline model, and then we will derive the wait-free based model and priority based model where the parameter synchronization is leveraged.

% \subsection{Distributed machine learning training models}
% \label{sec:model}

% In this section, we first provide a brief overview of three different training models based on whether and what parameter synchronization mechanism is employed, and then the training speed model will be established and analyzed.
% Remarkably, these models can be unified to the same theoretical model.

\textbf{Notation:} We let $N[i]$ denote the number of layers in the DNN model of job $i$. 
We use $r_j[i]$ to denote the time spent for sending gradients to or receiving parameters from the PSs for layer $j$ of job $i$ (assuming equal time for pushing gradients and pulling updated parameters).
Let $b_j[i]$ and $f_j[i]$ be the BP and FP computation times for layer $j$ of job $i$, respectively.
We use $\kappa_j[i]$ to denote the start time of sending gradients for layer $j$ of job $i$, and we let $s_j[i]$ be the start time of receiving parameters for layer $j$ of job $i$.
% We let $e_j[i]$ be the end time of communication for layer $j$ of job $i$ in the priority based model.
Let $\tau_j[i]$ be the start time of FP for layer $j$ of job $i$.
% We denote $\kappa_j[i]$ as the start time of sending gradients at layer $j$ of job $i$.
%We let $H_{ib}$ as the summation of all layers' BP time of job $i$.
%Let $H_{ir}$ be the summation of all layers' communication time, including upload and download time of job $i$.
%Let $H_{if}$ be the summation of all layers' FP time of job $i$.
%We denote $t_i$ as the training time per worker per iteration of job $i$.
For lighter notation, we will omit the job index ``$[i]$'' in the subsequent training speed modeling if there is only one training job involved in the context (e.g., $r_j:=r_j[i]$. We will revive ``$[i]$'' if confusion may arise).
Key notation is summarized in Table~\ref{table:notation} for ease of reference.
\begin{table}[t!]
\centering
\caption{Notation.}
\label{table:notation}
\begin{tabular}{| c | l|}
\hline
$\mathcal{I}$ & The set of active jobs in the scheduling interval \\ \hline
$\mathcal{P}[i]$/$\mathcal{R}$ & The set of PSs of job $i$ / The set of resource types\\ \hline
$t[i]$ &  \begin{tabular}[c]{@{}l@{}l@{}}Time of training one sample on a worker of job $i$\end{tabular}\\ \hline
$t_m[i]$ &  \begin{tabular}[c]{@{}l@{}l@{}}Time of one training step on a worker of job $i$\end{tabular}\\ \hline
% $t$ & Unit time of processing one slice \\  \hline
% $\l_{ij}$ & The layer $j$'s height of job $i$ after slicing \\ \hline
%$b_{i}(h,p)$ &\begin{tabular}[c]{@{}l@{}l@{}}Bandwidth consumed by a worker of job $i$, where \\ %$b_{i}(h,p) =b_{i}^{(e)}$, if $h\neq p$ or $b_{i}^{(i)}$, otherwise. \end{tabular} \\ \hline
$r_j[i]$ & \begin{tabular}[c]{@{}l@{}l@{}}\# of unit time for sending/receiving gradients\\ to and  parameters from the PSs at layer $j$ of job $i$ 
\end{tabular}\\ \hline
$q_{ij}$ & \begin{tabular}[c]{@{}l@{}l@{}} \# of unit time for parameter update  at layer $j$ of job $i$
\end{tabular}\\ \hline
$b_j[i]$ & \begin{tabular}[c]{@{}l@{}l@{}}\# of unit time for BP  at layer $j$ of job $i$ 
\end{tabular}\\ \hline
$f_j[i]$ & \begin{tabular}[c]{@{}l@{}l@{}} \# of unit time for FP at layer $j$ of job $j$
\end{tabular}\\ \hline
$\kappa_j[i]$ & Start time of sending gradients of job $i$ at layer $j$\\ \hline
$e_j[i]$ & 
\begin{tabular}[c]{@{}l@{}l@{}}End time of communication of job $i$ at layer $j$ in the \\priority based model
\end{tabular}\\ \hline
% $s_j[i]$ & Start time of receiving parameters of job $i$ at layer $j$\\ \hline
$s_j[i]$ & \begin{tabular}[c]{@{}l@{}l@{}}
Start time of receiving parameters of job $i$ at layer $j$\end{tabular}\\ \hline
$\tau_j[i]$ & \begin{tabular}[c]{@{}l@{}l@{}}
Start time of FP of job $i$ at layer $j$\end{tabular}\\ \hline
$N[i]$ & \# of layers in DNN model of job $i$\\ \hline
%$\mathcal{R}$ & The set of resource types \\ \hline
$E[i]$/$g[i]$ & \# of training iterations for job $i$  / Model size of job $i$\\ \hline
$K[i]$/$m[i]$ & Global batch size of job $i$ / One mini-batch size of job $i$ \\ \hline
$C^{r}$ & Capacity of type-$r$ resource in the DL cluster \\ \hline
$O^r[i]$ & Type-$r$ resource required by a worker in job $i$ \\ \hline
$G^r[i]$ & \begin{tabular}[c]{@{}l@{}l@{}}
Type-$r$ resource required by a PS in job $i$ \end{tabular}\\ \hline
$v^r[i]$ & The resource limit specified by the user for job $i$ \\ \hline
$w[i]$/$p[i]$ & \# of workers of job $i$ / \# of PSs of job $i$ \\ \hline
%$p[i]$ & Number of PSs of job $i$ \\ \hline
$B[i]$ & Bandwidth capacity of each PS of job $i$\\ \hline
%$g[i]$ & Model size of job $i$ \\ \hline
$t_f[i]$ & \begin{tabular}[c]{@{}l@{}l@{}} Average time of processing a sample  in the FP of job $i$
\end{tabular} \\ \hline
$t_b[i]$ & \begin{tabular}[c]{@{}l@{}l@{}} Average time of processing a sample in the BP of job $i$
\end{tabular} \\ \hline
$t_r[i]$ & \begin{tabular}[c]{@{}l@{}l@{}} Average time of processing a sample in the \\communication time of job $i$
\end{tabular} \\ \hline
%$m[i]$ & The size of one mini-batch of job $i$\\ \hline
\end{tabular}
\vspace{-.1in}
\end{table}

% In practice, it is normal in image classification models (e.g., ResNet-50) that they have skewed parameter size distribution since the final layer usually is fully connected and may introduce higher queuing delays compared to the initial lighter layers.\\
% \begin{figure}[h]
%   \centering
%   \includegraphics[width=\linewidth]{Figures/sequential_based.eps}
%   \caption{Illustration of sequential based model}
%   \label{fig:seq}
%   \end{figure} 

%\begin{figure}[h]
%    \centering
%    \includegraphics[width=1\columnwidth]{Figures/sequential-based.eps}
%    \caption{Sequential-based model.}\label{fig:seq}
%\end{figure}

\begin{figure}[t!]
    \centering
    \includegraphics[width=1\columnwidth]{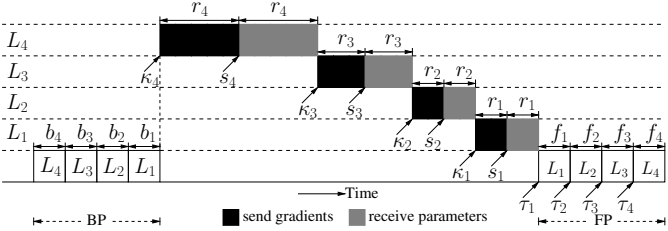}
    \caption{Sequential-based model.}\label{fig:seq}
    \vspace{-.2in}
\end{figure}

The sequential model~\cite{Zhang17:Poseidon} is illustrated in Fig.~\ref{fig:seq}, where push/pull of gradients/parameters start after BP of all layers are done, the push/pull of layers are done sequentially from the highest layer to the first layer, and FP of the next iteration starts after all push/pull are done. % $j$ starts after push/pull of the previous layer $j+1$ is done.
%Note that, here, the BP phase corresponds to the current iteration, while the FP phase is for the next iteration. This is because the gradients are available for transmission only after the BP rather than FP. Also, it can be seen that the BP, the communications for gradients/parameters exchanges, and the FP are arranged in a sequential way, hence the name ``sequential model.''
For an $N$-layer DNN model, it is easy to see that the per-sample training time $t$ can be computed as $\sum_{j=1}^N b_j + 2\sum_{j=1}^N r_j + \sum_{j=1}^N f_j$.
%\begin{align} \label{seq_T} 
%    t=\sum_{j=1}^N b_j + 2\sum_{j=1}^N r_j + \sum_{j=1}^N f_j.
%\end{align}
We let $t_{b}\triangleq\sum_{j=1}^N b_j$, $t_r\triangleq 2\sum_{j=1}^N r_j$, and $t_f\triangleq\sum_{j=1}^N f_j$ denote the BP time, the communication time, and the FP time for processing a sample of job $i$, respectively.
%For convenience, we let $H_{b} \triangleq \sum_{j=1}^N b_j$, $H_{r} \triangleq 2\sum_{j=1}^N r_j$, and $H_{f} \triangleq \sum_{j=1}^N f_j$ denote the BP time, the communication time, and the FP time of job $i$, respectively.
%Let $H_{ir}$ be the summation of all layers' communication time, including upload and download time of job $i$.
%Let $H_{if}$ be the summation of all layers' FP time of job $i$.
Note that the key feature of the sequential model is that computation phases (i.e., BP and FP) and communication phases are conducted in a {\em sequential} fashion, which underutilizes the channel.
%For example, in Fig.~\ref{fig:seq}, gradients of layer $L_4$ can immediately start communications when BP of $L_4$ is finished rather than waiting for the whole BP phase to be completed.
%As a result, the sequential model incurs unnecessary delay.

\begin{figure*}[t!]
   \begin{minipage}[t]{0.285\linewidth}
        \centering
        \includegraphics[width=1.3\textwidth]{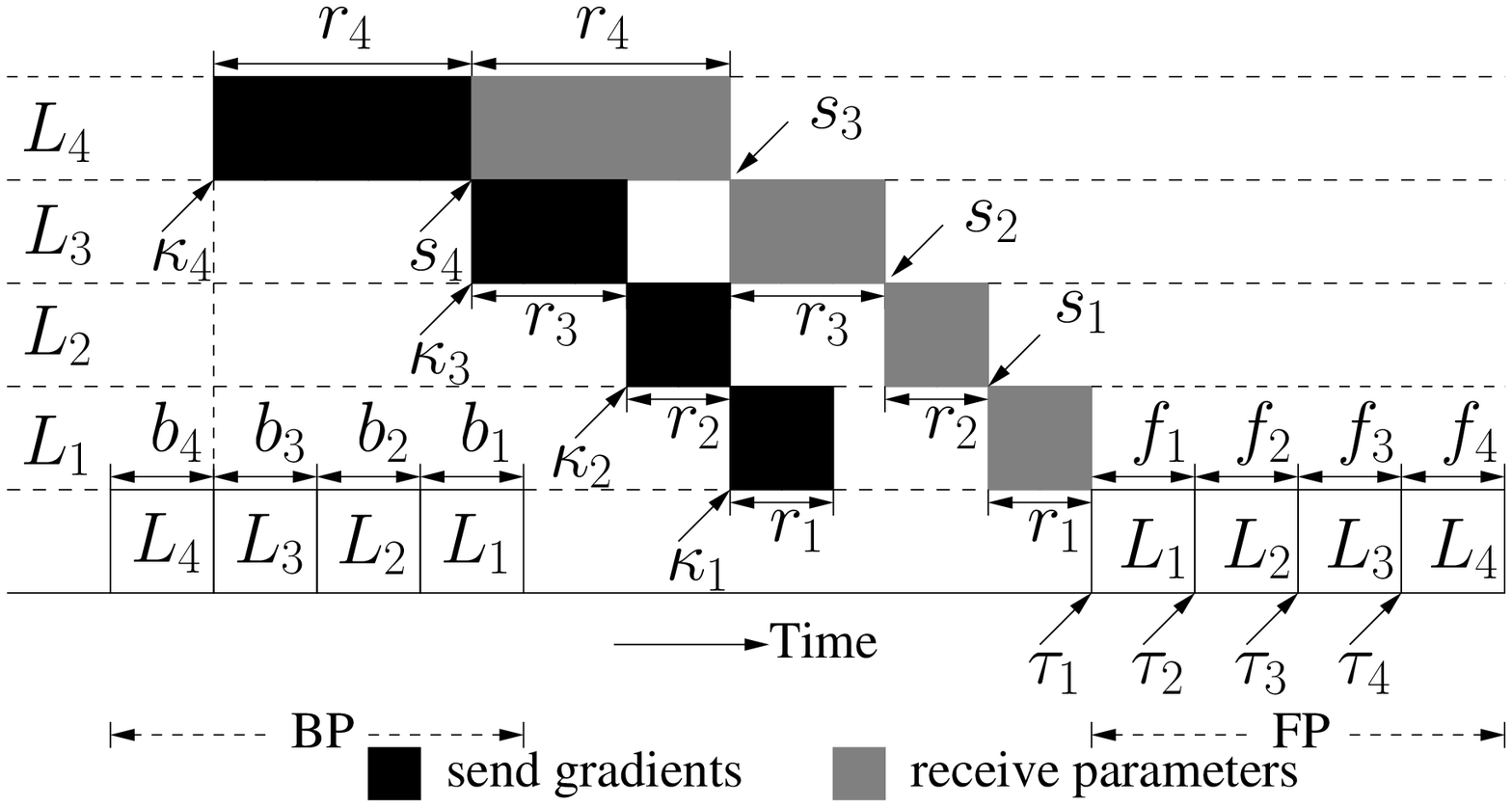}      
        \caption{Wait-free based model.} \label{fig:waitfree}
    \end{minipage}% 
    % \vspace{-.2in}
       \hspace{.074\textwidth}
    %   \vspace{-.1in}
         \begin{minipage}[t]{0.365\linewidth}
        \centering
        \includegraphics[width=0.74\textwidth]{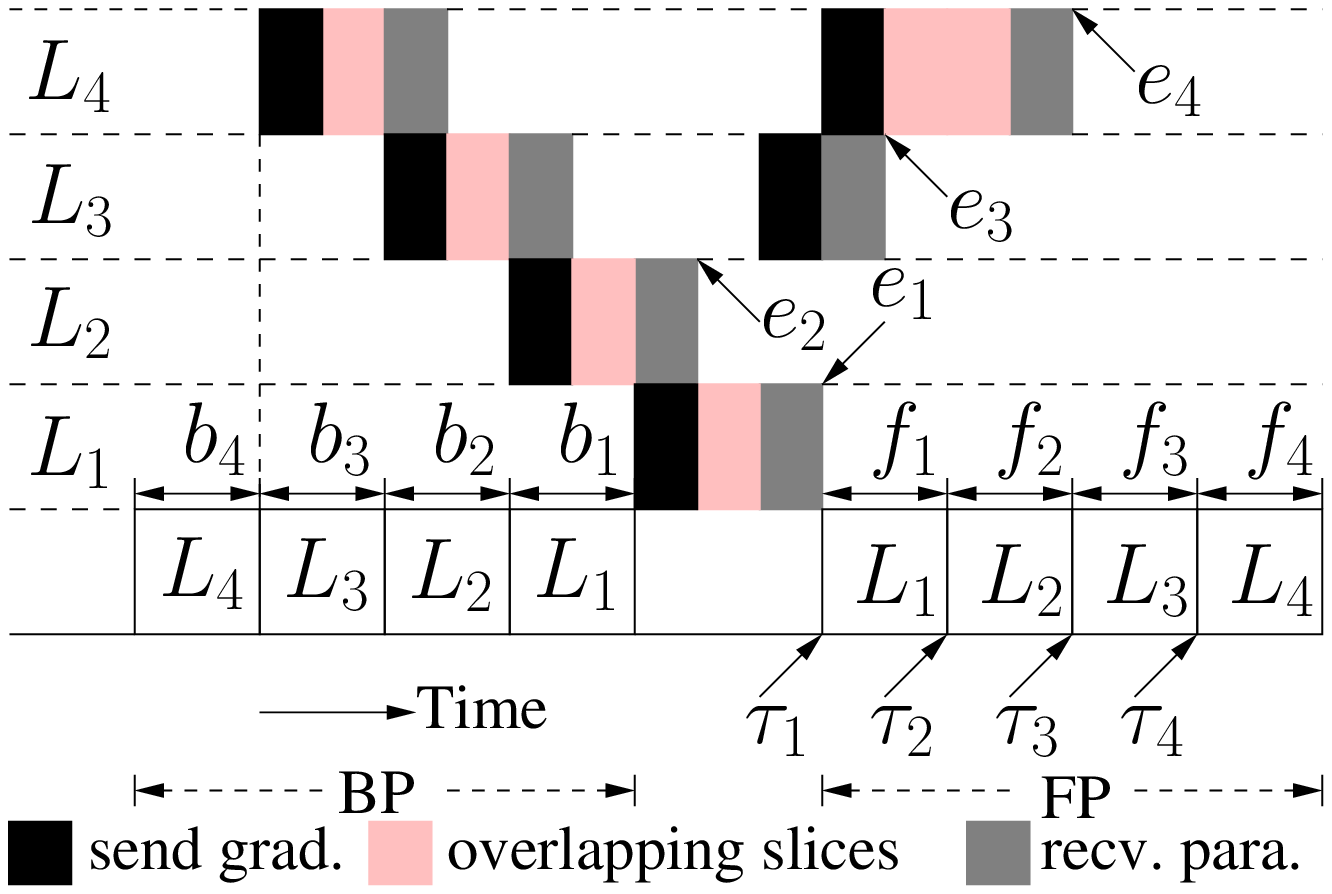}
        \caption{Priority based model.} \label{fig:priority}
    \end{minipage}%
        % \hspace{.02\textwidth}
    \begin{minipage}[t]{0.27\linewidth}
        \centering
        \includegraphics[width=.98\textwidth]{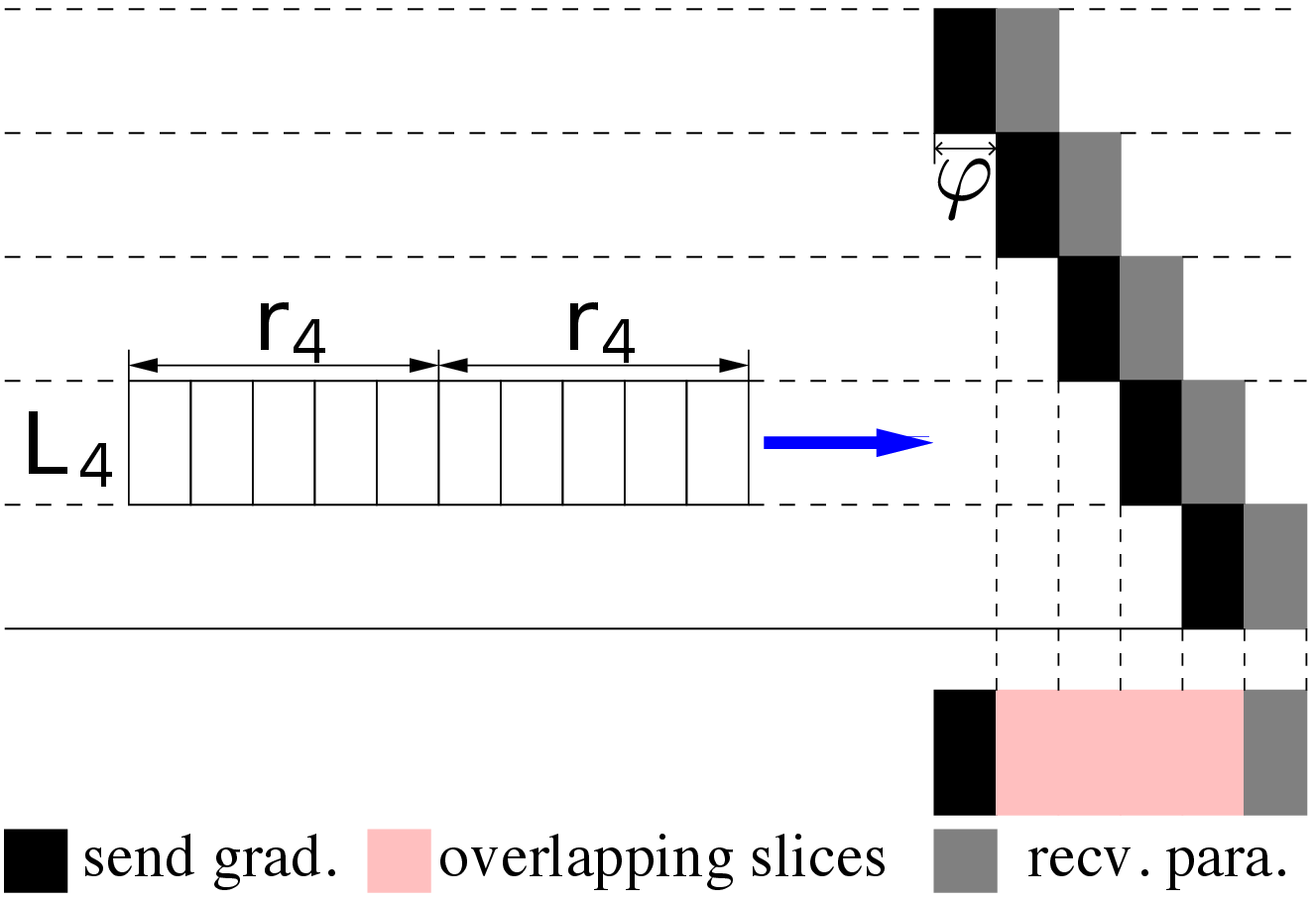}
        \caption{Zoom-in view of slicing.}\label{fig:slicing}
    \end{minipage}%
\vspace{-.2in}
\end{figure*}

Consider a job with $p$ PSs and $w$ workers.
We use $B$ to denote the bandwidth between each pair of PS and worker. 
Let $g$ be the model size (i.e., the number of elements in its gradient vector).
Let $m$ be the mini-batch size.
%Let $t_f$ denote the time for processing a sample in FP, i.e., $t_f=\sum_{j=1}^N f_j$.
Then the FP time for processing a mini-batch can be calculated as $m\cdot t_f$.
The BP time for processing a mini-batch does not depend on $m$ and can be computed as $t_{b}$.
%The BP time does not depend on $m$ and can be computed as $t_{b}=\sum_{j=1}^N b_j$.
% We assume the gradients are evenly divided into parameter servers, then the size of gradients of each parameter server is $\frac{S}{P}$.
%In practice, the communication between worker and parameter server is usually over a network and often becomes the bottleneck to scale the data parallel training linearly.
We use $w'_\rho$ to denote the average number of workers that send the computed gradients simultaneously to a PS $\rho \in \mathcal{P}$, where $\mathcal{P}$ denotes the set of PSs.
Then, the bandwidth occupied by each worker is $\frac{B}{w'_\rho}.$\footnote{In order to guarantee data transfer performance of each instance, it is common to reserve bandwidth for a VM/container for the accelerated computing in the cluster.
For example, the reserved bandwidth of Amazon EC2 GPU instance P2 on AWS is 10Gbps or 25Gbps \cite{AmazonEC2Instances}. Thus we can safely assume that the workers have the same bandwidth.}
%since the parameter server is often the bottleneck in the communication.
% If no parameter synchronization is leveraged, the data transfer time under the symmetric assumption can be computed as $2\sum_{j=1}^{N}r_j$.
% The communication time is related to the transmitted model size and the channel bandwidth, i.e., $2\frac{S/p}{B/w'_\rho}$.
We use $g_j$ to denote the gradient size of layer $j$ and let $p_j$ be the corresponding number of PSs that layer $j$ will send its gradients to and receive its parameters from.
For DNN models in practice, the number of neurons of a layer is usually much larger than the number of PSs, so it holds that $g_j \gg p_j$.
We also assume that $g_j$ is evenly divided into all PSs (implying $p_j = p$).\footnote{For lower implementation complexity and managed overhead, most distributed ML frameworks (e.g., Tensorflow \cite{TensorFlow}) adopt roughly equal parameter allocation by default.} 
% {(\color{red} reasonable in practice?, or we can make $p_j=1,\forall j$ since MXNet stores each layer's gradients in one parameter server)}, i.e., $p_j=p,\forall j$.
Then, the time for sending gradients/receiving parameters under the symmetric assumption (equal communication speed for uplink and downlink) for each layer $j$ can be computed as: $\frac{g_j/p_j}{B/w'_\rho}=\frac{g_j/p}{B/w'_\rho}$.
% If no parameter synchronization is leveraged (i.e., sequential based model), 
It then follows that the data communication time can be computed as $2\sum_{j=1}^N\frac{g_j/p}{B/w'_\rho}=2\frac{\sum_{j=1}^Ng_j/p}{B/w'_\rho}=2\frac{g/p}{B/w'_\rho}$.

% {\color{red}(without assumption, it becomes $2\sum_{j=1}^N\frac{s_j/p_j}{B/w'_\rho}\geq 2\frac{\sum_{j=1}^Ns_j/\sum_{j=1}^Np_j}{B/w'_\rho}=2\frac{S/p}{B/w'_\rho}$: minimizing lower bound makes little sense)}
% $2\frac{S/p}{B/w'_\rho}$, which captures the communication cost 
% We also let $T_{update}$ be the average time to update the parameters with size $S$ in the parameter server's side capturing $\frac{q_1}{r_1}$.
% Then the parameter update time on a parameter server is $\frac{T_{update}\cdot w'_\rho}{p}$ on average.
%Let $N$ be the number of layers in the DNN model.
% We assume the time of FP, BP and parameter update at each layer is roughly the same.
In addition to the communication and computation time, there exists extra communication overhead (e.g., establishing TCP connections) that increases linearly with the number of workers and PSs, which can be computed as $\beta_1 w+\beta_2 p$, where $\beta_1$ and $\beta_2$ are constants that depend on the underlying system~\cite{Peng18:Eurosys}.
Thus, the per-iteration-per-mini-batch training time can be computed as $t_m=\max_{\rho\in\mathcal{P}} \big[ mt_f+t_{b}+2\frac{g/p}{B/w'_\rho}+\beta_1 w+\beta_2 p\big].$
%\begin{align*}
%    % \label{time_per_iter} T&=\max_\rho[m{T_f}+\eta_1T_{b}+2\eta_2\frac{S/p}{B/w'_\rho}+\beta_1 w+\beta_2 p],
%    {\color{red}t_m}&=\max_{\rho\in\mathcal{P}} \bigg[ mt_f+t_{b}+2\frac{g/p}{B/w'_\rho}+\beta_1 w+\beta_2 p\bigg].
%\end{align*}
Next, we derive the training speed function (i.e., how many iterations can be completed per unit time).
We consider both synchronous and asynchronous trainings.

\smallskip
{\em 1) Synchronous Training:} In synchronous training, PSs perform an update only after receiving gradients from all workers in each iteration.
We let $K$ denote the global batch size, which is fixed throughout all iterations in synchronous training.
Keeping a fixed global batch size is conformal to the standard SGD implementation, which is important for ensuring the same model result in the training process~\cite{Goyal17:SGD,Bazaraa_Sherali_Shetty_93:NLP}.
%The fixed global batch size is important for ensuring the same model result and compliant with the standard SGD implementation .
We assume that $K$ is equally divided among all workers, which implies that the local batch size is $m = \frac{K}{w'_{\rho}} = \frac{K}{w}$.
As a result, the training speed is equal to $\frac{1}{t_m}$ and can be modeled as~\cite{Peng18:Eurosys}:
\begin{align*}
     f(p,w) &=\bigg(\frac{K}{w}    t_f+t_{b}+2\frac{g/p}{B/w}+\beta_1w+\beta_2p\bigg)^{-1}.
    % T_f+\eta_1T_{b}+2\eta_2\frac{S/p}{B/w}+\beta_1w+\beta_2p)
\end{align*}
%where $\boldsymbol{\theta}$ are coefficients for each job.\\

{\em 2) Asynchronous Training:} In asynchronous training, once the PSs receive gradients from a worker, they immediately update their parameters.
The expected number of steps completed by each worker in one time unit is $\frac{1}{t_m}$.
Thus, we can estimate the total number of steps performed by all workers in one time unit as $\frac{w}{t_m}$.
Since $w'_\rho$ is proportional to the number of workers $w$ (i.e., the number of concurrent workers that send gradients to the same PS also increases as $w$ increases), we have $w'_\rho = \alpha w$ for some $\alpha\in(0,1)$. 
Hence, the training speed function can be modeled as~\cite{Peng18:Eurosys}:
\begin{align*}
    f(p,w) &=
     w \bigg(mt_f+t_{b}+2\alpha\frac{g/p}{B/w}+\beta_1w+\beta_2 p \bigg)^{-1}.
    %  w/(mT_f+\eta_1T_{b}+2\eta_2\alpha\frac{S/p}{B/w}+\beta_1w+\beta_2 p)
\end{align*}

%%%%%%%%%%%%%%%%%%%%%%%%%%%%%%%%%%%%%%%%%%%%%%%%%%
\subsection{Generalization to Advanced Training Models} \label{sec:advanced_models}

%As mentioned earlier, the sequential model underutilizes the communication channels.
In this subsection, we introduce two more advanced training models, namely, i) the wait-free model~\cite{Zhang17:Poseidon} and ii) the priority-based model~\cite{Jayarajan19:SysML}, which overlap communication and computation to further reduce per-iteration delay. 
We remark that, although both models were not proposed by us, we are the {\em first} to quantitatively characterize the training speed for both models.
Interestingly, it turns out that our training speed modeling for the sequential model can be generalized to these two more advanced communication-computation models (proofs of Lemmas~\ref{lemma:wait} and \ref{lemma:priority} are omitted due to space limitation).\footnote{All missing proofs can be found in our Tech Report~\cite{Yu21:SMD_Tech_Report}.}
%In what follows, we first describe the basic ideas of both models and then characterize their training speeds.
% \begin{figure}[ht]
%     \centering
%     \includegraphics[width=0.35\textwidth]{Figures/wait-free-based.eps}
%     \caption{Wait-free based model.} \label{fig:waitfree}
% \end{figure}

\smallskip
{\em 1) Wait-free model~\cite{Zhang17:Poseidon}:}
As shown in Fig.~\ref{fig:waitfree}, starting from the final layer, each layer takes turn to send gradients to the PSs immediately after finishing its own BP and the completion of gradient pushing of its subsequent layer.
Compared to the sequential model, a key feature in this model is that part of the communication and computation can be conducted {\em simultaneously}.
%(see the concurrent BP and communication of $L_4$ and $L_3$ in Fig.~\ref{fig:waitfree}).
We  derive per-sample training time $t$ as follows:
%Note also that the example in Fig.~\ref{fig:waitfree} employs full-duplex communications.

%For the communication and computation pattern in the wait-free model, 
%For the wait-free model, we  derive the per-iteration training time $t$ as follows:
\begin{lem}[Wait-free model]
\label{lemma:wait}
For the wait-free model, the start time of gradient-sending $\kappa_j=\max\{\sum_{k=j}^Nb_k,\kappa_{j+1}+r_{j+1}\}$, for $j=1,\hdots,N\!-\!1$; otherwise it is $b_N$.
The start time of parameter-receiving $s_j=\max\{\kappa_j+r_j,s_{j+1}+r_{j+1}\}$, for $j=1,\hdots,N\!-\!1$; otherwise it is $b_N+r_N$.
The FP start time $\tau_j=\tau_{j-1}+f_{j-1}$, for $j=2,\hdots,N$; otherwise it is $s_1+r_1$.
%For the wait-free model, the start time of gradient-sending $\kappa_j$, the start time of parameter-receiving $s_j$, and the FP start time $\tau_j$ can be computed as:
%\begin{equation}
%    \label{wait_free_kappa}\kappa_j=\begin{cases}
%         b_N,&j=N,\\
%         \max\{\sum_{k=j}^Nb_k,\kappa_{j+1}+r_{j+1}\}, & 1\leq j<N,
%    \end{cases}
%\end{equation}
%\begin{equation}
%    \label{wait_free_s}s_j=\begin{cases}
%         b_N+r_N,&j=N, \\
%         \max\{\kappa_j+r_j,s_{j+1}+r_{j+1}\}, & 1\leq j<N,
%    \end{cases}
%\end{equation}
%\begin{equation}
%    \label{wait_free_tau}\tau_j=\begin{cases}
%         s_1+r_1,&j=1,\\
%        %  \max\{t_j+r_j,\tau_{j-1}+f_{j-1}\}, & 2\leq j\leq N
%        \tau_{j-1}+f_{j-1}, & 1< j\leq N.
%    \end{cases}
%\end{equation}
It thus follows that the per-sample training time is $t\!=\!\tau_N\!+\!f_N$.
\end{lem}

%The proof of Lemma~\ref{lemma:wait} can be found in Appendix~\ref{appdx:lem_wait}.
Some important remarks for Lemma~\ref{lemma:wait} are in order:
1) If the parameter size is skewed in the wait-free model, layers of larger sizes can introduce larger delay to layers of smaller sizes due to non-preemption.
2) In the wait-free model, only after receiving the updated parameters for the first layer, the FP of the next iteration can get started (see $\tau_1$ in Fig.~\ref{fig:waitfree}); thus the gradient sending of the subsequent layers causes delays to the gradient sending of the initial layer, which in turn induces delay for the next iteration.
Also, since the BP progresses in the reverse order of layers (i.e., from the last (output) layer to the first (input) layer), the gradients are also generated and sent in that order.
As a result, no overlap between computation and communication is possible during FP (see Fig.~\ref{fig:waitfree}).
%The priority-based model introduced next overcomes this limitation.

% {\color{red}Note, however, that since the BP progresses in the reverse order of layers (i.e., from the last (output) layer to the first (input) layer), the gradients are also generated and sent in that order, which makes the network idle during FP.}

% \begin{figure}[t]
%     \centering
%     \includegraphics[width=0.3\textwidth]{Figures/priority-based.eps}
%     \caption{Priority based model.} \label{fig:priority}
% \end{figure}

\smallskip
{\em 2) Priority-based model~\cite{Jayarajan19:SysML}:}
An insight from the discussions above is that the closer a layer is to the input layer, the higher priority its gradient/parameter communication should have.
The reason is that once the transmission of this layer is completed, the corresponding FP of this layer can start without waiting for all communications to be finished, thus significantly reducing delay.
As shown in Fig.~\ref{fig:priority}, the layer with a smaller index (i.e., closer to the input layer) always has a higher priority and its communication can preempt that of layers with larger indices.
%For example, in Fig.~\ref{fig:priority}, $L_3$ preempts $L_4$, $L_2$ preempts $L_3$ and $L_4$, and so on.
%This is exactly the basic idea of the priority-based model~\cite{Jayarajan19:SysML}: one should preempt all layers with larger indices once a layer with a smaller layer index finishes its BP.
This induces further communication-computation overlapping and helps the next iteration get started as soon as possible.

% \begin{figure}[h]
%     \centering
%     \includegraphics[width=0.3\textwidth]{Figures/slicing.eps}
%     \caption{Zoom-in view of slicing.}\label{fig:slicing}
% \end{figure}

Also, to better align communication and computation, the idea of ``parameter slicing'' can be used to mitigate the impact of skewed layer sizes.
% As shown in Fig.~\ref{fig:slicing}(d), we split each layer into smaller slices of size $\nu$ and these slices can further do concurrent communication.
% Fig.~\ref{}
As shown in Fig.~\ref{fig:slicing}, 
%the red blocks (which indicate overlapped push and pull), 
we split each layer into smaller slices of size $\varphi$, whose communications can be further overlapped.
%, which can further reduce the training time of each layer. 
%since we allow the concurrent sending gradients and receiving parameters.
Slices of the same layer have the same priority and the communication order of each slice within the same layer could be arbitrary.

We let $e_j[i]$ be the end time of communication for layer $j$ of job $i$.
%Note that $e_1=\sum_{k=1}^{N}b_k + r_1+ \nu$ and $\tau_1=e_1$.
%Based on the communication-computation pattern of the priority-based model,
We derive the per-sample training time $t$ as follows:

\begin{lem}[Priority-based model]\label{lemma:priority}
For the priority-based model, the communication end time $e_j=\sum_{k=2}^jr_k - \sum_{k=1}^{j-1}b_k + \underset{1\leq k\leq j-1}{\max} e_k$, if $\sum_{k=2}^j r_k > \sum_{k=1}^{j-1}b_k$; otherwise $e_j = 0$ for $j=2,\ldots,N$ and $e_1=\sum_{k=1}^{N}b_k + r_1+ \varphi$.
The FP start time $\tau_j=\max\{\tau_{j-1} + f_{j-1}, e_j\}$ for $j=2,\ldots,N$ and $\tau_1 = e_1$.
%and the FP start time $\tau_j$ can be computed as:
%\begin{align}
%&    \label{priority_e}e_j = \begin{cases}
%        0,& \sum_{k=2}^j r_k \leq \sum_{k=1}^{j-1}b_k, \\
%        \underset{1\leq k\leq j-1}{\max} e_k+\sum\limits_{k=2}^jr_k - \sum\limits_{k=1}^{j-1}b_k,& \text{otherwise}, 
%    \end{cases} \\
%& \label{priority_tau} \tau_j = \max\{\tau_{j-1} + f_{j-1}, e_j\}.       
%\end{align}
% \begin{equation}
%     \label{priority_tau}\tau_j=\begin{cases}
%          \tau_{j-1} +   f_{j-1},&e_j=0\\
%          \max\{\tau_{j-1} +   f_{j-1}, e_j\}, & e_j \neq    0
%         % \tau_{j-1}+f_{j-1}, & 1< j\leq N
%     \end{cases}
% \end{equation}
%\begin{equation}
%    \label{priority_tau}\tau_j=\max\{\tau_{j-1} +   f_{j-1}, e_j\},
%\end{equation}
It thus follows that the per-sample training time is $t = \tau_N + f_N$.
\end{lem}
%The proof of Lemma~\ref{lemma:priority} is relegated to Appendix~\ref{appdx:lem_priority}.
% {\color{red}Note that $e_j$ is not necessary the exect end communication time at layer $j$ (see Appendix~\ref{appdx:lem_priority})}.
% Several important remarks for Lemma~\ref{lemma:priority}:
% \begin{comment}
% \begin{enumerate}
%     \item Note that $e_j$ is just an intermediate variable to get our final training time, it is not necessarily the exact end communication time at layer $j$.
%     \item This model's advantages will become more noticeable when the parameter size is skewed.
% \end{enumerate}
% \end{comment}
% which can be estimated by experiments (e.g.,~\cite{Jayarajan19:SysML}).\\
% \subsection{Training time per iteration}
% At this point, we can have the training time for an iteration $t=\tau_N+f_N$, where $\tau_N$ can be replaced by Eqn.~(\ref{wait_free_tau}),(\ref{priority_tau}) for the wait-free based model and the priority based model, respectively.
%We can see from \eqref{priority_e} and \eqref{priority_tau} t
Note that $t$ has a recursive property.
Depending on whether the system is computation or communication dominant in each layer, Lemma~\ref{lemma:priority} could lead to different expressions.

\smallskip
{\em 3) Unified expression for per-iteration training time:}
Note that, in both wait-free and priority-based model, the per-sample training time $t$ is DNN-dependent and %the $s_j$,$\tau_j$,$e_j$ and $\kappa_j$ are all in recursive format.
determined by each layer's size.
%In other words, there is {\em no} simple closed-form expression for $t$ in the wait-free and the priority-based models.
Nonetheless, as long as the DNN of a training job is given, we can compute $t$ by using Lemmas~\ref{lemma:wait} and \ref{lemma:priority}.
%We can see that by utilizing the parameter synchronizations, we can reduce the training time significantly.
Also, we note that the communication-computation overlapping in wait-free and priority-based models effectively reduces the BP time, communication time, and FP time to certain fractions of those in the sequential model.
Thus, the per-iteration training time of both models can be expressed as $ t_m=\max_\rho \big[\eta_1m{t_f}+\eta_2t_{b}+2\eta_3\frac{g/p}{B/w'_\rho}+\beta_1 w+\beta_2 p\big]$,
%\begin{align}
%    % \label{time_per_iter} T&=\max_\rho[m{T_f}+\eta_1T_{b}+2\eta_2\frac{S/p}{B/w'_\rho}+\beta_1 w+\beta_2 p],
%    t_m&\!=\!\max_\rho \bigg[\eta_1m{t_f}+\eta_2t_{b}+2\eta_3\frac{g/p}{B/w'_\rho}+\beta_1 w+\beta_2 p\bigg],
%\end{align}
where the coefficients $\eta_1,\eta_2,\eta_3 \in(0,1]$ are DNN-model-dependent.

We let $H_f$, $H_b$ and $H_r$ be the FP time, BP time and communication time for processing a sample, respectively.
Then these coefficients are defined as $\eta_1\triangleq \frac{H_f}{\sum_{k=1}^Nf_k}$, $\eta_2\triangleq\frac{H_b}{\sum_{k=1}^Nb_k}$, and
$\eta_3\triangleq\frac{H_r}{2\sum_{k=1}^Nr_k}$.
%Specifically, these coefficients are defined as $\eta_1\triangleq \frac{H_f}{\sum_{k=1}^Nf_k}$, $\eta_2\triangleq\frac{H_b}{\sum_{k=1}^Nb_k}$, and
%$\eta_3\triangleq\frac{H_r}{2\sum_{k=1}^Nr_k}$.
% Note that $\eta_1,\eta_2,\eta_3$ are model-dependent.
For instance, for the wait-free instance in Fig.~\ref{fig:waitfree}, we can compute the coefficients as $\eta_1=\frac{\sum_{k=1}^4f_k}{\sum_{k=1}^4f_k}=1$, $\eta_2=\frac{b_4}{\sum_{k=1}^4b_k}$, and $ \eta_3=\frac{2r_4+r_3+r_2+r_1}{2\sum_{k=1}^4r_k}$.
%\begin{align*}
%    \eta_1=\frac{\sum_{k=1}^4f_k}{\sum_{k=1}^4f_k}=1, 
%    \eta_2=\frac{b_4}{\sum_{k=1}^4b_k}, 
%    \eta_3=\frac{2r_4+r_3+r_2+r_1}{2\sum_{k=1}^4r_k}.
%\end{align*}
% if we assume each layer has similar forward and backward computation time in DNN model, we can set $\eta_1=\frac{1}{4},\eta_3=\frac{1}{4}$ and $\eta_2=1$ for model~(\ref{eg}).\\
With this approach, the training speed function of a job under both synchronous training and asynchronous trainings can be generalized as:
\begin{align}
     \label{sync_speed_fn}f(p,w) &\!=\!1/(\eta_1\frac{K}{w}    t_f \!+\! \eta_2t_{b} \!+\! 2\eta_3\frac{g/p}{B/w} \!+\! \beta_1w \!+\! \beta_2p), \\
     \label{async_speed_fn} f(p,w) &\!=\!
     w/(\eta_1mt_f \!+\! \eta_2t_{b} \!+\! 2\eta_3\alpha\frac{g/p}{B/w} \!+\! \beta_1w \!+\! \beta_2 p).
    %  w/(mT_f+\eta_1T_{b}+2\eta_2\alpha\frac{S/p}{B/w}+\beta_1w+\beta_2 p)
\end{align}
Note that the sequential model is a special case with $\eta_1=1$, $\eta_2=1$ and $\eta_3=1$.
In the next section, we will see that these unified expressions in \eqref{sync_speed_fn} and \eqref{async_speed_fn} enable us to design a suite of approximation algorithms to solve Problem~\eqref{general_formulation}.
%Remarkably, these models can be unified to the same theoretical model.

%% file: Sec4_Algorithm/Sec4_Algorithm.tex
% !TEX root = ../ML_Networking_INFOCOM20.tex

\section{Solution Approach} \label{sec:algorithm}

%It can be seen from \eqref{general_formulation}--\eqref{resr_constr} that the resource scheduling problem is a {\em mixed-integer nonlinear programming} (MINLP) problem, which is NP-Hard~\cite{Freund01:GloPt}.
%Moreover, due to the sum-of-ratio structure in \eqref{sync_formulation} and \eqref{async_formulation}, the problem is {\em non-convex} even with continuous relaxation, which introduces yet another layer of difficulty.
Due to the fundamental hardness of Problem~\eqref{general_formulation} (to be shown soon), in this section, we will propose an approximation algorithmic approach, which we term \underline{s}um-of-ratio \underline{m}ulti-dimensional-knapsack \underline{d}ecomposition (SMD), to solve this problem.
In what follows, we will organize and present our approximation algorithmic approach in three main steps:

%In this section, we first provide some background of the sum-of-ratios problem under certain assumptions and MKP.
%Then we will introduce some scheduling algorithm and rounding technique to obtain the final scheduling with performance guarantee.

\smallskip
{\em Step 1) Sum-of-Ratios Multi-Dimensional-Knapsack Decomposition (SMD):}
First, we note that in Problem~\eqref{general_formulation}, the resource scheduling decision variables $(w[i],p[i])$ are independent across jobs due to the fact that each summand in \eqref{general_formulation} only depends on each job $i$.
Therefore, we can decompose the problem into an inner subproblem and an outer subproblem as follows.
%The two main steps to solve the Problem~(\ref{general_formulation}) includes designing the algorithm to resolve the maximization problem and then choosing active job set.
First, by setting $x_i=1,\forall i$, the inner resource allocation sub-problem for job $i$ can be written as: 
%(we again temporally remove the job index ``$[i]$'' and the subscript ``$i$'' in the inner subproblem for lighter notation since there is only one job involved here):
%\begin{align} \label{decomposed_formula}
%\hspace{-.1in}\begin{array}{l}
%\underset{w,p}{\text{Maximize}} \,\,\, \mu\Big(\frac{E}{f(p,w)}\Big) \\
%%
%\text{subject to }\  (O^r w + G^r p)\leq v^r,\quad\forall r\in\mathcal{R}, \\
%%
%\hspace{.6in} p\in\ \mathbb{Z}^{++}, w\in \mathbb{Z}^{++}.
%\end{array}
%\end{align}
\begin{align}
    & \label{decomposed_formula}\underset{w,p}{\text{Maximize }} &&\hspace{-.4in}\mu\Big(\frac{E}{f(p,w)}\Big)\\
    & \label{res_cap}\text{subject to } && \hspace{-.4in} (O^r w+G^r p)\leq v^r, \quad \forall r\in\mathcal{R}, \\
    \label{pos_int}& && \hspace{-.4in} p\in\mathbb{Z}^{++}, w\in \mathbb{Z}^{++}.
    % \nonumber&\text{Constraintef{resr_constr})-(\ref{int}),
\end{align}

%Note that maximize the utility is equivalent to minimize the completion time since the utility function is non-increasing w.r.t the completion time.
Recall that the training speed function $f(p,w)$ has different forms for synchronous and asynchronous trainings.
Hence, Problem~\eqref{decomposed_formula} can be further specialized as follows:

\smallskip
{\em a) Synchronous training:} In this case, we have:
% From the analytical model~(\ref{reformulation}), we can have the following formulation:
\begin{align}
    &\label{sync_formulation}\underset{w,p}{\text{Maximize }} &&\hspace{-.4in} \mu \bigg(\theta^1 w+\theta^2p
    +\theta^3+\frac{\theta^4w}{p}+\frac{\theta^5}{w} \bigg)\\
%    & \label{res_cap}\text{subject to } && \hspace{-.4in} (O^r w+G^r p)\leq v^r, \quad \forall r\in\mathcal{R}, \\
	\nonumber &\text{subject to } &&\hspace{-.4in} \text{Constraints}~(\ref{res_cap})-(\ref{pos_int}),
%
%    \label{pos_int}& && \hspace{-.4in} p\in\mathbb{Z}^{++}, w\in \mathbb{Z}^{++},
    % \nonumber&\text{Constraintef{resr_constr})-(\ref{int}),
\end{align}
where $\theta^1=E\beta_1$, $\theta^2=E\beta_2$, $\theta^3=E\eta_2t_b$, $\theta^4=2E\eta_3 g/B$, and $\theta^5=\eta_1 E K t_f$.

% where $\theta_i^1=E[i]\beta_i^1$, $\theta_i^2=E[i]\beta_i^2$, $\theta_i^3=E[i]\eta_i^1t_{back_i}$, $\theta_i^4=\frac{2E[i]\eta_i^2 S_i}{B[i]}$, and $\theta_j^5=E[i]K[i]t_{forward_i}$.

\smallskip
{\em b) Asynchronous training:} In this case, we have:
% From the analytical model~(\ref{reformulation}), we can have the following formulation:
\begin{align}
    \label{async_formulation}&\underset{w,p}{\text{Maximize}} && \hspace{-.6in} \mu \bigg(\theta'^1+\frac{\theta'^2p}{w} +\frac{\theta'^3}{w}+\frac{\theta'^4}{p} \bigg)\\
    \nonumber &\text{subject to } &&\hspace{-.6in} \text{Constraints}~(\ref{res_cap})-(\ref{pos_int}),
\end{align}
where $\theta'^1=E \beta_1$, $\theta'^2=E \beta_2$, $\theta'^3=E(\eta_1 m t_f +\eta_2 t_b)$, and $\theta'^4=2E \alpha \eta_3 g/B$.

It is clear that Problem~(\ref{decomposed_formula}) is a mixed-integer nonlinear programming (MINLP) problem, which is NP-Hard in general~\cite{Freund01:GloPt}.
In addition, even with continuous relaxation, it remains in the class of sum-of-ratios optimization problems, which is well-known to be NP-complete~\cite{Schaible77:SOR}.
%Overcoming these issues constitute the rest of the section.
But thanks to the low dimensionality of the inner subproblem (a consequence of SMD), we will propose an $\epsilon$-approximation algorithm for solving the inner subproblem.
%Firstly, we introduce the approximation algorithm to resolve the sum-of-ratios problem; then reduce the job selection to the multidimensional knapsack problem (MKP).
Upon solving \eqref{decomposed_formula}, the outer subproblem reduces to selecting active jobs to be run in the current scheduling interval:
\begin{align}
    \label{reformulation}\underset{\mathbf{x}}{\text{Maximize }} &\sum_{i\in\mathcal{I}}\mu_i \bigg(\frac{E[i]}{f(p[i],w[i])} \bigg)x_i\\
    \nonumber \text{subject to } &\text{Constraint~(\ref{selet})}, x_i\in\{0,1\}, \forall i\in\mathcal{I}.
\end{align}
Since $\mu_i (\frac{E[i]}{f(p[i],w[i])} )$ is known after solving the inner subproblem,
it is clear that the outer subproblem is a multi-dimensional knapsack problem (MKP) in essence.
Therefore, in what follows, we will consider solving the inner sum-of-ratios subproblem and the outer MKP problem separately.

\smallskip
{\em Step 2) Solving the inner sum-of-ratios subproblem:}
We first consider the inner subproblems (\ref{sync_formulation}) and (\ref{async_formulation}) after relaxing the integrality constraint \eqref{pos_int}.
Recall that the utility function $\mu_{i}(\cdot)$ is non-increasing.
Thus, both problems can be equivalently reformulated as a sum-of-ratios problem with affine constraints, which can be written in the following general form:
\begin{align}
    \label{P}\underset{\mathbf{x}}{\text{Minimize }}&\zeta(\mathbf{x})=\sum_{j\in\mathcal{J}}\frac{\mathbf{a}_j^{\top} \mathbf{x} +q_j}{\mathbf{c}_j^{\top} \mathbf{x} + d_j}=\sum_{j\in\mathcal{J}}\zeta_j(\mathbf{x})\\
    \text{subject to } &\mathbf{Ax}\leq \mathbf{C}, \mathbf{x}\geq \mathbf{0}, \nonumber
\end{align}
where $\mathbf{a}_j,\mathbf{c}_j\in\mathbb{R}^n,q_j,d_j\in\mathbb{R}$, $\forall j\in\mathcal{J}$, and $\mathbf{A}\in\mathbb{R}^{m\times n}, C\in\mathbb{R}^m$.
We also note that in both problems $\mathbf{a}_j^{\top} \mathbf{x}+q_j > 0, \mathbf{c}_j^{\top} \mathbf{x}+d_j > 0, \forall j\in\mathcal{J},\forall\mathbf{x}\in\Omega$,
where $\Omega=\{\mathbf{x}\in\mathbb{R}^n|\mathbf{Ax}\leq \mathbf{C},\mathbf{x}\geq \mathbf{0}\}$ denote the feasible domain of the problem.
The sum-of-ratios problem with affine constraints is known to be NP-complete~\cite{Schaible77:SOR} but $\epsilon$-approximation approach~\cite{Shen19:sum-of-ratios} exists. %~\cite{Shen19:sum-of-ratios}

%Our $\epsilon$-approximation algorithm design follows a similar token as in ~\cite{Shen19:sum-of-ratios} but with the following key difference:
Unlike the problem in~\cite{Shen19:sum-of-ratios} with covering constraints (may lead to unbounded search space), we exploit the special {\em packing-like} constraint structure in (\ref{res_cap}) to first obtain a tight upper bound for each ratio term to significantly reduce the search space.
%approximation algorithms exist.
%Here, we consider the $\epsilon$-approximation approach in ~\cite{Shen19:sum-of-ratios}. 
%In what follows, we first give a brief overview of this $\epsilon$-approximation approach.
Specifically, we let $\zeta_j(\mathbf{x})\triangleq \frac{\mathbf{a}_j^{\top} \mathbf{x} +q_j}{\mathbf{c}_j^{\top} \mathbf{x} + d_j}$, which is a linear fractional programming with non-empty and bounded feasible region $\Omega$.
We choose the lower and upper bounds as $l_j=\min_{\mathbf{x}\in\Omega} \zeta_j(\mathbf{x})$ and $\phi_j=\max_{\mathbf{x}\in\Omega} \zeta_j(\mathbf{x})$.
%\begin{align*}
%    &l_j=\min_{\mathbf{x}\in\Omega} \zeta_j(\mathbf{x})=\min_{\mathbf{x}\in\Omega}\frac{\mathbf{a}_j^{\top} \mathbf{x} +b_j}{\mathbf{c}_j^{\top} \mathbf{x} + d_j}, \\
%    &\phi_j=\max_{\mathbf{x}\in\Omega} \zeta_j(\mathbf{x})=\max_{\mathbf{x}\in\Omega}\frac{\mathbf{a}_j^{\top} \mathbf{x} +b_j}{\mathbf{c}_j^{\top} \mathbf{x} + d_j}.
%\end{align*}
%Note that $\zeta_j(\mathbf{x})$ is a linear fractional programming with non-empty and bounded feasible region $\Omega$. 
% To calculate $l_j,\phi_j$, note that the feasible region $\Omega$ is non-empty and bounded for each summand.
Then transform the problem into a linear program using the Charnes-Cooper transformation~\cite{Chames62:FractionalProgramming}, which then can be solved efficiently.
%Problem~(\ref{P}) sums over $J$ terms, and we first reduce it to $J\!-\!1$ terms to reduce the complexity.
Without loss of generality, we assume that the last summand $J$ has the largest ratio between the upper and lower bounds, i.e., $J = \arg\max_{j\in\mathcal{J}}\Big\{\frac{\phi_j}{l_j}\Big\}$.
% In order to project the original Problem~(\ref{P}) with $J$-dimensional space onto the $J-1$-dimensional space to reduce the complexity, we next pick the term whose ratio between the upper bound and lower bound is maximal.
% Based on the definitions of $l_j$, $\phi_j$, without loss of generality, we assume that
%\begin{align*}
%    J = \arg\max_{j\in\mathcal{J}}\Big\{\frac{\phi_j}{l_j}\Big\}.
%\end{align*}
Then, the feasible domain for the $J\!-\!1$ summands is a polytope characterized as $ \mathcal{H}=[l_1,\phi_1]\times[l_2,\phi_2]\times\hdots\times[l_{J-1},\phi_{J-1}]$.
% and define a rectangle $\mathcal{H}$ as
%\begin{align*}
%  \mathcal{H}=[l_1,\phi_1]\times[l_2,\phi_2]\times\hdots\times[l_{J-1},\phi_{J-1}].  
%\end{align*}
We let $\chi\in\mathbb{R}^{J-1}$ and $z\in\mathbb{R}$. 
Then, Problem~(\ref{P}) can be transformed into the following equivalent formulation:
\begin{align}\label{EP}
 \underset{\mathbf{x}\in\Omega,z}{\text{Minimize }}&\sum_{j=1}^{J-1}\chi_j+z\\
    \text{subject to }
    &\zeta_j(\mathbf{x})\leq \chi_j, j=1,..,J-1,\nonumber\\
    &\zeta_J(\mathbf{x})=z,\nonumber\\
    &\chi=(\chi_1,..,\chi_{J-1})\in\mathcal{H}.\nonumber
    % &\mathbf{Ax}\leq \mathbf{C}, \mathbf{x}\geq \mathbf{0}, \nonumber   
\end{align}
% where $\mathbf{y}\in\mathbb{R}^{J-1}$ and $z\in\mathbb{R}$.
We can see from the reformulated Problem~(\ref{EP}) that, if a point $\chi\in\mathcal{H}$ is given, there is only one variable $z$ associated with the summand $\zeta_J(\mathbf{x})$ to be solved.
Thus, the complexity of the problem is reduced significantly.

% Compared to the original Problem~(\ref{P}), Problem~(\ref{EP}) has a relatively low-rank decomposition structure since the objective function only involves the term $\zeta_J(\mathbf{x})$ if we fix a point $\mathbf{y}\in\mathcal{H}$.
{\em Step 2.1): Determining the Set of Grid Points:} After finding the range for each summand in the objective function, we divide the polytope $\mathcal{H}$ into smaller polytopes to perform a grid search, where the granularity is controlled by a precision parameter $\epsilon$.
% We define the sets
We first find the largest integer number that does not exceed the upper bound $\phi_j$ of each summand when searching from the lower bound $l_j$, i.e., $\lambda_j=\arg\max\{n\in\mathbb{N}|l_j(1+\epsilon)^n\leq \phi_j\}, j=1,...,J-1$.
%\begin{align*}
% &\lambda_j=\arg\max\{n\in\mathbb{N}|l_j(1+\epsilon)^n\leq \phi_j\}, j=1,...,J-1.
%\end{align*}
Then, we have the grid points set for each summand as $\mathcal{Q}_j^\epsilon=\{l_j,l_j(1 +   \epsilon),\hdots,l_j(1+\epsilon)^{\lambda_j}\}, j=1,...,J-1$.
% \begin{align*}
% &\mathcal{Q}_j^\epsilon=\{l_j,l_j(1 +   \epsilon),\hdots,l_j(1+\epsilon)^{\lambda_j}\}, j=1,...,J-1.
%\end{align*}
Next, by searching all the $J-1$ summands, we can obtain the search grid set as $\mathcal{T}^\epsilon=\{(\nu_1,\nu_2,\hdots,\nu_{J-1})|\nu_j\in\mathcal{Q}_j^\epsilon,j=1,...,J-1\}$.
%\begin{align*}
%\mathcal{T}^\epsilon=\{(\nu_1,\nu_2,\hdots,\nu_{J-1})|\nu_j\in\mathcal{Q}_j^\epsilon,j=1,...,J-1\}.
%\end{align*}
It is clear that for any $(\chi_1,\chi_2,\hdots,\chi_{J-1})\in\mathcal{H}$, we can always find a point $(\nu_1,\nu_2,..,\nu_{J-1})\in\mathcal{T}^\epsilon$, such that $\chi_j\in[\nu_j,(1+\epsilon)\nu_j], j=1,...,J-1$, thus $\mathcal{H}$ can be approximated by the set $\mathcal{T}^\epsilon$.

Hence, Problem~(\ref{EP}) can be solved by iterating over $\mathbf{\nu}\in\mathcal{T}^\epsilon$. 
For a given $\mathbf{\nu}$, the subproblem needs to be solved is as follows:
% solving Problem~(\ref{EP}) can be transformed into solving a series of subproblems, which is defined as follows for each $\mathbf{\nu}\in\mathcal{T}^\epsilon$:
\begin{align}\label{Q(v)}
  \underset{\mathbf{x}\in\Omega,z}{\text{Minimize }}&\Psi(\mathbf{\nu})=\sum_{j=1}^{J-1}\nu_j+z\\
  \text{subject to }
%   &\mathbf{Ax}\leq\mathbf{C}, \mathbf{x}\geq 0,\nonumber\\
  &\zeta_j(\mathbf{x})\leq \nu_j, j=1,...,J-1,\nonumber\\
  &\zeta_J(\mathbf{x})=z.\nonumber
%   &\mathbf{x}\in\Omega\nonumber.
\end{align}

% For given $\nu\in\mathcal{T}^\epsilon$, Problem~(\ref{Q(v)}) can be solved using the following equivalent formulation:
Notice that for a given $\nu\in\mathcal{T}^\epsilon$, the term $\sum_{j=1}^{J-1}\nu_j$ becomes a constant, then the equivalent formulation to Problem~(\ref{Q(v)}) is:
\begin{align}\label{P(v)}
 \underset{\mathbf{x}\in\Omega}{\text{Minimize }}&\zeta_J(\mathbf{x})=\frac{\mathbf{a}_J^{\top} \mathbf{x} +q_j}{\mathbf{c}_J^{\top} \mathbf{x} + d_J}\\
 \text{subject to }
 &\zeta_j(\mathbf{x})\leq \nu_j, j=1,..,J-1,\nonumber
\end{align}
which again can be transformed into a linear program using the Charnes-Cooper transformation~\cite{Chames62:FractionalProgramming} and solved efficiently.

The basic idea of the algorithm is first to perform the dimensionality reduction to reduce $J$ summands to $J-1$ terms, 
%project the original Problem~(\ref{P}) onto the $J-1$-dimensional space, 
and then divide the feasible domain into smaller nonuniform grids.
%By finding the upper and lower bound of each summand to exploit the search grids, the original sum-of-ratios problem can be transformed and decomposed into a series of linear programming
%subproblems, in which each subproblem is connected with a grid point. 
The feasible polytope domain obtained by finding the lower and upper bound of each summand is used to confine the space of grid points.
Then, the original sum-of-ratios problem can be transformed and decomposed into a set of linear programming (LP) subproblems, each of which is associated with a grid point.
As a result, the computational cost boils down to solving LP subproblems related to points in $\mathcal{T}^{\epsilon}$.
%, {\color{red} and the granularity is controlled by $\epsilon$}. 

Note, however, that the total number of grid points still increases exponentially as the number of summands in \eqref{P} increases, which is intractable as the problem size gets large.
Fortunately, we note that both inner problems~(\ref{sync_formulation}) and (\ref{async_formulation}) have just a few summands (four and three, respectively) thanks to SMD.
Thus, it remains affordable to adopt a grid-search-based approach.
%The above observation reveals an interesting insight: the {\em physical nature} of the computing cluster that each job has its reserved resources actually helps decouple the resource constraints and reduce the difficulty of the problem.
By leveraging the special feature of cloud systems that each job has its reserved resources, we can reduce the high dimensionality of the problem and decompose it as in Problem~(\ref{decomposed_formula}).
Thus, the problem is reduced to solving $I$ times of sum-of-linear-ratios with a small number of terms, which can be solved efficiently (Algorithm~\ref{alg:sum-of-ratios}). %(four in Sync-SGD and three in Async-SGD)
%We summarize the procedure in Algorithm~\ref{alg:sum-of-ratios}.

\begin{algorithm}[t!]
% \label{alg:sum-of-ratios}
\SetAlgoLined
% \KwResult{The $\epsilon$-approximation fractional solution for each job $i$. }
\textbf{Initialization:} Set $\tilde{L}=+\infty$.
Let $\epsilon\in(0,1)$,
$\tilde{\mathbf{x}}=\emptyset$\label{line:intialization}\;
Obtain the set $\mathcal{T}^\epsilon$ as described in Step~2.1)\label{T}\;
%  $\zeta_j(\mathbf{x}) \triangleq   \frac{\mathbf{a}_j^{\top} x+b_j}{\mathbf{c}_j^{\top} x+d_j}$,
%   $l_j=\underset{\mathbf{x}}{\min} \zeta_j(\mathbf{x})$,
%  $\phi_j=\underset{\mathbf{x}}{\max} \zeta_j(\mathbf{x})$,
% %  Define $H=[l_1,u_1]\times[l_2,u_2]\times\hdots[l_J,u_J]$\;
%  set $\lambda_j=\arg\max\{n \in \mathbb{N}|l_j(1+\epsilon)^n\leq \phi_j\},\forall j \in \mathcal{J}$\;
%  \label{line:gridset} Define the set $\mathcal{Q}_j^\epsilon=\{l_j,l_j(1 +   \epsilon),\hdots,l_j(1 +   \epsilon)^{\lambda_j}\}$ to obtain the search grid: $\mathcal{T}^\epsilon=\{(y_1,y_2,\hdots,y_J)|y_j \in \mathcal{Q}_j^\epsilon,\forall j \in \mathcal{J}\}$\;
 \label{line:iterate}\For{$\mathbf{\nu}\in\mathcal{T}^\epsilon$}{
  Solve Problem~(\ref{Q(v)}) to obtain the solution $\mathbf{x}^\nu$ with the objective value $\Psi(\mathbf{\nu})=\sum_{j=1}^{J-1}\nu_j+\zeta_J(\mathbf{x}^\nu)$\;
  \label{line:if}\If{$\Psi(\mathbf{\nu})<\tilde{L}$}{
   $\label{line:L}\tilde{L}=\Psi(\mathbf{\nu})$, $\tilde{\mathbf{x}}=\mathbf{x}^\nu$\;
%   $\tilde{\mathbf{x}}=\mathbf{x}^\nu$\label{line:x}\;
   }
%   {
%   instructions3\;
%   }
  }
  \Return $\tilde{\mathbf{x}}$\;
 \caption{$\epsilon$-Approximation for the Continuous Relaxation of the Inner Subproblems (\ref{sync_formulation}) and (\ref{async_formulation}).}
 \label{alg:sum-of-ratios}
\end{algorithm}

\begin{algorithm}[t!]
\SetAlgoLined
 Pick some $\delta\in(0,1]$. Pick some integer $F\geq1$. Let $cnt\leftarrow 0$.
Let $\mathbf{x}' = M_\delta\bar{\mathbf{x}}$ for some $0<M_\delta\leq1$ ($M_\delta$ signifies its dependence on $\delta$ and is to be specified)\label{line:rounding_init}\;
Randomly round $\mathbf{x}'$ to $\hat{\mathbf{x}} \in\mathbb{Z}_{+}^{n}$ as:
$\hat{x}_{j} =\lceil x'_{j} \rceil$ w.p. $x'_j - \lfloor x'_j \rfloor$ and $\hat{x}_{j} = \lfloor x'_j \rfloor$ w.p. $\lceil \bar{x}_j \rceil - \bar{x}_j$, otherwise\label{line:rounding_scheme}\;
%If $\hat{\mathbf{x}}$ is infeasible, go to line~\ref{line:rounding_scheme}\;
 If $\hat{\mathbf{x}}$ is infeasible or $cnt\!<\!F$, then $cnt\!\!\leftarrow\!\!cnt\!+\!1$, go to \ref{line:rounding_scheme}.
\caption{Randomized Rounding Scheme.}
\label{alg:randomizied_rounding}
\end{algorithm}

% To calculate $l_j,\phi_j$ in Line~\ref{line:intialization}, note that the feasible region is non-empty and bounded for each summand.
% Thus, $\zeta_j(\mathbf{x})$ can be transformed into a linear program using the Charnes-Cooper transformation~\cite{Chames62:FractionalProgramming}.
 %\vspace{-0.05in}

In Algorithm~\ref{alg:sum-of-ratios}, we first obtain the grid points over the feasible domain in Line~\ref{T}.
We then iterate each point and update the objective $\tilde{L}$ and the fractional solution $\tilde{\mathbf{x}}$ in Line~\ref{line:L} if the current objective value is smaller.
Finally, we return the solution with the smallest objective value.
Following similar analysis as in \cite{Shen19:sum-of-ratios} (hence omitted for brevity), we can show that Algorithm~\ref{alg:sum-of-ratios} is an $\epsilon$-approximation.

%\textbf{Rounding technique: } 
Upon solving the continuous relaxation of inner sum-of-ratios subproblems (\ref{sync_formulation}) and (\ref{async_formulation}), 
%using Algorithm~\ref{alg:sum-of-ratios},
it remains to obtain an integer solution to calculate the utility.
This is still an NP-Hard integer programming problem with generalized packing-type constraints in (\ref{res_cap}).
%(i.e., integer-valued variables rather than 0-1 variables) in (\ref{resr_constr}).
%Thanks again to the low dimensionality by SMD and the monotonicity of the utility function, 
We propose the following solution approach:
First, we solve the continuous relaxation of $\min \Big\{\sum_{l=1}^L \frac{\mathbf{a}_l^{\top}\mathbf{x}}{\mathbf{d}_l^{\top}\mathbf{x}}+\mathbf{c}^{\top} \mathbf{x} :\mathbf{Bx} \leq \mathbf{b}, \mathbf{x} \in\mathbb{Z}_{+}^{n} \Big \}$, where $\mathbf{B} \in\mathbb{R}_{+}^{r\times n}$,
% $\mathbf{a}  \in  \mathbb{R}_{+}^{m}$,
$\mathbf{b} \in\mathbb{R}_{+}^{r}$, and $\mathbf{a}_l,\mathbf{d}_l, \mathbf{c} \in\mathbb{R}_{+}^{n},\forall l$.
Let $\bar{\mathbf{x}}$ be the obtained fractional optimal solution.
Then, we propose a randomized rounding scheme as shown in Alg.~\ref{alg:randomizied_rounding} to round $\bar{\x}$ and arrive at an integer solution.
%by Algorithm~\ref{alg:sum-of-ratios}.
% Consider the randomized rounding scheme: 
% Let $\mathbf{x}' = G\bar{\mathbf{x}}$ for some $0 < G \leq  1$ (to be specified later).
% Randomly round $\mathbf{x}'$ to $\hat{\mathbf{x}}  \in  \mathbb{Z}_{+}^{n}$ as:
% $\hat{x}_{j} = \lceil x'_{j} \rceil$ w.p. $x'_j  -    \lfloor x'_j \rfloor$ and $\hat{x}_{j} = \lfloor x'_j \rfloor$ otherwise. 
% \begin{comment}
% % First, we solve the linear programming relaxation of Problem~(\ref{decomposed_formula}).
% Let $\{\bar{w}_i, \bar{p}_i, \forall i \}$ be the fractional solution, and then we randomly round $\{ \bar{w}_i, \bar{p}_i, \forall i \}$ to generate an integer solution: 
% \begin{align}
% \label{eqn_rounding_w} w[i] &= \begin{cases}
% \lceil \bar{w}_i \rceil, & \hspace{-.1in} \text{with probability } \bar{w}_i - \lfloor \bar{w}_i \rfloor,\\
% \lfloor \bar{w}_i \rfloor, & \hspace{-.1in} \text{with probability } \lceil \bar{w}_i \rceil - \bar{w}_i,
% \end{cases} \!\!\!\\
% %
% \label{eqn_rounding_p} p[i] &= \begin{cases}
% \lceil \bar{p}_i \rceil, & \hspace{-.1in} \text{with probability } \bar{p}_i - \lfloor \bar{p}_i \rfloor,\\
% \lfloor \bar{p}_i \rfloor, & \hspace{-.1in} \text{with probability } \lceil \bar{p}_i \rceil - \bar{p}_i.
% \end{cases}
% \end{align}
% \end{comment}
%We will later prove in Theorem~\ref{thm_Alg4} that Algorithm~\ref{alg:randomizied_rounding} enjoys an approximation ratio guarantee that is {\em independent} on the problem size.

% \subsection{Multidimensional Knapsack Problem}

\smallskip
{\em Step 3) Solving the outer MKP subproblem:}
%In order to select jobs to be processed in the current scheduling interval due to the resource capacity limit in the DL cluster, we reduce it to MKP.
The general formulation of the outer MKP subproblem can be written as:
\begin{align}
    \label{MKP}&\underset{\mathbf{x}}{\text{Maximize }} \underset{i\in\mathcal{I}}{\sum}u_ix_i\\
    % &\underset{i\in\mathcal{I}}{\sum}v_i^rx_i\leq C^r, \forall r\in\mathcal{R},\\
    % &0\leq x_i\leq 1, \forall i\in\mathcal{I},\\
    &\nonumber \text{subject to }
    % \hspace{-1.5in}
    \text{Constraints~(\ref{selet})}, x_i\in\{0,1\}, \forall i\in\mathcal{I},
\end{align}
where $u_i\triangleq\mu_i(\frac{E[i]}{f(p[i],w[i])})$ is the utility value of each job $i$. 
Our goal is to select jobs that maximize the total utility among all the submitted jobs $\mathcal{I}$.
The MKP problem is still a well-known NP-Hard problem~\cite{Chu98:MKP} but admits an $\epsilon$-approximation solution\footnote{This $\epsilon$-value is different and should not be confused with the $\epsilon$ in Step~2.} that runs in polynomial time if $\epsilon$ and $I$ are fixed.
%The problem has been proved by Korte and Schrader~\cite{Korte82:FAS} that no FPTAS (fully polynomial time approximation scheme, i.e., runtime polynomial in $1/\epsilon$) algorithm exists unless P = NP.
Here, we adopt the $\epsilon$-approximation scheme~\cite{FRIEZE84:MKP} to our problem setting. 
We let $T(\mathcal{S})=\{t\in\mathcal{I}\setminus S:u_t >\min(u_i:i\in\mathcal{S})\}$, for $\mathcal{S}\subset\mathcal{I}$.
 Let $LP(\mathcal{S})$ be the linear program obtained from Problem~(\ref{MKP}) by setting $x_i\!=\!1$, if $i \in \mathcal{S}$; and set it to $0$, if $i \!\in\! T(\mathcal{S})$.
 Let $\mathbf{x}^B(\mathcal{S})$ be an optimal basic feasible solution to $LP(\mathcal{S})$.
%as summarized in Algorithm~\ref{alg:MKP}.
The main idea %of Algorithm~\ref{alg:MKP}
of this approach is to solve $LP(\mathcal{S})$ for all $\mathcal{S}\!\subset\!\mathcal{I}$.
% in Line~\ref{line:LP(S)}.
Then, we round down the solution $\mathbf{x}^B(\mathcal{S})$.
%in Line~\ref{line:round}.
Finally, we return the best solution with the largest objective value.
The difference between $\mathbf{x}^B(\mathcal{S})$ and $\lfloor\mathbf{x}^B(\mathcal{S})\rfloor$ is small since $R$ is a fixed small number in our case.

\subsection{Performance Analysis} \label{subsec:performance_analysis}
We now examine the overall approximation ratio of our proposed algorithms.
Note that the key component in our algorithm is the proposed randomized rounding scheme in Algorithm~\ref{alg:randomizied_rounding} for Problem~(\ref{decomposed_formula}).
Thus, we first prove the following result for the randomized rounding algorithm (proofs are omitted due to space limitation).
%(see proof in Appendix~\ref{appdx:lem_ILP}):
\begin{lem}[Rounding] \label{lem_ILP}
% Let $W_{b} \! \triangleq   \! \min \{ b_{i}/[\mathbf{B}]_{ij}\!: [\mathbf{B}]_{ij} \! > \! 0 \}$.
% Let $\delta \in (0,1]$. 
% and define $G$ as:
% \begin{align*} 
% G \triangleq 1 + \frac{3\ln(2r/\delta)}{2W_b} -     \sqrt{ \bigg(\frac{3\ln(2r/\delta)}{2W_b} \bigg)^{2} + \frac{3\ln(2r/\delta)}{W_b} }. \!\!\!
% \end{align*}
Let $W_{b} \triangleq \min \{ b_{i}/[\mathbf{B}]_{ij}: [\mathbf{B}]_{ij} > 0 \}$.
Let L be the number of sum-of-ratios terms.
Pick some constant $\delta \in (0,1]$, and define $M_\delta$ as:
%in Eqn.~(\ref{eqn_L_def}).
\begin{align*}
M_\delta \triangleq 1 +    \frac{3\ln(2r/\delta)}{2W_b} -     \sqrt{ \bigg(\frac{3\ln(2r/\delta)}{2W_b} \bigg)^{2} +    \frac{3\ln(2r/\delta)}{W_b} }. \!\!\!
\end{align*}
With probability greater than
$1\! - \! \delta$, $\hat{\mathbf{x}}$ achieves a
cost at most $\frac{8L/M_\delta+4}{\delta}$ times the
cost of $\bar{\mathbf{x}}$, and
% $\hat{\mathbf{x}}$ satisfies $\mathrm{Pr}\{
% (\mathbf{B}\hat{\mathbf{x}})_{i}
% \frac{W_{b}}{b_{i}} > W_{b},\exists i \}\!
% \leq\! \frac{\delta}{2r}$.
$\mathrm{Pr}\{
(\mathbf{B}\hat{\mathbf{x}})_{i}
 \!>\! b_i,\exists i \}
\!\leq\! \frac{\delta}{2r}$.
\end{lem}

% Several remarks for Lemma~\ref{lem_ILP} as follows:

% 1) The theoretical approximation ratio $\frac{8L/M_\delta + 4}{\delta}$ is conservative. 
% %(due to the uses of Markov inequality and union bound in our proof).
% Our numerical studies show that the approximation ratio performance in reality is much smaller than $\frac{8L/M_\delta + 4}{\delta}$.
% 2) 
% The probability parameter $\delta$ controls the trade-off between approximation ratio and efficiency in finding a feasible rounding solution:
% A larger $\delta$ implies a smaller approximation ratio, but the probability of obtaining a feasible solution of this ratio is also smaller (i.e., more rounds of rounding needed).
% %This implies taking more rounds of rounding.
Note that $\delta$ is used for characterizing the randomized rounding algorithm's performance.
Lemma~\ref{lem_ILP} indicates that with probability $1-\delta$, one achieves an approximation ratio at most $\frac{8L/M_\delta+4}{\delta}$ with the stated probabilistic feasibility guarantee.
From the statement, we can see that the approximation ratio is ultimately determined by $\delta$, since $M_\delta$ increases as $\delta$ increases. 
%(cf. Eqn~(\ref{eqn_L_def})).
Thus, if one desires a better approximation ratio, then a larger $\delta$ should be picked.
% In other words, the approximation ratio is determined by $\delta$ ultimately.
% In other words, the statement means that the probability of getting a better approximation ratio is smaller under randomized rounding (i.e., better approximation ratio $\Rightarrow$ smaller $\frac{8L/G+4G}{\delta}$ $\Rightarrow$ larger $\delta$ $\Rightarrow$ smaller probability $1-\delta$).
%(the smaller the value of $\frac{3G}{\delta}$, the better the approximation ratio). 
That is, there exists a trade-off between the approximation ratio value and its achieving probability, both of which are quantified by $\delta$.
% Interestingly, for $\delta=1$, Lemma~\ref{lem_ILP} indicates that there is still {\em non-zero} probability to achieve an approximation ratio not exceeding $8L/M_\delta + 4$.
% 3) 
%The results in Lemma~\ref{lem_ILP} are in fact applicable for general linear sum-of-ratios problem with packing constraints.
%Hence, these results could be of independent theoretical interest.

The approximation ratio of our algorithm is the worst-case upper bound of the ratio between the overall utility of admitted jobs obtained by the optimal solution of Problem~(\ref{general_formulation}) and the total utility achieved by Algorithm~SMD in the overall time horizon.
By specializing Lemma~\ref{lem_ILP} with parameters in Problem~(\ref{decomposed_formula}), we have the
following result for Algorithm~\ref{alg:randomizied_rounding}:
\begin{thm}[Approximation Ratio of Rounding in Alg.~\ref{alg:randomizied_rounding}]\label{thm_Alg4}
% Let $W_{2} \! \triangleq   \! \min\{
% \hat{C}_{r}/O^r[i],
% \hat{C}_{r}/G^r[i], \forall i,r \}$.
% Let $\delta \in (0,1]$. 
% Define $\epsilon \triangleq   \frac{3}{W_2}ln\frac{2r}{\delta}$
% {\small
% \begin{align*}
% &G \triangleq 1 + \frac{3W_3}{2W_2}-\sqrt{ \bigg(\frac{3W_3}{2W_2} \bigg)^{2} \! +   \! \frac{3W_3}{W_2} }.
% \end{align*}
% }%
% Let $W_{1} \triangleq \min\{
% v^{r}[i]/O^r[i],
% v^{r}[i]/G^r[i], \forall i,r \}$.
Let $\delta$ be selected as in Lemma~\ref{lem_ILP}, and let $M_\delta$ be defined as in Lemma~\ref{lem_ILP}.
%Eqn~(\ref{eqn_L_def}).
With probability greater than $1 -   \delta$, Algorithm \ref{alg:randomizied_rounding} obtains a schedule $\{ w[i], p[i], \forall i \}$ that has an approximation ratio at most $\frac{24/M_\delta+4}{\delta}$ with $\mathrm{Pr}\{LHS (\ref{res_cap})  >  v^r[i]\} \leq \frac{\delta}{8}$.
\end{thm}
%Theorem~\ref{thm_Alg4} follows directly from Lemma~\ref{lem_ILP}, and we omit the proof for brevity.

Let $\epsilon_1$ and  $\epsilon_2$ be the performance ratios of solving the sum-of-ratios problem and MKP, respectively, where $\epsilon_1, \epsilon_2 \in (0,1)$.
%Let $\delta$ be as defined in Theorem~\ref{thm_Alg4}.
Let $\delta$ and $M_\delta$ be defined as in Lemma~\ref{lem_ILP}.
Let $\tau_i$ be the completion time of job $i$ and $\tau_i^*$ be the optimal completion time of job $i$.
%We use $\mu \triangleq   \max_{i}\{\frac{\mu_i(\tau_i^*)}{\mu_i((1+\epsilon_1)\tau_i^*)}\}$ to denote the upper bound of the utility gap.
%We use $\mu'_i \triangleq   \frac{\mu_i(\tau_i^*)}{\mu_i(\tau_i^*(1+\epsilon_1)/\delta)}$ to denote the utility gap between the optimal utility and the utility obtained by our algorithm.
Let $F$ be chosen as in Alg.~\ref{alg:randomizied_rounding}.
We let $\mu^*\triangleq \max_i\{\mu_i(\tau_i^*)\}$, and let $\mu'\triangleq\min_i\{\mu_i\big(\tau_i^*(1+\epsilon_1)(24/M_\delta+4)/\delta\big)\}$.
Following Theorem~\ref{thm_Alg4}, we can establish the overall approximation ratio and running time complexity of the SMD approach as follows:
\begin{thm}[Overall Approximation Ratio of SMD]
\label{ratio}
%Let $\delta$ be defined as in Lemma~\ref{lem_ILP}.
% Let $W_1$ be defined as in Theorem~\ref{thm_Alg4}.
With probability greater than $(1 -   (\delta/8)^F)^I$, the proposed SMD-based method returns a feasible solution with $\frac{\mu' (1-\epsilon_2)}{\mu^*}$-approximation performance guarantee. 
%algorithm, with $\mathrm{Pr}\{LHS (\ref{res_cap})\}  >  v^r[i]\} \leq \frac{\delta}{8}$.}
\end{thm}

\begin{thm}[Polynomial Running Time]
\label{runningtime} 
%Let $\delta$ be defined as in Lemma~\ref{lem_ILP}.
% Let $W_1$ be defined as in Theorem~\ref{thm_Alg4}.
%According to Lemmas~\ref{lem:sum-of-ratios} and \ref{lem:MKP}, we can have the overall time complexity of Algorithm~SMD as follows:
%% We denote $T(R,I)$ as a polynomial upper bound of the time to evaluate the objective function at some point.
%% We let $\iota\triangleq\sum_{r=1}^R\sum_{i=1}^{I}l(v^r[i])  +    \sum_{r=1}^Rl(C^r)  +    \sum_{i=1}^{I}l(\mu_i(\frac{E[i]}{f(p[i],w[i])}))$ with $l(\xi)\triangleq[\log_2(\xi+1)]$.
%% With probability greater than $1 -   \delta$,
%% the average time complexity of our proposed algorithm~SMD is 
%% $O(\frac{I}{\delta}\frac{T(R,I)}{\epsilon_2^4}\iota\log \iota\log\log \iota)$, with $\mathrm{Pr}\{LHS (\ref{res_cap})\}  >  W_1, \exists i\} \leq \frac{\delta}{8}$.
%% We let $T(R+4,3)$ be the time taken to solve a linear program in $3$ variables and $R+4$ constraints.
%% Let $\omega=\min\{\theta^1w,\theta^2p,\theta^4w,\theta^5,\theta^6,\theta^\}$
%% We let $\iota\triangleq\sum_{r=1}^R\sum_{i=1}^{I}l(v^r[i])  +    \sum_{r=1}^Rl(C^r)  +    \sum_{i=1}^{I}l(\mu_i(\frac{E[i]}{f(p[i],w[i])}))$ with $l(\xi)\triangleq[\log_2(\xi+1)]$.
%with probability greater than $1 -   \delta$,
%the average time complexity of our proposed algorithm~SMD is $O(\frac{I}{\delta}T_1T_2)$
%% $O(\frac{I}{\delta}\frac{T(R,I)}{\epsilon_2^4}\iota\log \iota\log\log \iota)$,
%with $\mathrm{Pr}\{LHS (\ref{res_cap})\} > v^r[i]\} \leq \frac{\delta}{8}$.
We let $T_i^s$ and $T_i^a$ be the time complexity for solving the sum-of-ratios problem under synchronous and asynchronous training of job $i$, respectively, which can be solved in polynomial time~\cite{Shen19:sum-of-ratios}.
Let  $T_i=\max\{T_i^s,T_i^a\}$.
Let $T_2$ be the time complexity for MKP problem, which is polynomial~\cite{FRIEZE84:MKP}.
%According to Lemmas~\ref{lem:sum-of-ratios} and \ref{lem:MKP}, 
Then, the overall time complexity of Algorithm~SMD is $O(\sum_{i\in\mathcal{I}}(T_i+F)+T_2)$.

\end{thm}

\begin{figure*}[t!]
   % \hspace{0.005\linewidth}
       \begin{minipage}[t]{0.24\linewidth}
        \centering
        \includegraphics[width=1.05\textwidth]{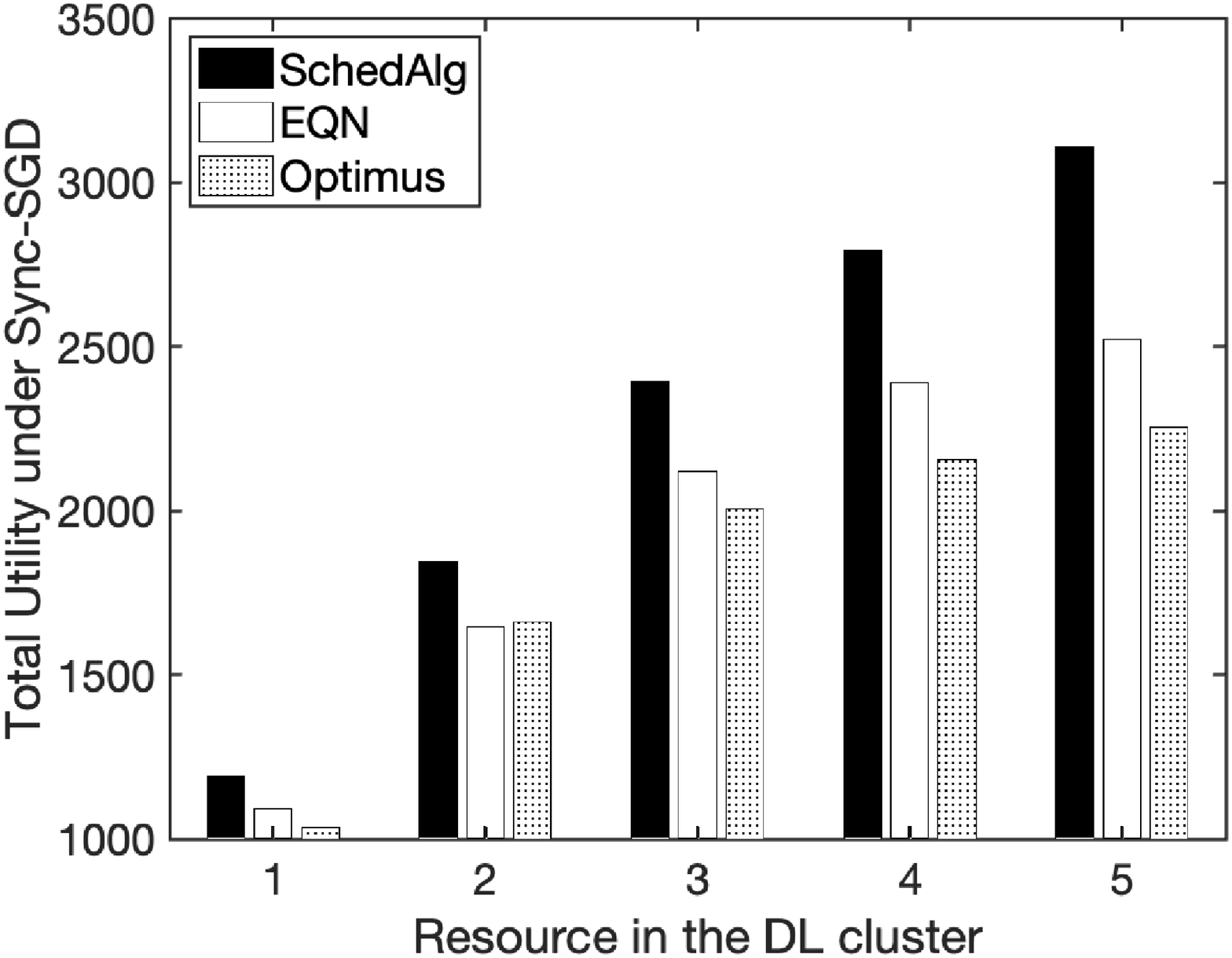}
        %\captionsetup{width=0.75\textwidth}       
        \caption{Total utility vs. cluster resources (Async-SGD).}\label{fig:rsr_async} 
        % with unit resource (cpu:3400,gpu:600,memory:1400GB,storage:1200GB)
    \end{minipage}%
    \hspace{0.006\linewidth}
    \begin{minipage}[t]{0.24\linewidth}
        \centering
        \includegraphics[width=1.04\textwidth]{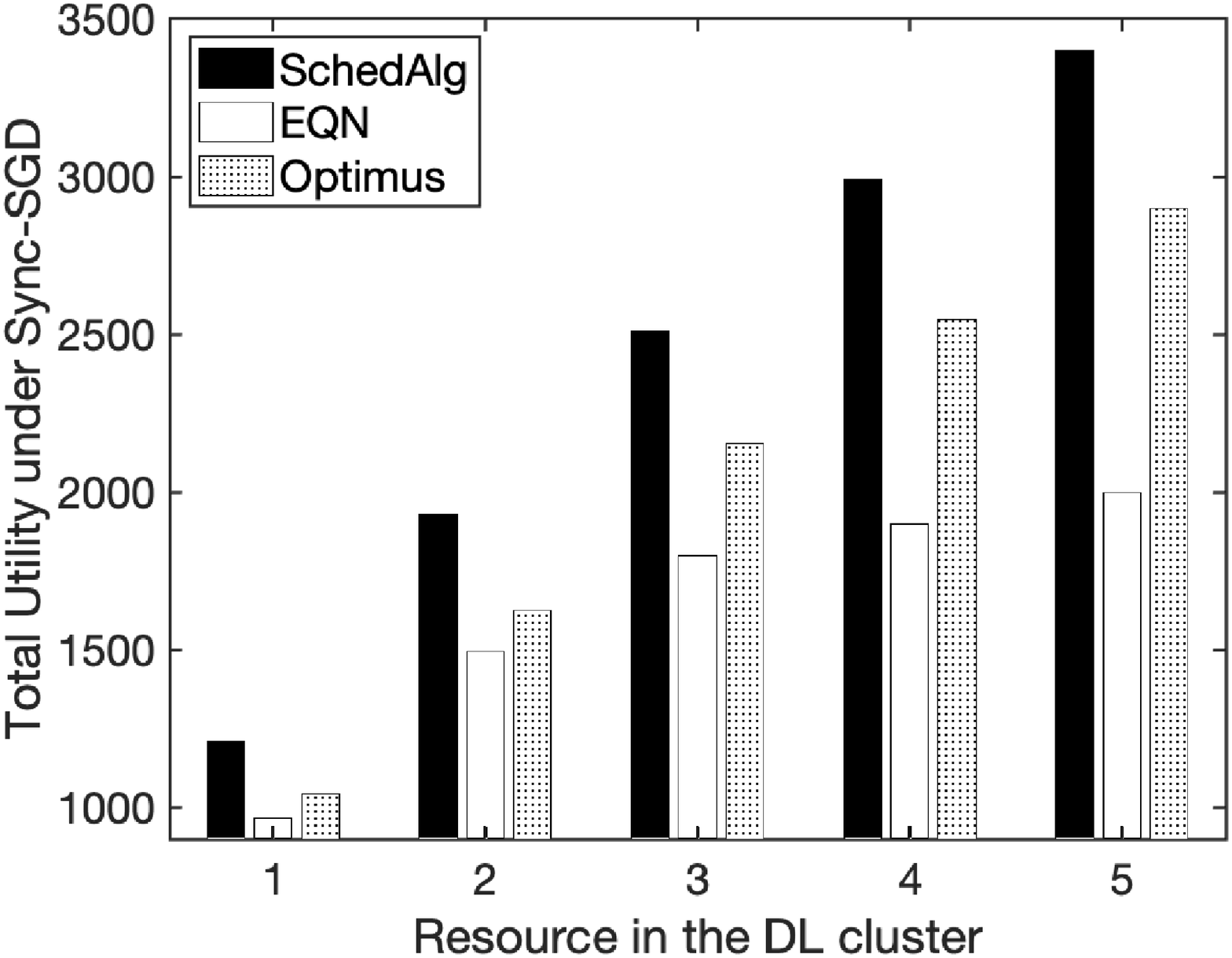}      
        \caption{Total utility vs. cluster resources (Sync-SGD).} \label{fig:rsr_sync}
    \end{minipage}%  
      \hspace{0.01\textwidth}
         \begin{minipage}[t]{0.24\linewidth}
        \centering
        \includegraphics[width=1.04\textwidth]{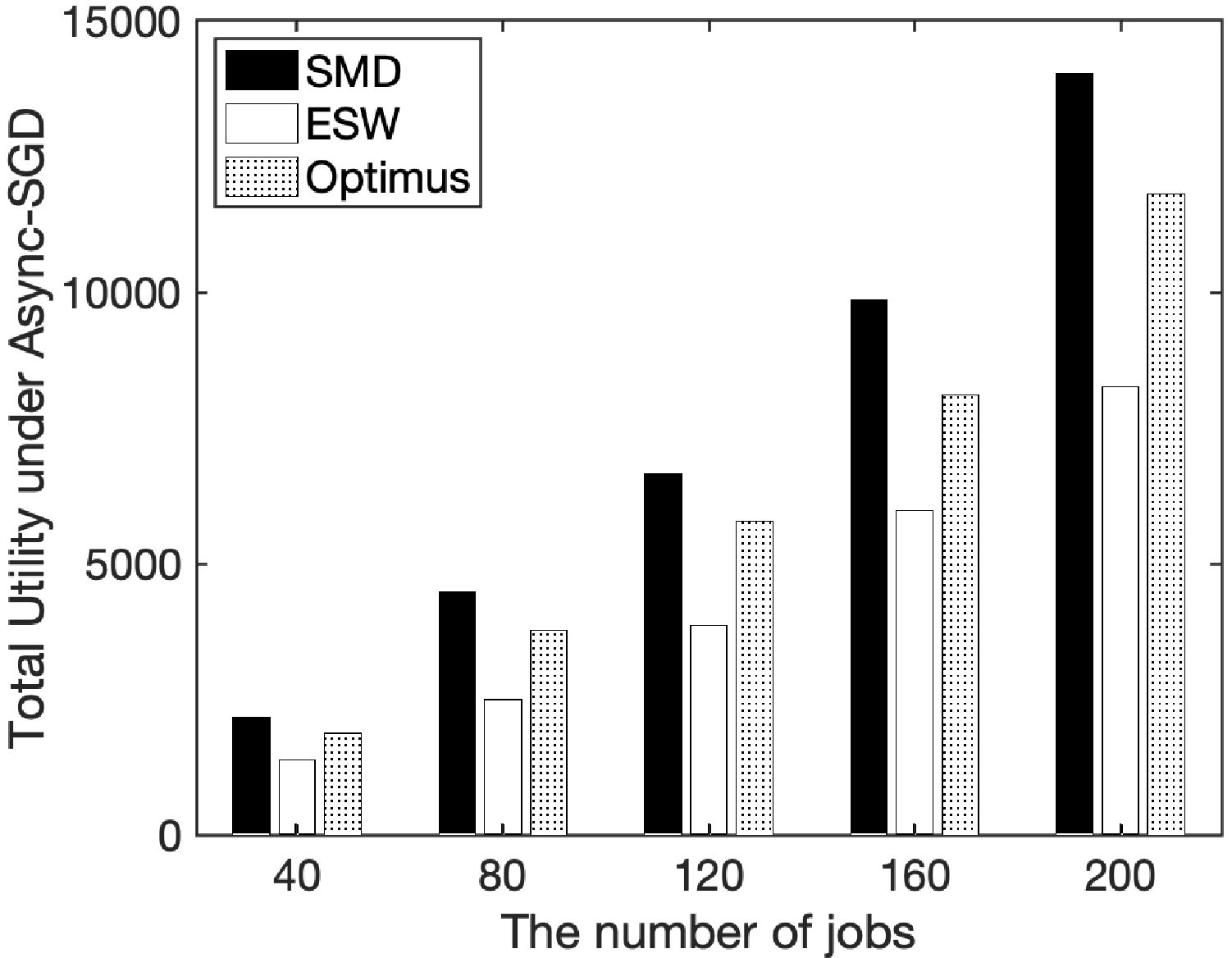}
        \caption{Total utility vs. number of jobs (Async-SGD).} \label{fig:job_async}
    \end{minipage}%
           \hspace{0.007\textwidth}
    \begin{minipage}[t]{0.24\linewidth}
        \centering
        \includegraphics[width=1.01\textwidth]{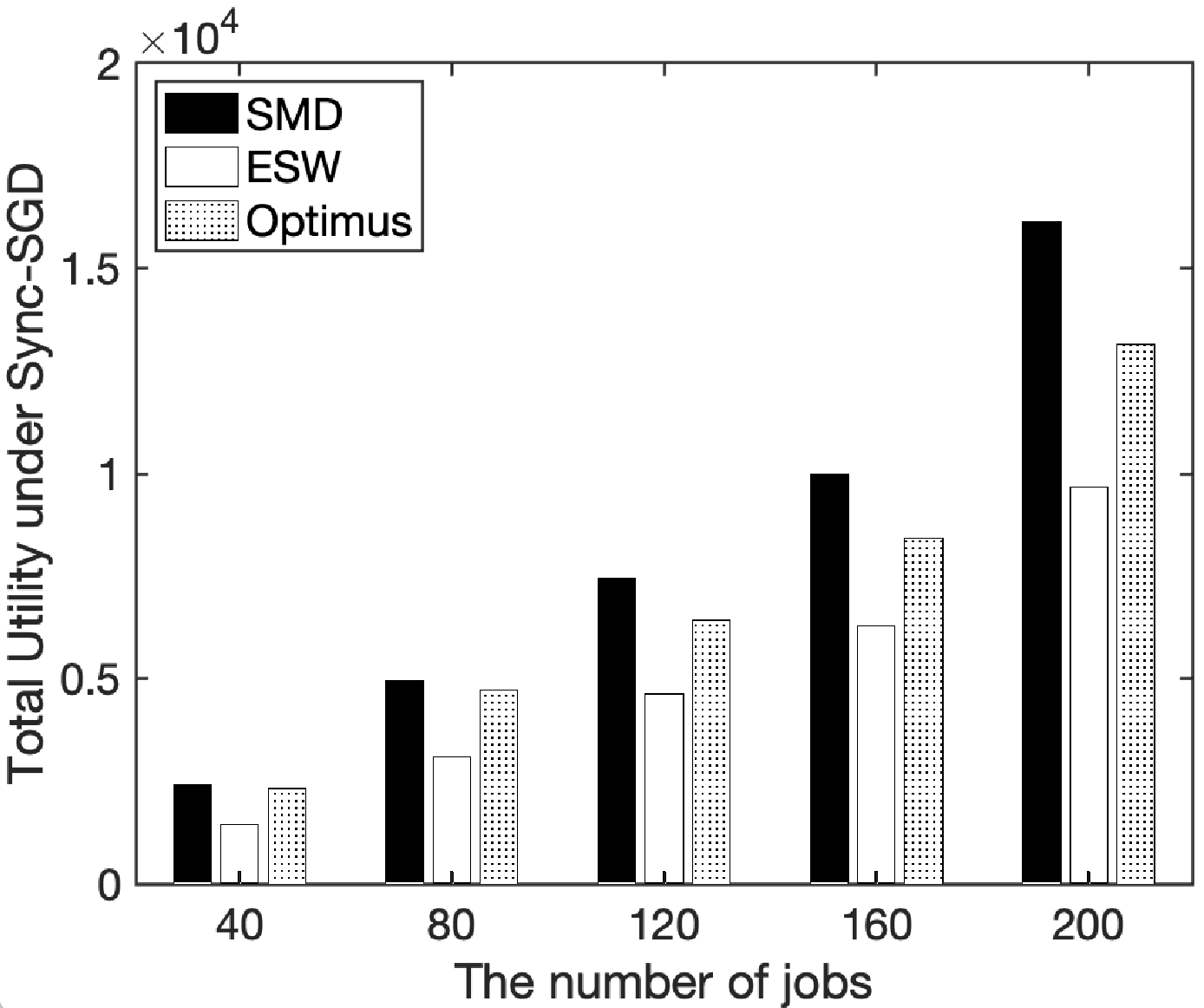}
        \caption{Total utility vs. number of jobs (Sync-SGD).}\label{fig:job_sync}
    \end{minipage}%
\vspace{-.2in}
\end{figure*}

\begin{figure}[!t]
\centering
\begin{minipage}{.24\textwidth}
   \centering
  \includegraphics[height=0.15\textheight]{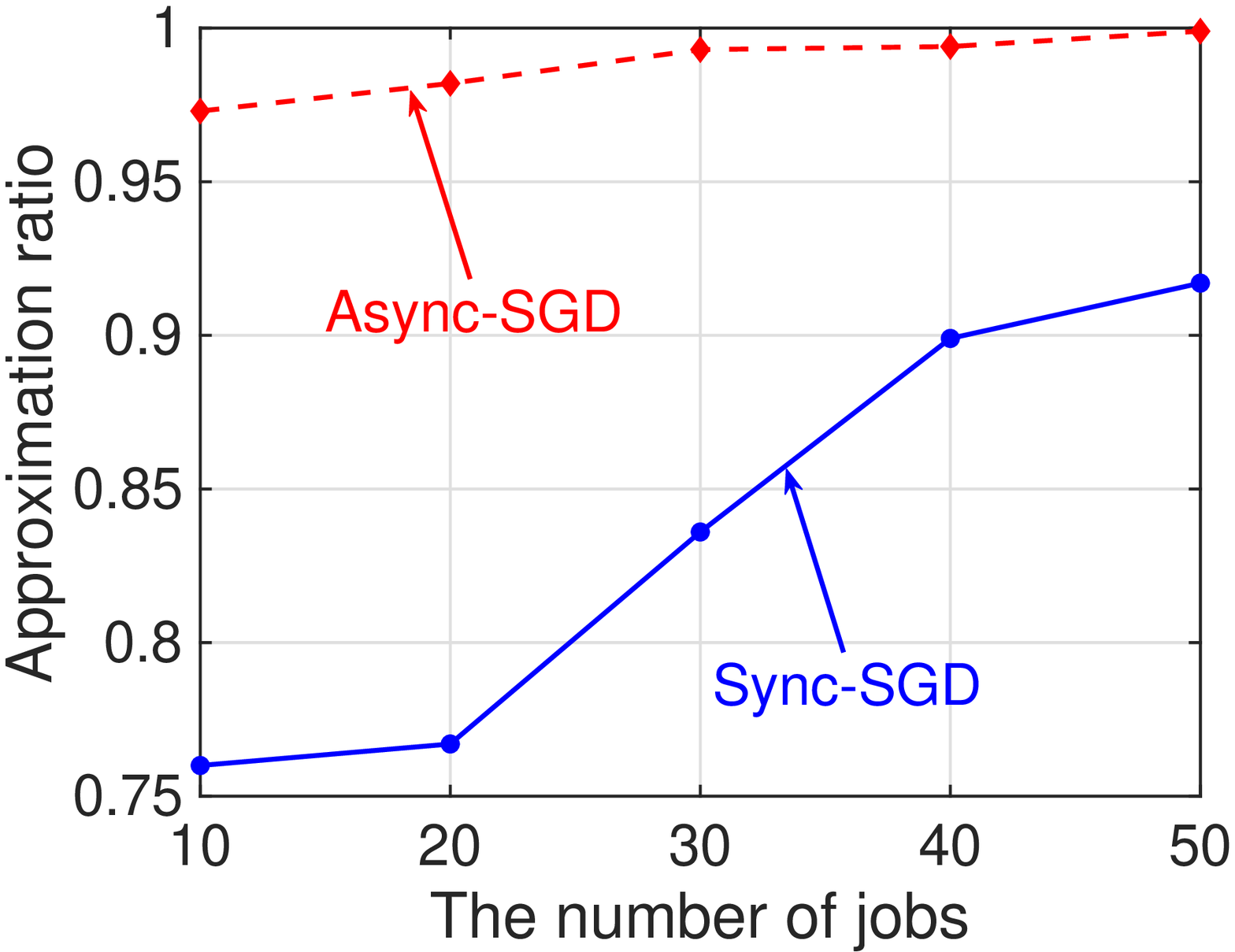}
   \vspace{-0.16in}
  \caption{The comparison of approximation ratios.}
  \label{fig:ratio}
\end{minipage}%
\hspace{0.01\columnwidth}%
\begin{minipage}{.24\textwidth}
   \centering
  \includegraphics[height=0.15\textheight]{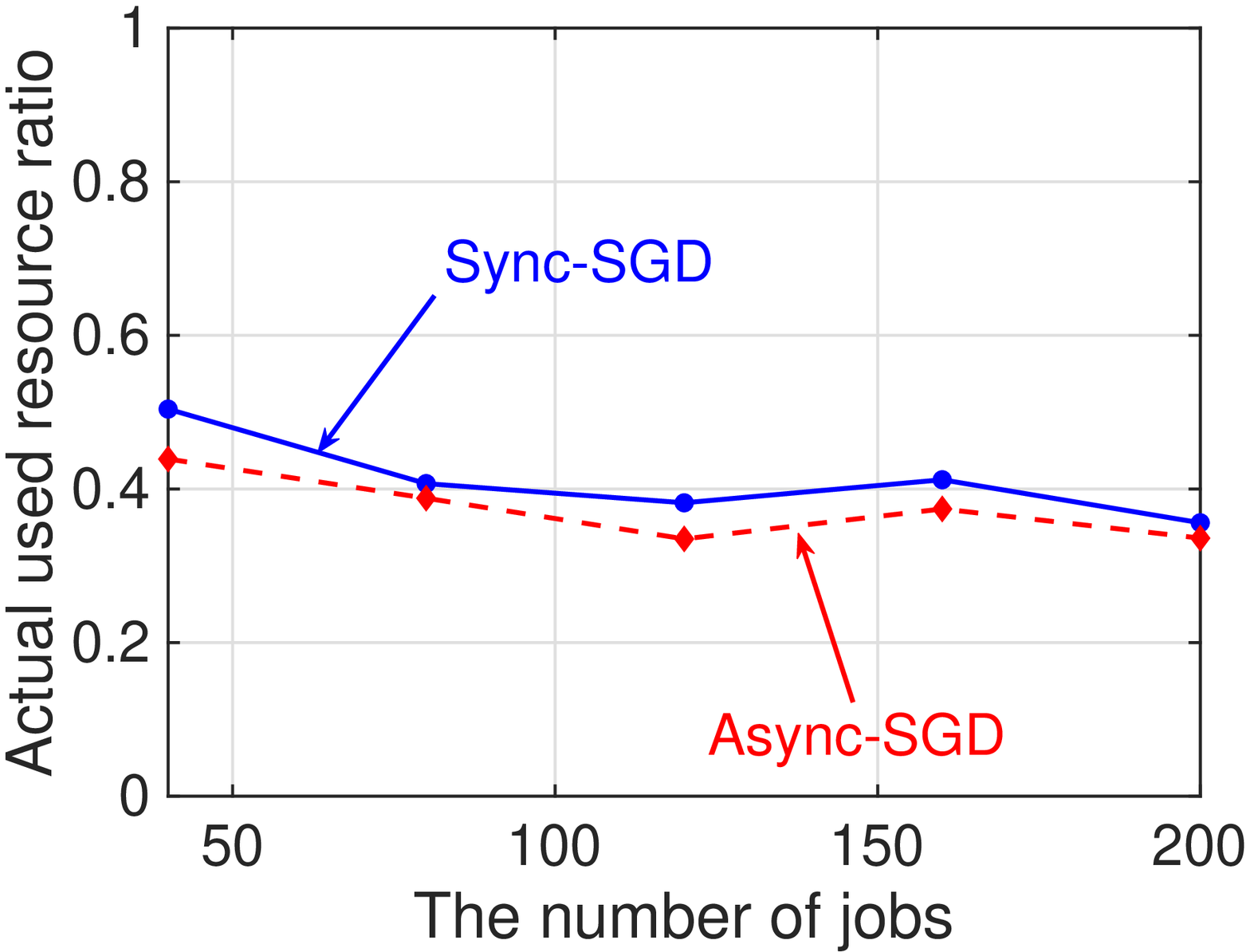}
%   \vspace{0.02in}
  \caption{The actual used resources ratios.}
  \label{fig:actualusedresource}
\end{minipage}
\vspace{-.2in}
\end{figure}

%% file: Sec5_Numerical/Sec5_Numerical.tex
% !TEX root = ../ML_Networking_INFOCOM20.tex

\section{Numerical Evaluation}
\label{sec:numerical}
% The simulator uses the
% following from the traces collected from our testbed experiments:
% training losses of each kind of jobs, training speeds under different
% resource configurations, resource capacities of each server, job
% configurations (e.g., resource requirements of workers/parameter
% servers), DL model details (e.g., parameter size).
We conduct simulation studies to evaluate the efficacy of our proposed algorithms.
In our simulation, the computing cluster follows the real-world system in\cite{Iandola16:FireCaffe} with job parameters generated uniformly at random from the following intervals: 
$E[i] \in [50,200]$, $g[i] \in [30,575]$ MB, $m[i] \in [10,100]$,
$K[i] \in [1,100] \cdot m[i]$, $N[i] \in [10,100]$.
% and $B[i] \in [5,20]Gbps$ 
We consider four types of resources: GPU, CPU, memory and storage.
For fair comparisons, we use similar settings as in~\cite{Li14:OSDI}~\cite{Chilimbi14:OSDI}~\cite{Sun17:Dorm} and set resource demands of each worker as: 0--4 GPUs, 1--10 vCPUs, 2--32 GB memory, and 5--10GB storage.
We set resource configuration of each PS as: 1--10 vCPUs, 2--32GB memory and 5-10GB storage.
We set bandwidth capacity of each PS to $B[i] \in [5,20]$ Gbps.
Capacities of virtual instances to run workers/PSs are set according to resource configuration of Amazon EC2 C4 instances. We set the resource limit of each job to $\vartheta$ times of the resource limit of each instance with $\vartheta \in [1,20]$ (according to the reality that each user is limited to a maximum of 20 instances per region in Amazon EC2).
% We also set the DL cluster resource capacity to 50 times of the total resource of EC2 instances.
We set $b_j[i] \in [1,300]$ ms, $f_j[i] \in [1,500]$ ms and $r_j[i] \in [80,500]$ ms following the traces collected from the experiments in
 Optimus~\cite{Peng18:Eurosys} based on Google cluster trace~\cite{Googletrace15:Github},
% {\color{red}where they used Google cluster workload traces\cite{???} 
 which include jobs' training losses, training speeds with various resource configuration, each server's resource capacities, job configuration such as requirements of workers/PSs, as well as DL model specifications like parameter size.
%training losses of each kind of jobs, training speeds under different
%resource configurations, resource capacities of each server, job
%configurations (e.g., resource requirements of workers/parameter
%servers), DL model details (e.g., parameter size).
Then, we set $t_b[i]=\sum_{j=1}^{N[i]}b_j[i]$ and $t_f[i]=\sum_{j=1}^{N[i]}f_j[i]$.
We set $\beta_1[i] \in [3,4]$, $\beta_2[i] \in [0,0.01]$ and $\alpha[i] \in [0,1]$ following the tested values from Optimus~\cite{Peng18:Eurosys}.
We use a Sigmoid utility function ~\cite{Huang15:CORA,Peng18:Eurosys}, $\mu_i(\pi_i)=\frac{\gamma_1}{1+e^{\gamma_2}(\pi_i-\gamma_3)}$ with $\gamma_1 \in [1,100]$, $\gamma_2 \in [4,6]$ and $\gamma_3 \in [1,15]$.
Note that this range of $\gamma_2$ corresponds to time-critical jobs ~\cite{Bao18:ML_INFOCOM}.

We first compare our SMD algorithm with two baseline resource allocation policies: (1) ESW (setting the ratio of number of workers to number of PSs to 1:1~\cite{Kubernetes17:blog} for each job); 
and (2) Optimus~\cite{Peng18:Eurosys} (compare the utility gain by adding one more worker and one more PS and choose the one with larger utility gain).
Since Optimus estimates the training speed function based on an online learning approach by monitoring the convergence rate, we use our own speed function to estimate the utility for each job.
We set $\epsilon_1=\epsilon_2=0.01$, $M_\delta=1$, and $I=50$.
To study how the total utility changes as the computing cluster resource capacity increases, we set one unit of resources as follows: ($\text{vCPU} = 3400$, $\text{GPU} = 600$, $\text{Memory} = 1400 \text{ GB}$, $\text{Storage}=1200 \text{ GB}$),
and vary the resource capacity using 1-5 times of the unit resource.
The comparison results under both Sync-SGD and Async-SGD are shown in Figs.~\ref{fig:rsr_async}--\ref{fig:job_sync}.
% We increase the resource capacity in the DL cluster by 1-5 times as indicated in the $x$ axis in Figs.~\ref{fig:rsr_async} and \ref{fig:rsr_sync}.
We can see that SMD significantly outperforms other policies and the gains in total utility becomes more pronounced as the number of jobs and resources in the computing cluster increases.

Next, we examine the approximation ratio of SMD.
We evaluate the performance in terms of the ratio between the total utility obtained by our algorithm and the optimal total utility.
The optimal utility is computed by enumerating all the possible combinations of numbers of workers and PSs for each job, and the combination with the largest utility will be returned.
We vary the number of jobs per scheduling interval from $10$ to $50$.
We also set the cluster resource capacity as $1000$ times of that of a virtual instance. %the total resource in Amazon EC2 C4 instances.
The results are shown in Fig.~\ref{fig:ratio}.
We can see that the ratio is much better than the theoretical bound and becomes larger as the number of jobs increases, which implies that our algorithm is scalable.
%{\color{red}
Further, Sync-SGD has a worse approximation ratio since it is more sensitive to the changes of numbers of workers and PSs based on Eqn.~(\ref{sync_formulation}) due to the linear term $\theta^1w+\theta^2p$.
In other words, the error introduced from the "grid search" and randomized rounding when solving the inner sum-of-ratios-subproblem could lead to more utility loss compared to asynchronous training.
%}
Recall that the randomized rounding scheme is the key of our proposed Algorithm~SMD.
The packing constraints~(\ref{res_cap}) are easier to satisfy with a smaller $M_\delta$.
%{\color{red}In our experiments, if the total rounds of randomized rounding before we find an integer feasible solution exceeds a pre-set threshold (e.g, $1000$), we will discard the corresponding job.}
Theorem~\ref{thm_Alg4} suggests that there is a trade-off: if we set $M_\delta$ to be close to one to pursue a better total utility result, the rounding time could be large to obtain a feasible solution.
As $M_\delta$ becomes larger, the probability of violating the packing constraints increases, meaning that we need to have more rounding attempts to obtain an integer feasible solution.
%xx \textbf{explain why by pointing to earlier section} .
However, according to our numerical experiences, if the machine's resource capacity is relatively large compared to the jobs' resource demands per worker/PS, the number of rounding attempts is small and not sensitive to $M_\delta$.

Lastly, we examine the actual used resources in our algorithm.
A key feature of sum-of-ratios problems is that optimality is not necessarily obtained when the resource capacity constraints are binding. 
In other words, 
%instead of allocating as more numbers of workers and parameter servers to the admitted jobs, 
compared to the number of workers and parameter servers, 
the ratio between the number of workers and parameter servers plays a more critical role to minimize the training completion time.
If such a better ratio can be found, it is possible that the system can save resources while having the same or even better performance in terms of the average training completion time.
Hence, compared to other resource allocation policies that use as much resources specified by the user as possible, our SMD method may use much less resources while achieving the optimal performance.
We let $\epsilon=0.01$ and vary the number of jobs per scheduling interval from $40$ to $200$.
We can see from Fig.~\ref{fig:actualusedresource} that the actual resources used is 30\%-50\% of that specified by the users.
%meaning that our algorithm actually use much less resources.
From the system's perspective, the unused resources can be released and allocated to other jobs. %, and as a result, more jobs can be admitted at the same time.

%% file: Sec6_Conclusion/Sec6_Conclusion.tex
% !TEX root = ../ML_Networking_INFOCOM20.tex

\section{Conclusions}
\label{sec:conclusion}
In this paper, we studied resource scheduling for DNN jobs in computing clusters.
%We consider the general setting that users submit their jobs to the cluster with resource limit specification.
We demonstrated that the problem can be formulated as a non-convex integer non-linear program with bin-packing constraints, which is NP-Hard.
We proposed an approximation scheduling algorithm based on a sum-of-ratios multi-dimensional knapsack (SMD) approach.
Specifically, we developed a performance model that considers the special layered structure of DNN under different parameter synchronization mechanisms.
Through careful investigation of the structure of the non-convex problem, we decomposed the problem based on SMD and proposed a suite of approximation techniques to solve the packing-type %constraints 
 integer program with performance guarantees. 
 Evaluation under realistic settings confirmed  superior performances of SMD over existing works.
 %{\color{red} 
DNN resource scheduling remains an under-explored area. 
Future extensions of this work may include, e.g., datacenter topology and communication contention among jobs.
 %}
%Collectively, our results fills the theoretical gap of non-convex optimization algorithm design for resource scheduling in DL clusters.